\documentclass[useAMS,usenatbib]{mn2e}

\usepackage{subfigure}
\usepackage{graphicx}
\usepackage{times}
\usepackage{tabularx}
\usepackage{natbib}
\usepackage{xcolor}
\usepackage{caption}
\usepackage{amsmath}
\usepackage{float}
\usepackage{multirow}

\newcommand{\enzo}{{\it {\small ENZO}}}
\newcommand{\dd}{\mathrm{d}}
\newcommand{\Mpc}{\mathrm{Mpc}}
\newcommand{\Msun}{\mathrm{M}_{\odot}}
\newcommand{\Mth}{\mathrm{M}_{200}}

\newcommand{\kpc}{\mathrm{kpc}}
\newcommand{\radio}{\mathrm{radio}}
\newcommand{\ph}{\mathrm{ph}}
\newcommand{\cm}{\mathrm{cm}}
\newcommand{\km}{\mathrm{km}}
\newcommand{\sek}{\mathrm{s}}
\newcommand{\keV}{\mathrm{keV}}
\newcommand{\MeV}{\mathrm{MeV}}
\newcommand{\GeV}{\mathrm{GeV}}

\newcommand{\TeV}{\mathrm{TeV}}
\newcommand{\G}{\mathrm{G}}
\newcommand{\K}{\mathrm{K}}
\newcommand{\Gyr}{\mathrm{Gyr}}

\newcommand{\all}{\mathrm{all}}
\newcommand{\Perp}{\mathrm{perp}}
\newcommand{\para}{\mathrm{para}}
\newcommand{\erg}{\mathrm{erg}}
\newcommand{\Hz}{\mathrm{Hz}}

\newcommand{\GHz}{\mathrm{GHz}}
\newcommand{\pre}{\mathrm{pre}}
\newcommand{\post}{\mathrm{post}}

\begin{document}
 \definecolor{myred}{rgb}{1,0,0}
 
 \title[Testing cosmic-ray acceleration with radio relics]{Testing cosmic-ray acceleration with radio relics: a high-resolution study using MHD and tracers}
 \author[D. Wittor, F. Vazza, M. Br\"{u}ggen]{D. Wittor$^{1}$\thanks{%
 E-mail: dwittor@hs.uni-hamburg.de}, F. Vazza$^{1}$, M. Br\"{u}ggen$^{1}$\\
 $^{1}$ Hamburger Sternwarte, Gojenbergsweg 112, 21029 Hamburg, Germany}
 \date{Accepted ???. Received ???; in original form ???}
 \maketitle

 \begin{abstract}
  Weak shocks in the intracluster medium may accelerate cosmic-ray protons and cosmic-ray electrons differently depending on the angle between the upstream magnetic field and the shock normal.  In this work, we investigate how shock obliquity affects the production of cosmic rays in high-resolution simulations of galaxy clusters.\\
  For this purpose, we performed a magneto-hydrodynamical simulation of a galaxy cluster using the mesh refinement code \enzo. We use Lagrangian tracers to follow the properties of the thermal gas, the cosmic rays and the magnetic fields over time. We tested a number of different acceleration scenarios by varying the obliquity-dependent acceleration efficiencies of protons and electrons, and by examining the resulting hadronic $\gamma$-ray and radio emission. \\
  We find that the radio emission does not change significantly if only quasi-perpendicular shocks are able to accelerate cosmic-ray electrons. Our analysis suggests that radio emitting electrons found in relics have been typically shocked many times before $z=0$. On the other hand, the hadronic $\gamma$-ray emission from clusters is found to decrease significantly if only quasi-parallel shocks are allowed to accelerate cosmic-ray protons. This might reduce the tension with the low upper limits on $\gamma$-ray emission from clusters set by the \textit{Fermi}-satellite.   
  
  \end{abstract}

 \label{firstpage}

 \section{Introduction}
 \label{sec:intro}
Galaxy clusters grow through the continuous accretion of matter and by merging with other clusters. In the process, shock waves and turbulent motions in the intracluster medium (ICM) can (re)accelerate cosmic rays \citep[see][and references therein]{Brunetti_Jones_2014_CR_in_GC}. Radio emission, observed as diffuse radio halos in the centre of clusters and as highly polarized radio relics at the cluster periphery, confirms the existence of cosmic-ray electrons \citep[e.g.][]{Ferrari_et_al_2007_obs_radio_em_clu}. However, cosmic-ray protons that would produce $\gamma$-rays or secondary electrons as a product of inelastic collisions with thermal protons, appear to be accelerated less efficiently than expected \citep[for example see][]{2014MNRAS.437.2291V, 2015MNRAS.451.2198V}. The \textit{Large Area Telescope} on board of the \textit{Fermi}-satellite \citep[from here on \textit{Fermi}-LAT, see][for a detailed description]{2009ApJ...697.1071A} has thoroughly searched for these $\gamma$-rays, yielding upper limits for the $\gamma$-ray flux above $500 \ \MeV$ is in the range of $0.5 - 22.2 \cdot 10^{-10} \ \mathrm{ph} \ \cm^{-2} \ \sek^{-1}$ \citep{2014ApJ78718A}. \citet{2013A&A...560A..64H} have analysed a collection of stacked \textit{Fermi}-LAT count maps and derived a flux upper limit of the order of a few $10^{-10} \ \mathrm{ph} \ \cm^{-2} \ \sek^{-1}$. Extended searches for $\gamma$-ray emission from the Coma cluster \citep[see][]{2016ApJ819149A} and the Virgo cluster \citep[see][]{2015ApJ812159A} have been performed. The limits for the $\gamma$-ray flux above $100 \ \MeV$ have been estimated to be $5.2 \cdot 10^{-9} \ \mathrm{ph} \ \cm^{-2} \ \sek^{-1}$ for Coma and  $1.2 \cdot 10^{-8} \ \mathrm{ph} \ \cm^{-2} \ \sek^{-1}$ for Virgo. Overall, these observations constrain the ratio of cosmic-ray to thermal pressure within the virial radius to be below a few percent. \\
 Recently, Particle-in-Cell (PIC) simulations have quantified how the acceleration efficiency varies with the Mach number and the obliquity, $\theta$, i.e. the angle between shock normal and upstream magnetic field vector (\citealt{Caprioli_Spitkovsky_2014_ion_accel_I_eff, Guo_eta_al_2014_I,Guo_eta_al_2014_II}). These studies have shown that cosmic-ray electrons have a higher acceleration efficiency in perpendicular shocks, while the acceleration of cosmic-ray protons is more efficient in parallel shocks (i.e $\theta < 50$, see Fig 3. in \citealt{Caprioli_Spitkovsky_2014_ion_accel_I_eff}). Protons are efficiently accelerated by diffusive shock acceleration (DSA) by crossing the shock multiple times where they are scattered off magneto-hydrodynamic waves in the up- and downstream region. Thermal electrons cannot be injected into DSA because their gyro-radius is too small compared to the thickness of the shock front, which is controlled by the gyro-radius of the protons. Therefore, electrons need to be pre-accelerated before they can be injected into the DSA cycle. Recent PIC simulations have shown that even in the case of the weak shocks typically found in galaxy clusters ($M<5$, where $M$ is the Mach number), electrons can be efficiently pre-accelerated by shock drift acceleration (SDA). In SDA electrons gain energy while drifting along magnetic field lines down the shock front (\citealt{Guo_eta_al_2014_I,Guo_eta_al_2014_II}). \\
In this paper, we investigate how linking the obliquity of shocks to the acceleration efficiency of cosmic rays may affect the radio and $\gamma$-ray emission in galaxy clusters. To this end, we developed a Lagrangian tracer code to track the injection and advection of cosmic rays in the cosmological simulations produced using the \enzo \ code. \\
 This paper is structured as follows. In Sec.~\ref{subsec:enzo} we describe our computational setup in \enzo. In Sec.~\ref{subsec:tracer} we introduce our Lagrangian tracer code. Our main results are presented in Sec.~\ref{sec:results}. First, we discuss the basic properties of our Lagrangian tracers in Sec.~\ref{subsec:GeneralResults}. General results on the distribution of shock obliquities are presented in Sec.~\ref{subsec:obliquity}. In Sec.~\ref{subsec:RadioEmission} and \ref{subsec:gammaEmission} we show how the radio and the $\gamma$-ray emission are changed by the modified acceleration efficiencies. A more detailed analysis of the simulated relics is presented in Sec.~\ref{subsec:closeup}. Finally, we summarize and discuss our results in Sec.~\ref{sec:summary}. Additional tests on our tracers and on a lower mass cluster are given in the Appendix. 
 \\
 \section{Simulation Setup}
 \label{sec:setup}
 \subsection{ENZO}
 \label{subsec:enzo}
 We simulated the formation of galaxy clusters with the cosmological magneto-hydrodynamical (MHD) code \enzo \ \citep{2014ApJS..211...19B}. \enzo \ uses a N-body particle-mesh solver to simulate the dark matter \citep{1988csup.book.....H}, and an adaptive mesh method to follow the baryonic matter \citep{1989JCoPh..82...64B}. \\
 In our simulations we used a piecewise linear method \citep{1985JCoPh..59..264C} with hyperbolic Dedner cleaning \citep{2002JCoPh.175..645D} to solve the MHD equations \citep[see Sec. 2.1 in][]{2014ApJS..211...19B}. \\
 We focus on the re-simulation of a single galaxy cluster extracted from a cosmological volume. This cluster has a final mass of $M_{200}\approx 9.745 \cdot 10^{14} \ \Msun$ at $z=0$ and it has been chosen for this study because it shows a major merger event at $z \approx 0.27$, leading to detectable radio relics (see Sec.~\ref{subsec:RadioEmission}). \\
 The simulation starts from a root grid of volume $ \approx (250 \ \Mpc)^3$ (comoving) sampled with $256^3$ cells and $256^3$ dark matter particles. The comoving volume of $\approx (25 \ \Mpc)^3$ centred around the most massive cluster in the box has been further refined $2^5$ times using 5 levels of AMR (up to a maximum resolution of $31.7$ kpc). In order to resolve the turbulent evolution of the intracluster magnetic field, we adopted the aggressive AMR criterion of refining all cells that are $\geq 10\%$ denser than their surrounding, beginning at the start of the simulation. From $z=1$, we additionally refined all cells with a 1-dimensional velocity jump $\Delta v / v \geq 1.5$, where $\Delta v$ is the velocity jump along any coordinate axis and $v$ is the local velocity. This procedure ensures that typically $\sim 80\%$ of the cluster volume is refined up to the highest resolution, and that the virial volume is sampled with  $\geq 200^3$ cells. 
 For the post-processing with our tracer algorithm (see Sec. \ref{subsec:tracer}) we saved all snapshots of the root grid timestep, for a total of $250$ snapshots. As in our previous work in \citet{2010NewA...15..695V}, we chose the cosmological parameters as: $H_0 = 72.0 \ \km \ \sek^{-1} \ \Mpc^{-1}$, $\Omega_{\mathrm{M}} = 0.258$, $\Omega_{\Lambda} = 0.742$ and $\sigma_8 = 0.8$. We seeded the large-scale magnetic field with a uniform primordial seed field at $z=30$, with a comoving value of $B_0 = 10^{-10}\ \G$ along each coordinate axis. \\
  \begin{figure*}\centering
   \subfigure[]{\includegraphics[width = 0.33\textwidth]{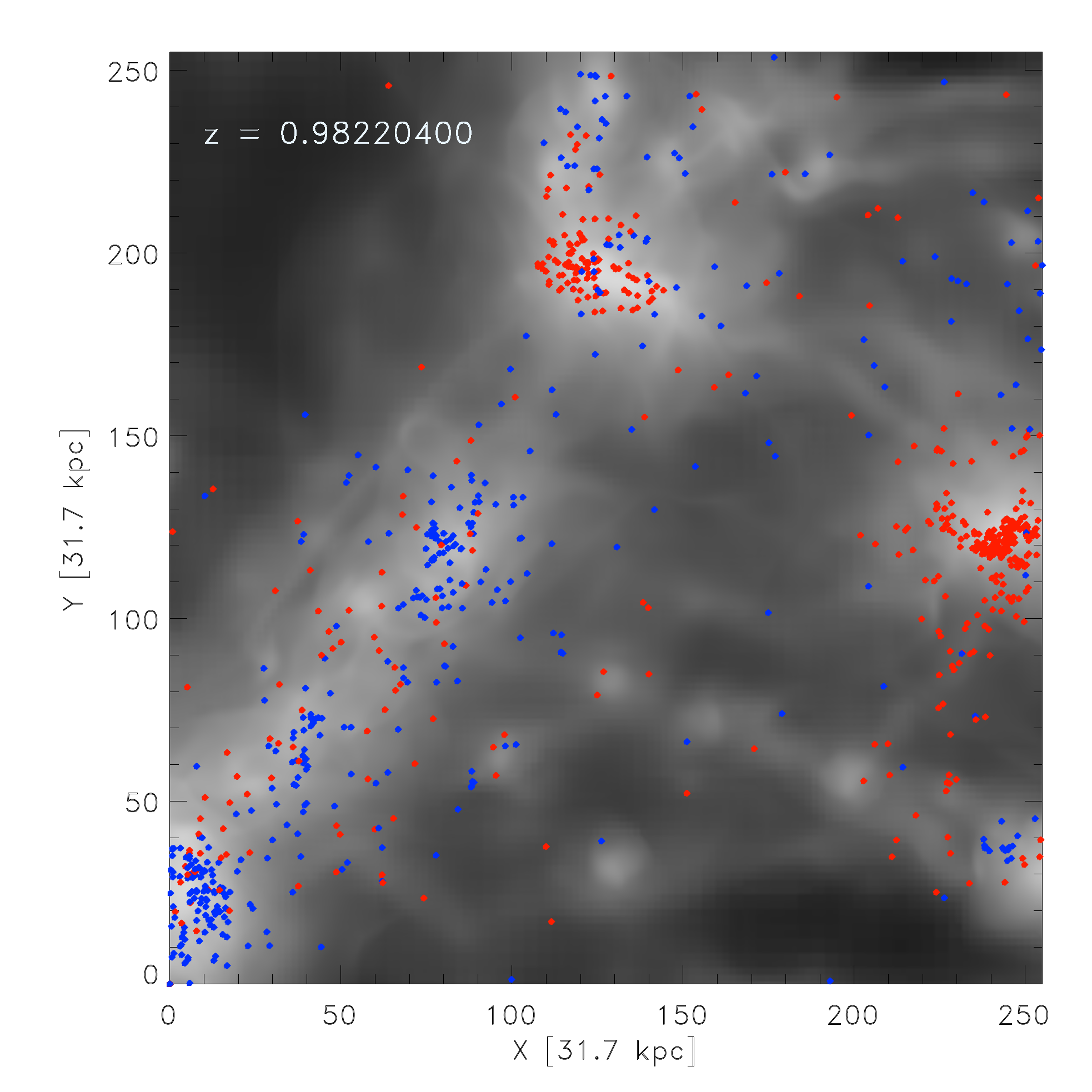}\label{subfig:tacc_post_sigma_a}}
   \subfigure[]{\includegraphics[width = 0.33\textwidth]{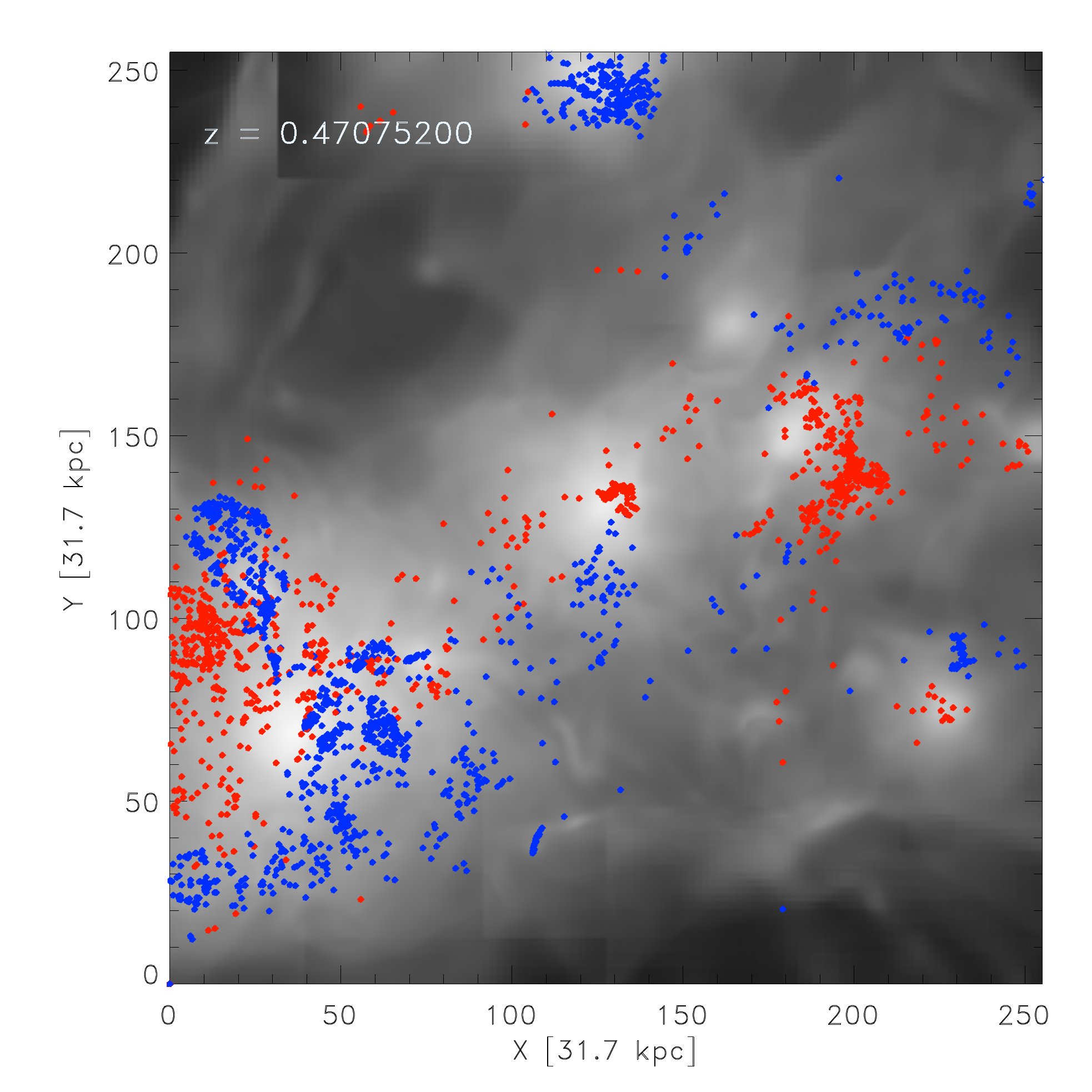}\label{subfig:tacc_post_sigma_b}}
   \subfigure[]{\includegraphics[width = 0.33\textwidth]{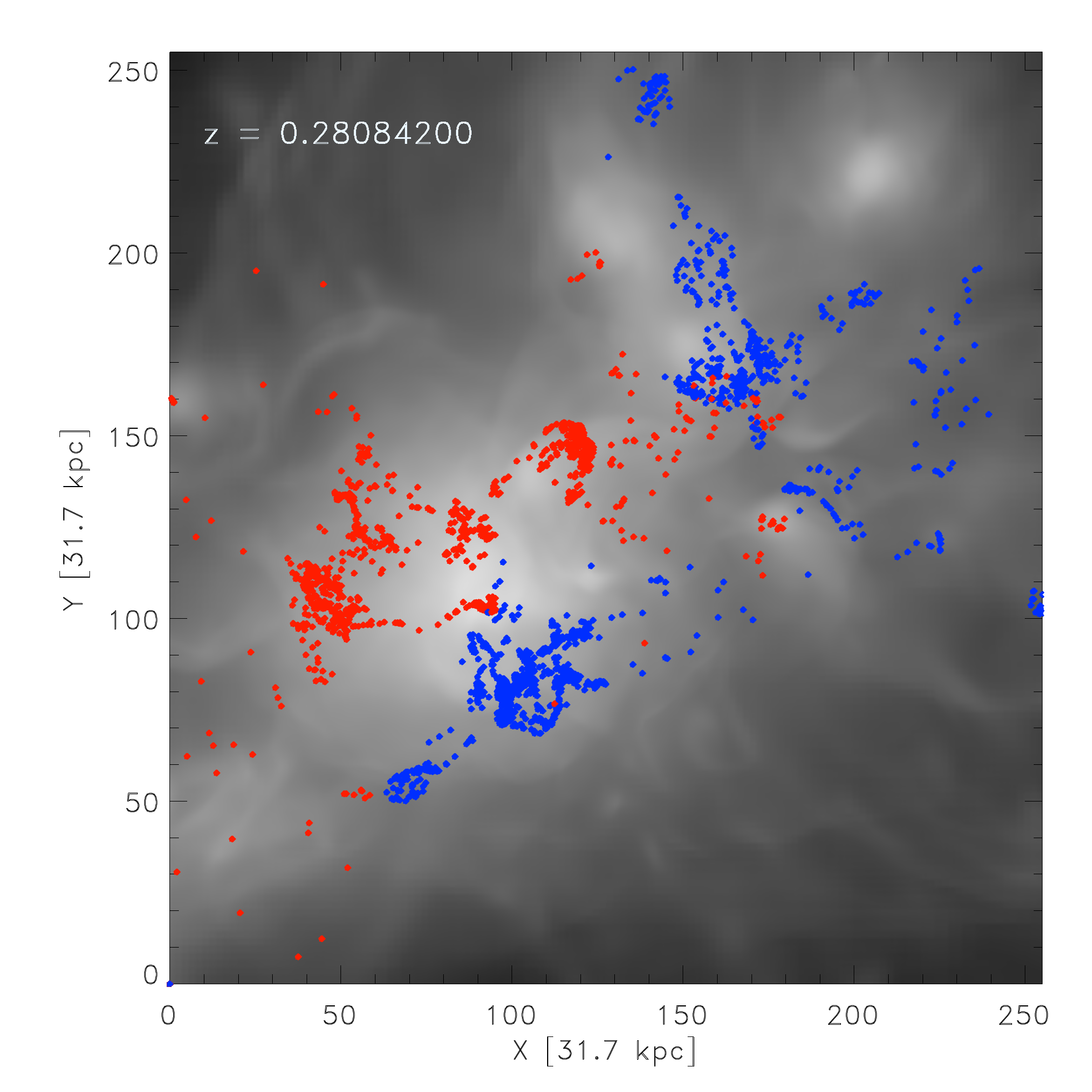}\label{subfig:tacc_post_sigma_c}}\\
   \subfigure[]{\includegraphics[width = 0.33\textwidth]{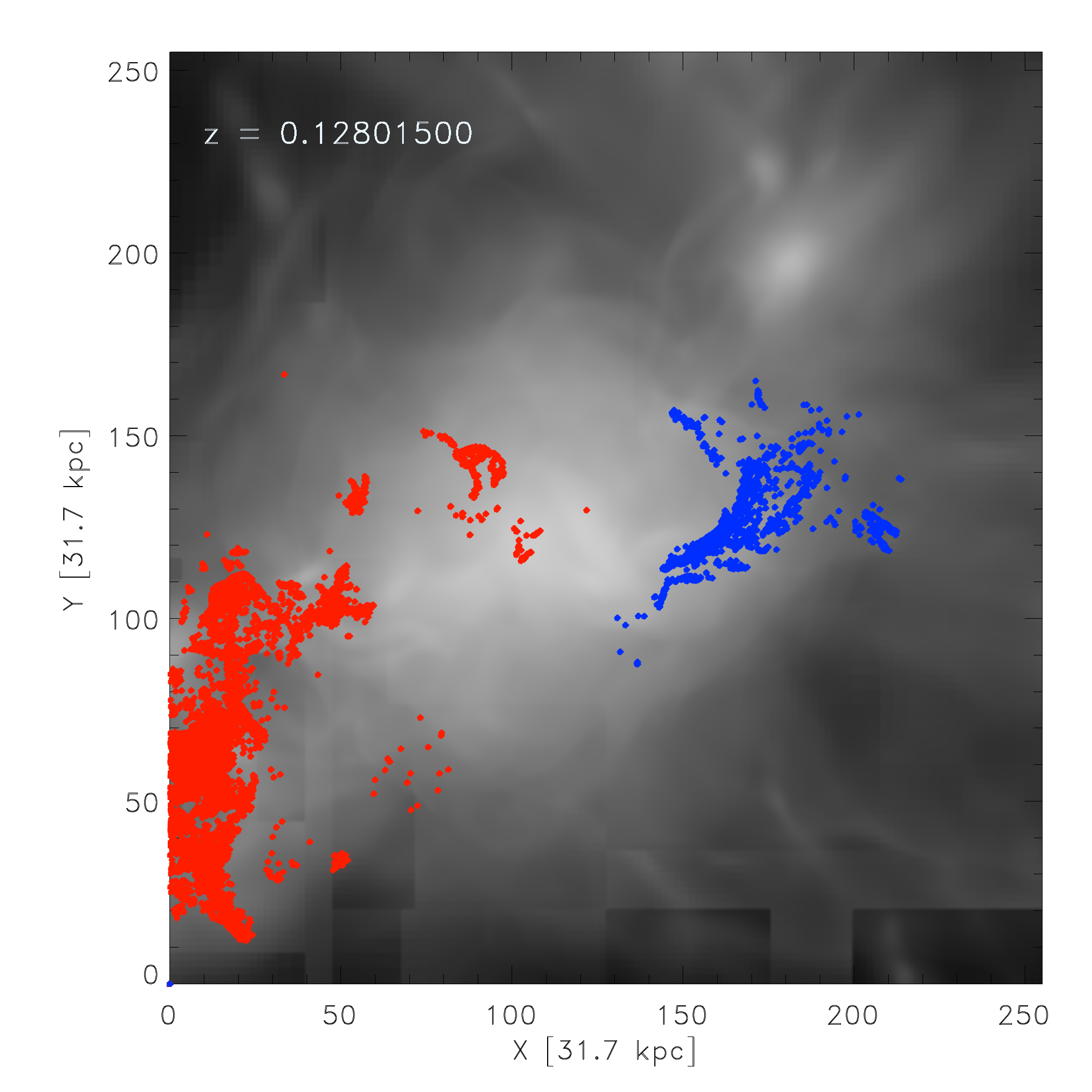}\label{subfig:tacc_post_sigma_d}}
   \subfigure[]{\includegraphics[width = 0.33\textwidth]{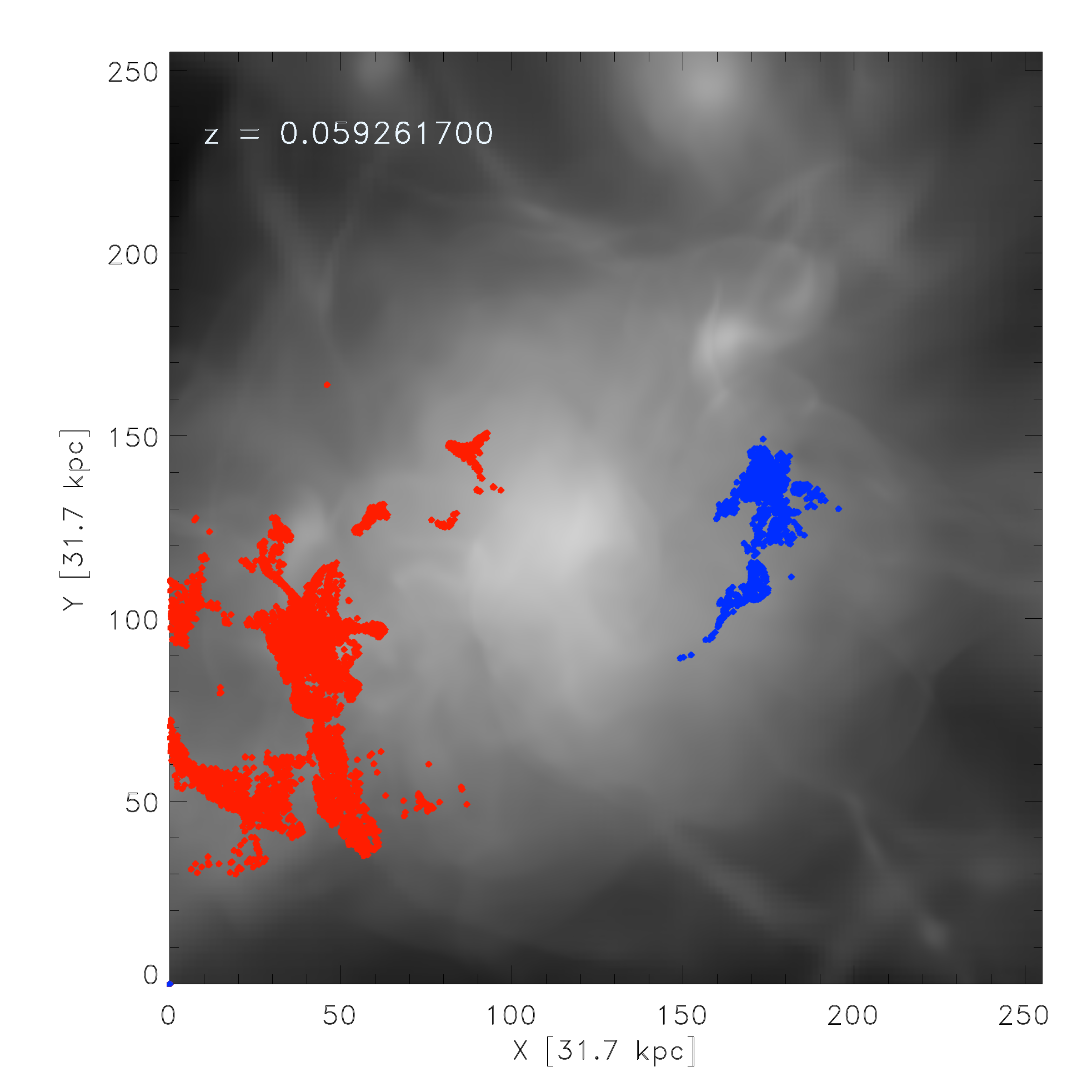}\label{subfig:tacc_post_sigma_e}}
   \subfigure[]{\includegraphics[width = 0.33\textwidth]{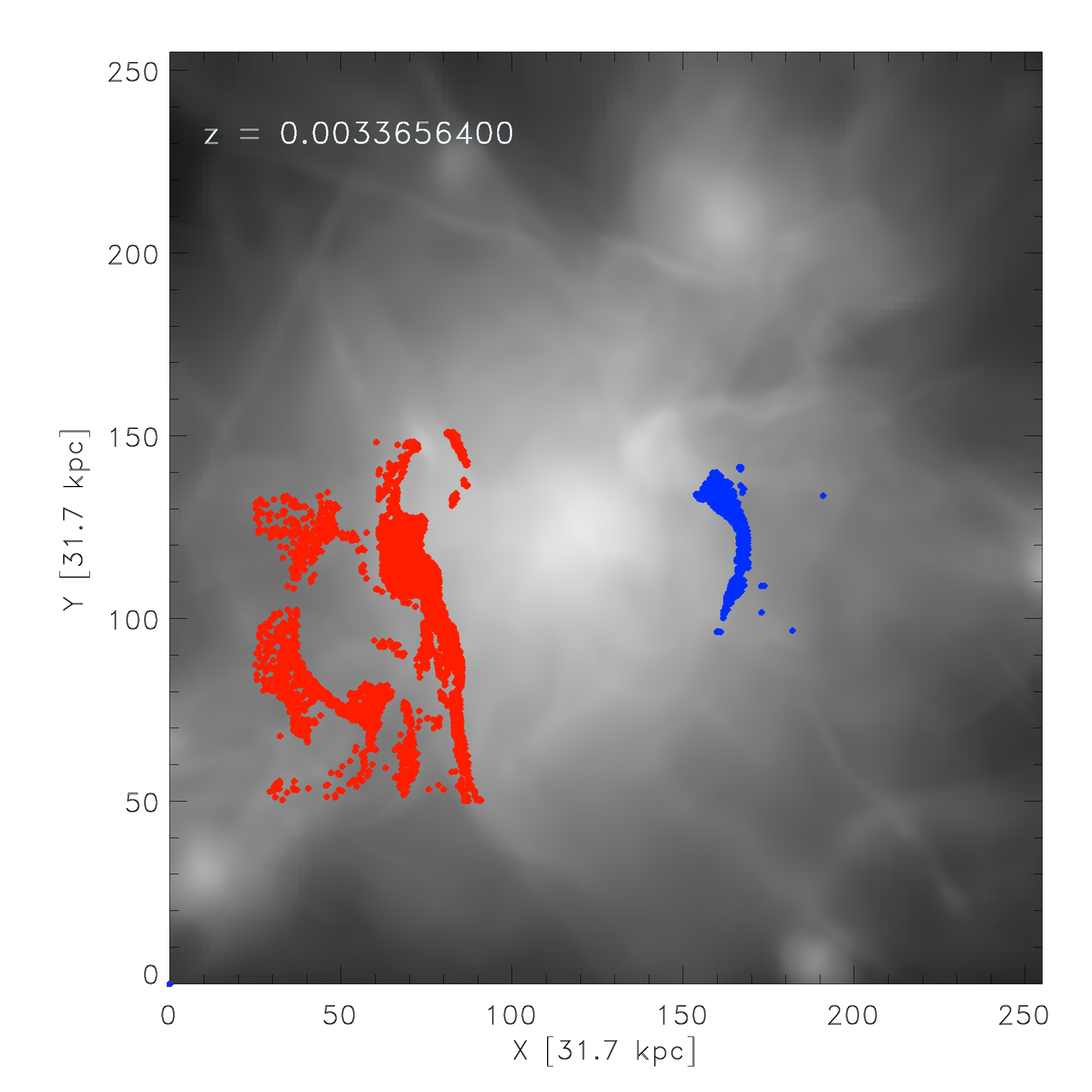}\label{subfig:tacc_post_sigma_f}}
  \caption{Evolution of the projected baryonic matter density (in grey) overlayed with the projected positions of the tracers. Only  the tracers ending up in the two relics (see Sec. \ref{subsec:RadioEmission}) are shown and are divided into two colours based on their final position.}
  \label{fig:tacc_post_sigma}
 \end{figure*}
 \subsection{Lagrangian Tracer}
 \label{subsec:tracer}
 We tracked the evolution of cosmic rays using Lagrangian tracer particles. These allow us, both, to accurately follow the advection of baryonic matter and to monitor the enrichment of shock-injected cosmic rays over time. The tracers are generated in post-processing, using the \enzo \ data at the highest spatial resolution (in the case in which the cell value is only available at lower resolution due to the AMR structure, they are linearly interpolated to the maximum resolution).\\
 In the simplest case, the tracers are advected using the velocities at their location, $\boldsymbol{v}=\boldsymbol{\tilde{v}}$, which are interpolated between the neighbouring cells \citep[e.g.][]{2010A&A...513A..32V}. However, in the case of complex flows this procedure might underestimate the amount of mixing due to fluid motions, and stochastic correction terms have been proposed to solve this problem \citep{Genel_at_al_2014_following_the_flow}. To cure for this effect we introduced a small correction term, which takes into account the small-scale velocity contribution from the neighbouring 27 cells, $\delta \boldsymbol{v}_{i,j,k}$: 
 \begin{align}
  \delta \boldsymbol{v}_{i,j,k} = \boldsymbol{v}_{i,j,k}- \frac{\sum\limits_{i = 0}^{2} \sum\limits_{j = 0}^{2} \sum\limits_{k = 0}^{2}  \boldsymbol{v}_{i-1, j-1, k-1}}{27} .
 \end{align}
 This term is added to the interpolated velocity, $\boldsymbol{v}=\boldsymbol{\tilde{v}}+\delta \boldsymbol{v}$. This procedure (unlike stochastic approaches) ensures that the thermodynamical jumps recorded by the tracers are due to gas dynamics. Our tests showed that the final distribution of tracers is not very sensitive to small variations (e.g. different interpolation schemes) in the advection procedure. 
 After computing their velocity, the tracers are advected linearly in time. During the advection, the local values of the gas on the \enzo \ grid are assigned to each tracer, and other properties (e.g. Mach number, obliquity etc.) are computed on the fly. The solenoidal and compressive modes of the velocity components for a tracer are computed using numerical stencils $\dd v_{(x,y,z)}$ as
 \begin{align}
  &v_{\mathrm{sol}} = \frac{\sqrt{ \left(\dd v_z - \dd v_y \right)^2 + \left(\dd v_x - \dd v_z \right)^2 + \left(\dd v_y - \dd v_x \right)^2 }}{2} \label{eq:curl} ,\\
  &v_{\mathrm{comp}} = \frac{\dd v_x}{2} + \frac{\dd v_y}{2} + \frac{\dd v_z}{2} \label{eq:div}.
 \end{align}
Subsequently, we apply a shock finding method based on temperature jumps as described in \citet{Ryu_et_al_2003_shock_waves_large_scale_universe}. However, instead of using the shock finder between two neighbouring grid cells, we applied it to the positions of a tracer at two consecutive timesteps. A tracer is considered to be shocked if the following requirements are matched:
 \begin{itemize}
  \item $T_{\mathrm{old}} > 100 \ \K$
  \item $\frac{T_{\mathrm{new}} }{ T_{\mathrm{old}}} > 1.00001$
  \item $\frac{S_{\mathrm{new}} }{ S_{\mathrm{old}}} > 0$
  \item $\nabla \cdot \boldsymbol{v} < 0$.
 \end{itemize}
 The Mach number is computed from the Rankine-Hugoniot relations, assuming $\gamma = 5/3$, as 
 \begin{align}
  M = \sqrt{\frac{4}{5} \frac{T_{\mathrm{new}}}{T_{\mathrm{old}}} \frac{\rho_{\mathrm{new}}}{\rho_{\mathrm{old}}} + 1}.
 \end{align}
Every time a shock is recorded, the obliquity is computed as the angle between the velocity jump $\Delta \boldsymbol{v} = \boldsymbol{v}_{\post} - \boldsymbol{v}_{\pre}$ and the pre-/post-shock magnetic field $\boldsymbol{B}_i$:
 \begin{align}
  \theta_i = \arccos \left( \frac{ \Delta \boldsymbol{v} \cdot \boldsymbol{B_i}}{\vert \Delta \boldsymbol{v}\vert \vert\boldsymbol{B_i}\vert} \right).
 \end{align}
 The index, $i$, refers to either the pre- or post-shock quantity. \\
 \begin{figure*}\centering
  \subfigure[]{\includegraphics[width = 0.48\textwidth]{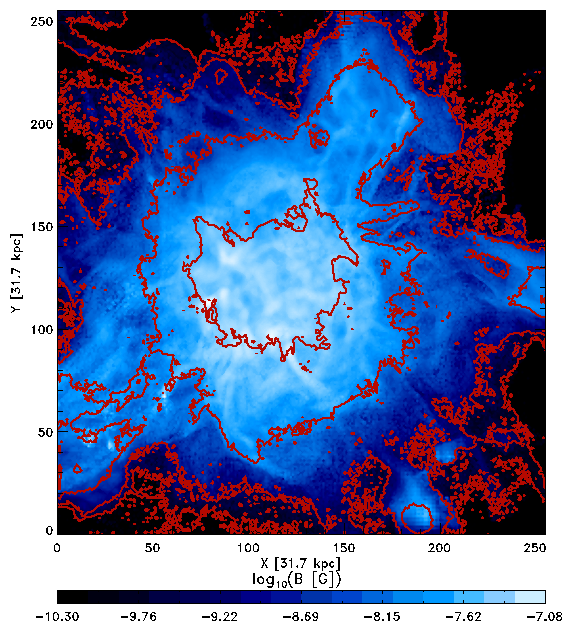}\label{fig:Bmap}}
  \subfigure[]{\includegraphics[width = 0.48\textwidth]{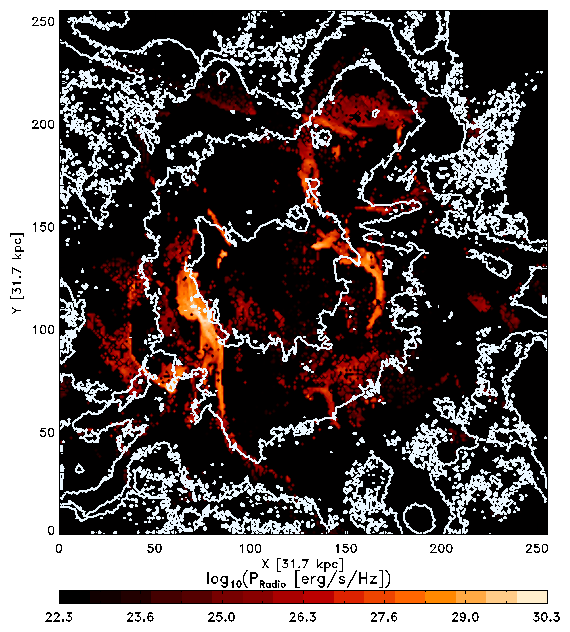}\label{fig:radiomap}}  
  \caption{Panel (a) shows the projected mass weighted magnetic field strength (colour) overlayed with the corresponding density contours (red contours) at $z \approx 0$. Panel (b) shows the projected radio emission (colour) and the corresponding density contours (white contours) at $z \approx 0$. Both plots have been produced from the tracer data. The outer regions are noisy owing to the lack of tracers in those areas (see also Appendix \ref{sec:density}).}
  \label{fig:bpmap}
 \end{figure*}
 The kinetic energy flux across shocks is $F_{\Psi} = 0.5 \cdot \rho_{\pre} v_{\mathrm{sh}}^3 $, where $\rho_{\pre}$ is the pre-shock density and $v_{\mathrm{sh}}$ is the shock velocity. Using the acceleration efficiencies $\delta \left(M\right)$ and $\eta \left( M \right)$ given in \citet{2013ApJ...764...95K}, the thermal energy flux is $F_{\mathrm{th}} = \delta(M)  F_{\psi}$ and the  cosmic-ray energy flux is $F_{\mathrm{CR}} = \eta(M) F_{\psi}$. The acceleration efficiency $\eta \left( M \right)$ also includes the effect of re-acceleration, in case the shock runs over a region which was previously enriched of cosmic rays. Following \citet{2014MNRAS.439.2662V}, we compute the effective acceleration efficiency by interpolating the acceleration efficiencies of the single injection $\eta_{\mathrm{acc}}(M)$ and of the re-acceleration $\eta_{\mathrm{re}}(M)$ case (given in \citealt{2013ApJ...764...95K}):
 \begin{align}
  \eta(M) = \frac{ (0.05 - \chi) \cdot \eta_{\mathrm{acc}}(M) + \chi \cdot \eta_{\mathrm{re}}(M)}{0.05}, \label{eq:eta_of_M}
 \end{align}
using the ratio of cosmic-ray to gas energy $\chi = E_{\mathrm{CR}} / E_{\mathrm{gas}}$.  \\
In order to maximise the number of tracers within the cluster, we injected tracers only in a $256^3$ sub-box of the \enzo-simulation, centred on the mass centre of our galaxy cluster at $z \approx 0$. The bulk of the cosmic-ray energy is expected to be generated by shocks at low redshifts \citep[e.g.][]{2016MNRAS.459...70V}, and therefore we start generating the tracers at $z \approx 1$.  \\
 In detail, the tracers were first initialised based on the gas mass distribution on the grid at $z \approx 1$. We assigned a fixed mass to each tracer and we set the number of tracers per cell according to:
 \begin{align}
  n_{\mathrm{tracers}} = \left\lfloor \frac{m_{\mathrm{cell}}}{m_{\mathrm{tracers}}} \right\rfloor \label{eq:ntracer},
 \end{align}
 where $m_{\mathrm{cell}}$ is the comoving gas mass within each high-resolution cell. The mass resolution of the tracers has to be high enough to ensure that the structure of the cluster is resolved accurately while it can still be handled computationally. In our case the trade-off is represented by a mass resolution of $m_{\mathrm{tracer}} (z =1) \approx  10^{8} \ \Msun$. \\
 Moreover, at each snapshot additional tracers were injected at the boundaries according to Eq. (\ref{eq:ntracer}). In total, we used $240$ snapshots from $z=1$ to $z=0$ and our procedure generated $N_p  \approx 1.33\cdot 10^7$ tracers during run-time. The final spatial distribution of tracers at $z\approx 0$ and their radial profile (compared to the gas profile directly simulated by ENZO) are shown in Appendix \ref{sec:density}. \\
 In order to speed up the computations and follow the largest possible number of tracers, we parallelized our advection routines using \textit{openMP}. Both, the tracers injected at $z=1$ and those generated at run-time are evenly spread among the threads, thus balancing the workload. \\
 This simulation used 48 threads minimizing the computational time to 6 hours for the Lagrangian tracer run\footnote{Compared to $\sim 60$ hours in the serial version. We obtain a non-perfect scaling of the speed-up because the bottle neck is the output of the tracer data. Further speed-up could be obtained using parallel I/O.}, running on the Intel Xeon E5-2680 v3 Haswell CPUs on the JURECA supercomputer in J\"ulich.\\
 \section{Results}
 \label{sec:results}
 \subsection{Thermal and magnetic properties}\label{subsec:GeneralResults}
 As an example for the application of our tracer-based approach, we show in Fig. \ref{fig:tacc_post_sigma} the trajectories of tracers found in the proximity of the two powerful radio relics (see Sec. \ref{subsec:RadioEmission}). The tracers ending up in the two relics at $z \approx 0$ are, both, coming from the first injection at $z=1$, as well from the injections at lower redshift. They follow the gas, mostly coming from filamentary and clumpy accretion that is heated by shock waves moving outwards after the major merger at $z \approx 0.27$. \\
 In Fig. \ref{fig:Bmap}, we show  the projected magnetic fields (mass-weighted along the line of sight) and the contours of the projected gas density probed by the tracers. The  magnetic field strengths range from $\sim 1.2 \cdot 10^{-7} \ \G$ in the cluster centre to $\sim 1.9 \cdot 10^{-10} \ \G$ in the cluster outskirts. While the average magnetic field strength within the virial volume is of the order of what is confirmed by observations $\sim 0.1-0.2 \ \mu \G$, the innermost field is significantly lower \citep[e.g. compared to the central $4.7 \ \mu \G$ value inferred for the Coma cluster by][]{2010A&A...513A..30B}. This is presumed to be caused by insufficient resolution to reproduce the small-scale dynamo, a common problem in MHD simulations \citep[e.g.][]{2014MNRAS.445.3706V}. However, the magnetic field values at the relic locations, of the order of $\sim 0.1 \ \mu \G$, are plausible values for peripheral relics \citep[e.g.][]{2007MNRAS.375...77H}. \\
 \begin{figure}
  \centering
  \includegraphics[width = 0.48\textwidth]{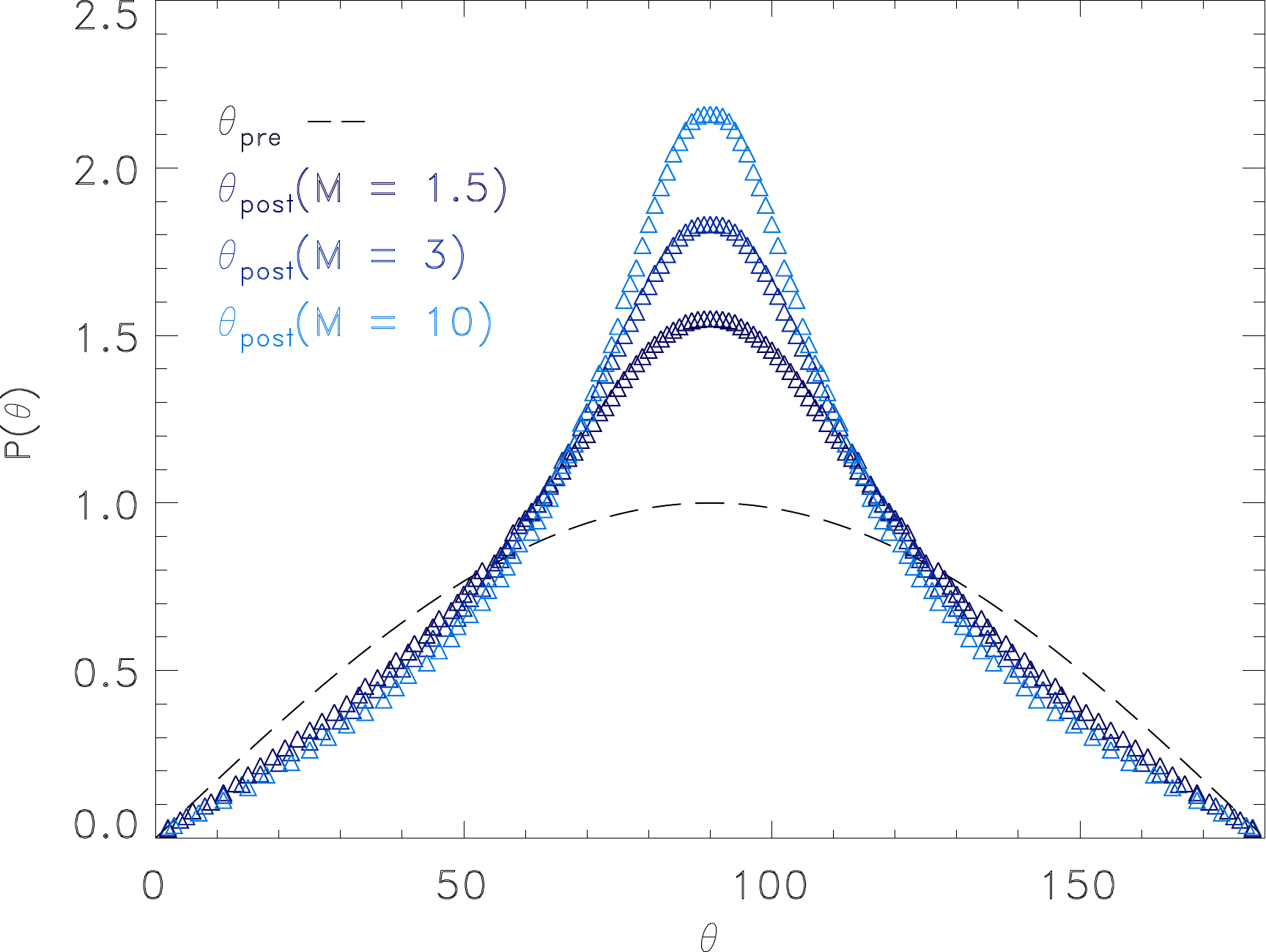}
  \caption{Expected distribution of random angles in a three-dimensional space (dashed black line). If a shock of a given Mach number $M$ crosses this distribution of angles the distribution is more concentrated towards $\theta = 90^{\circ}$ according to Eq. (\ref{eq:thetaoftheta}). The blue curves show these post-shock distributions for $M = 1.5$, $M = 3$ and $M = 10$.}
 \label{fig:thetashock}
 \end{figure}
 \subsection{Shock obliquity}
 \label{subsec:obliquity}
\begin{figure*}\centering
  \subfigure[]{\includegraphics[width = 0.49\textwidth]{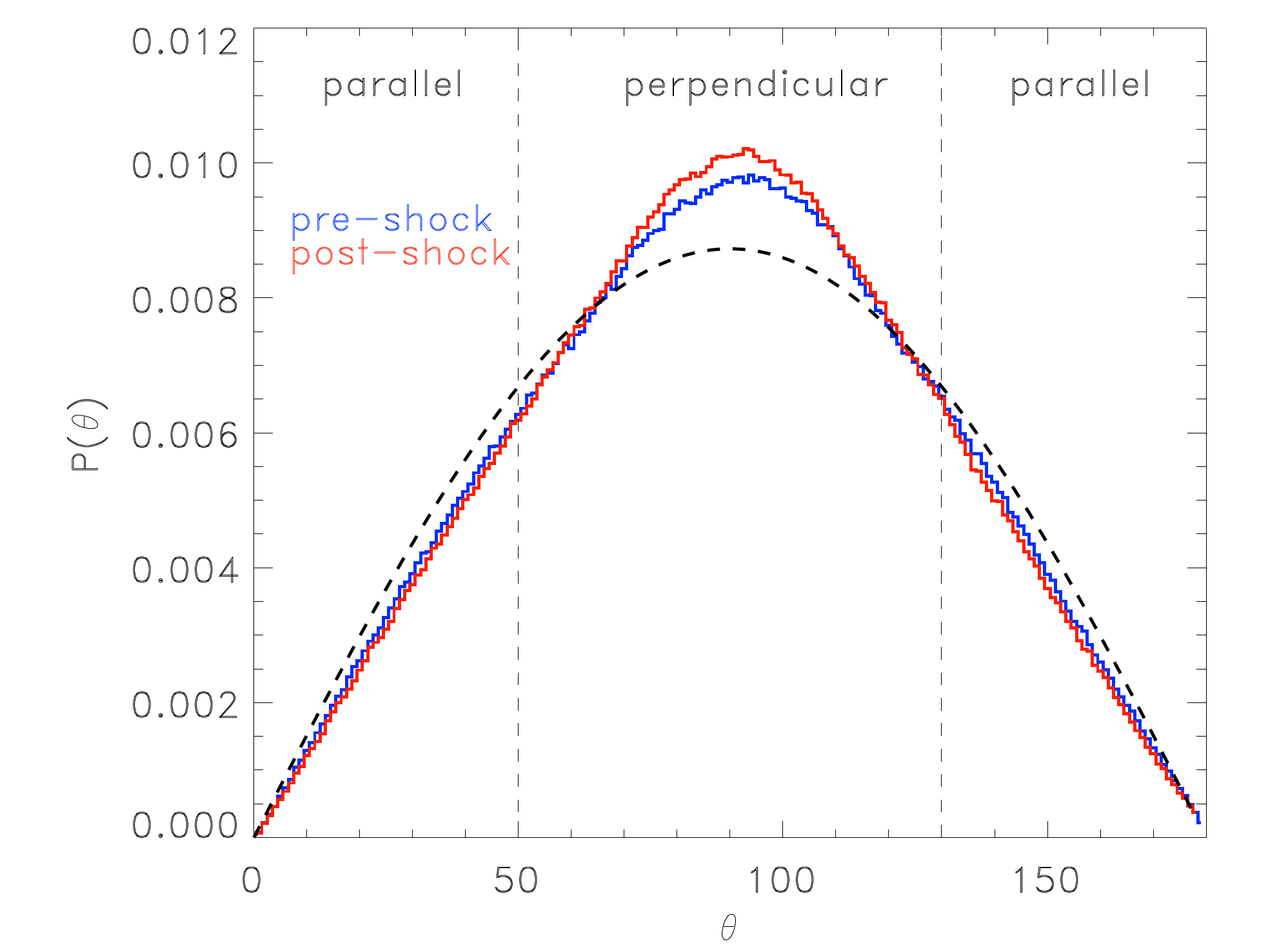}\label{fig:thetashock_a}} 
  \subfigure[]{\includegraphics[width = 0.49\textwidth]{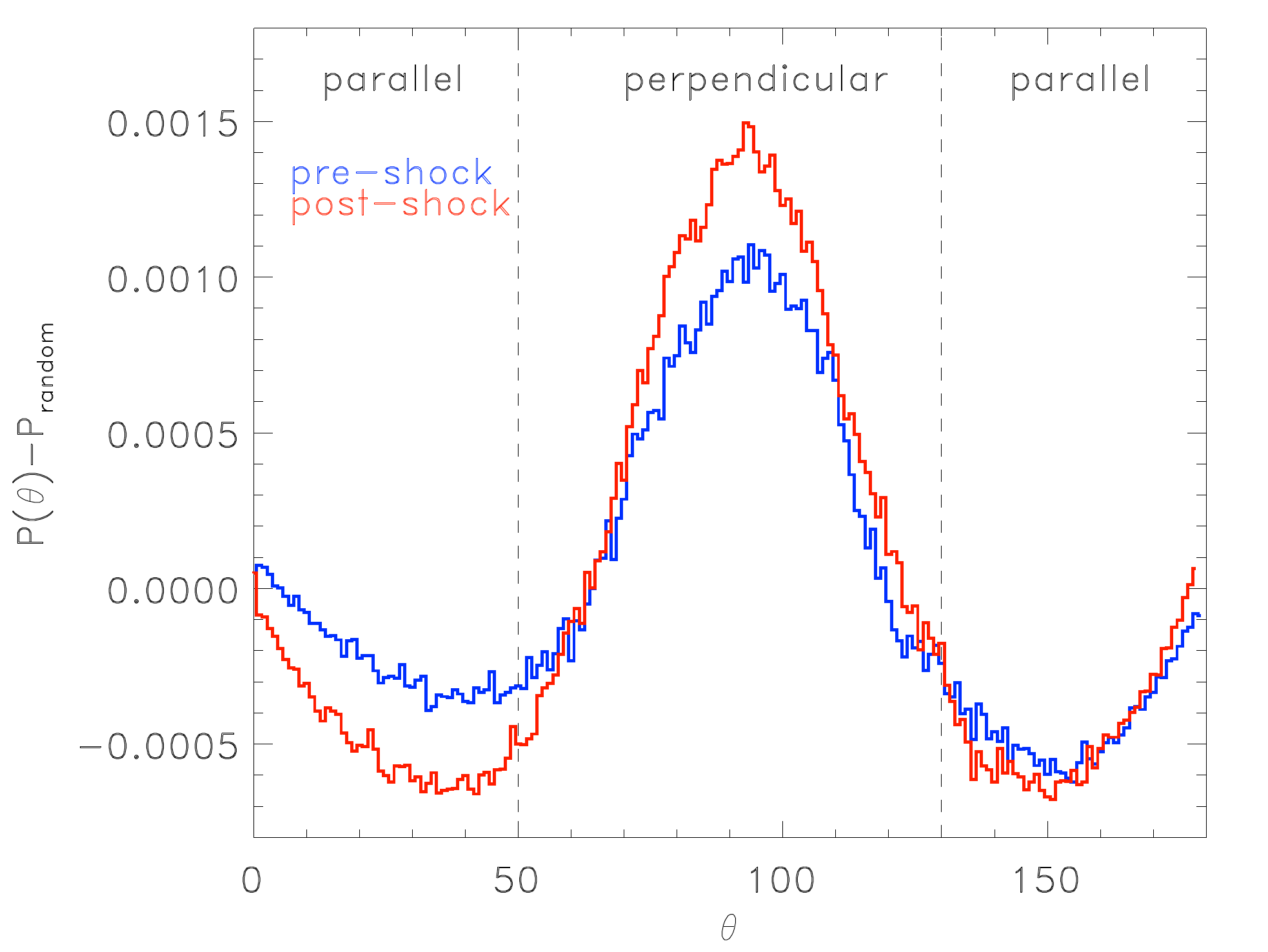}\label{fig:thetashock_b}} \\
  \subfigure[]{\includegraphics[width = 0.49\textwidth]{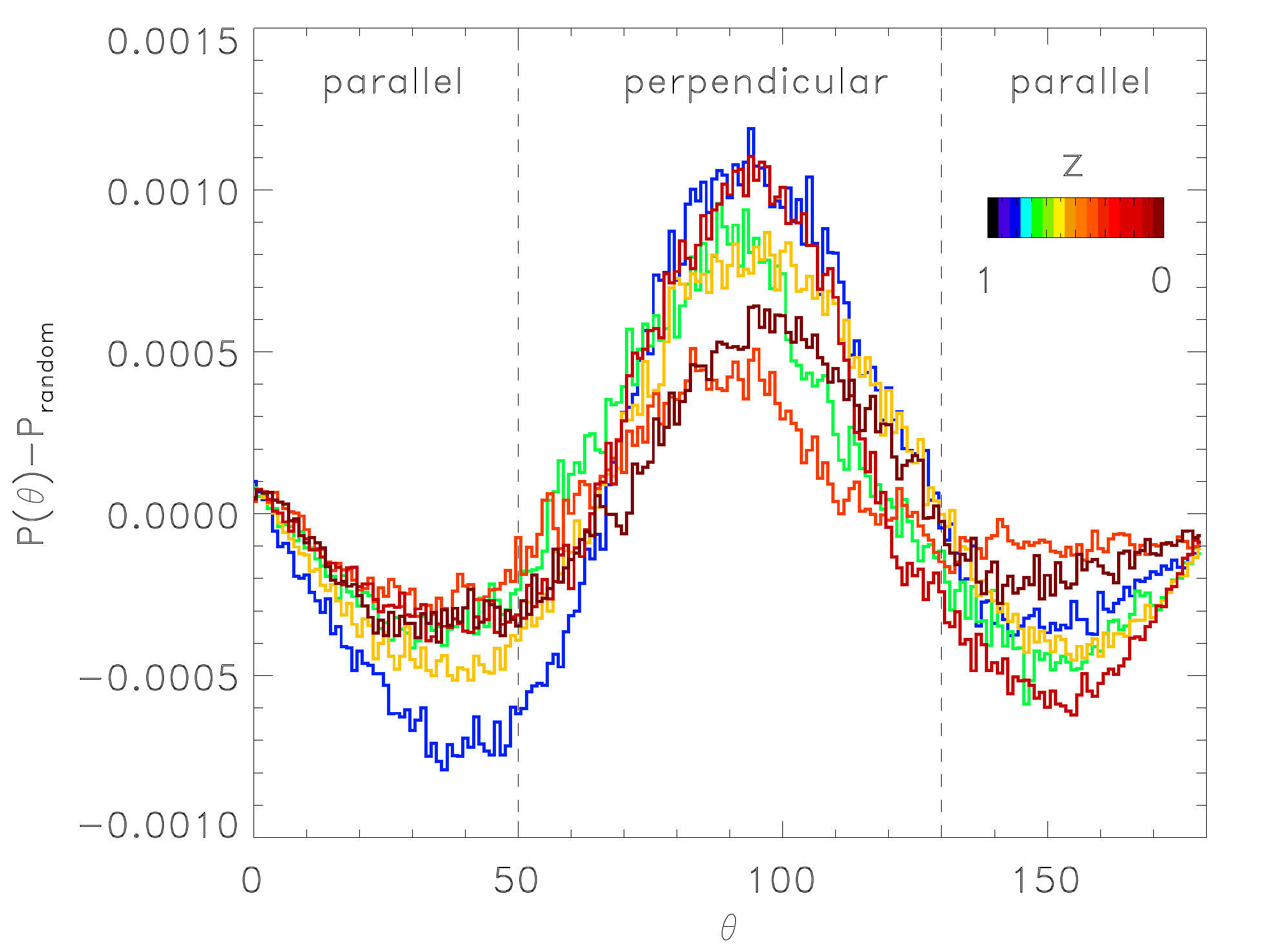}\label{fig:thetashock_c}}
  \subfigure[]{\includegraphics[width = 0.49\textwidth]{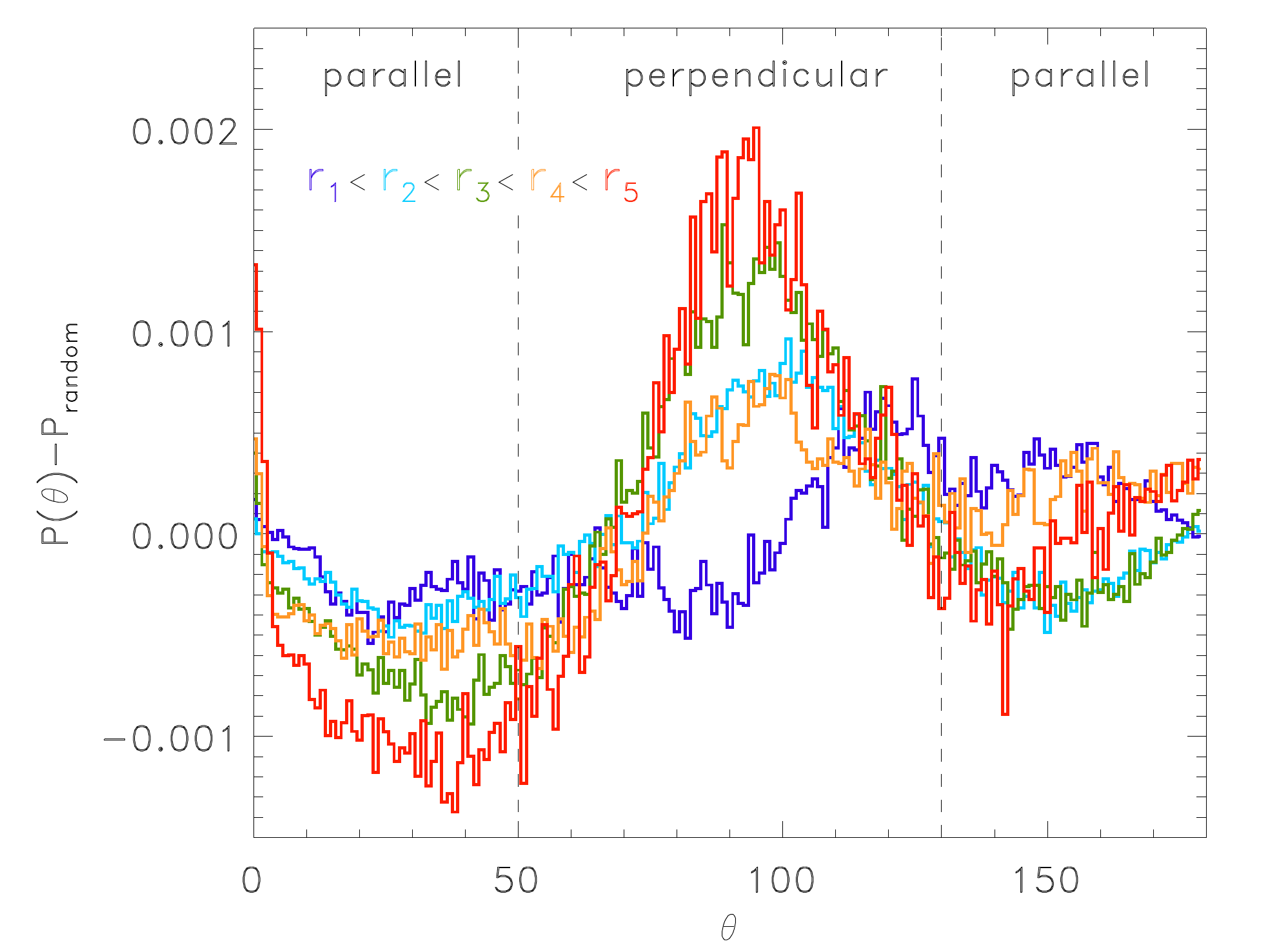}\label{fig:thetashock_d}}
  \caption{Distribution of pre- (blue) and post-shock (red) obliquities at redshift $z \approx 0.12$ are shown in panel (a). The dashed line shows the expected distribution of angles for a random distribution. Panel (b) shows the differences of the computed distributions from the expected distribution. It is observed that the post-shock distribution (red line) is more peaked towards $\theta = 90^{\circ}$ than the pre-shock distribution (blue line). Panel (c) shows the differences between the distribution of pre-shock obliquities at different redshifts and the expected distribution. Panel (d) shows the differences between pre-shock obliquities for different radial selections at $z \approx 0$ and the expected distribution. The radius of each region is $\frac{1}{5}$-th of the simulation box size.}
  \label{fig:thetahisto}
 \end{figure*}

 Before assessing the effect of shock obliquity on the acceleration of cosmic rays, we first study the distribution of shock obliquity across the simulated cluster volume. To interpret the results, it is useful to start by deriving an analytical relation between the shock Mach number and the change in the obliquity across shocks. \\
 As long as the upstream magnetic fields are isotropic,  the expected distribution of angles between the shock normal and the upstream magnetic fields follows the geometrical distribution of angles between two random vectors in a 3D space, that is $\propto \sin(\theta)$, as shown by the black dashed line in Fig. \ref{fig:thetashock} \citep[see][]{randomVectors}. \\
 Following \citet{FitzpatrickPlasma}, it is convenient to define a shock frame, where the shock lies in the $z$-plane and the shock normal is perpendicular to the $x$-direction and parallel to the $y$-direction and we transform into the de Hoffmann-Teller frame $\left(\left| \boldsymbol{v}_{\pre} \times \boldsymbol{B}_{\pre}\right| = 0 \right)$. This leads to the general MHD-jump conditions in the form
 \begin{align}
  \frac{\rho_2}{\rho_1}  = r \label{eq:rhojump}\\
  \frac{B_{x,\post}}{B_{x,\pre}} &= 1 \label{eq:bxjump}\\
  \frac{B_{y,\post}}{B_{y,\pre}} &= r \left( \frac{v_{x,\pre}^2 - \cos^2\theta_{\pre} v_{A, \pre}^2}{v_{x,\pre}^2 -r \cos^2\theta_{\pre} v_{A, \pre}^2} \right) \label{eq:byjump}\\
  \frac{v_{x,\post}}{v_{x,\pre}} &= r^{-1} \label{eq:vxjump}\\
  \frac{v_{y,\post}}{v_{y,\pre}} &= \frac{v_{x,\pre}^2 - \cos^2\theta_{\pre} v_{A, \pre}^2}{v_{x,\pre}^2 -r \cos^2\theta_{\pre} v_{A, \pre}^2} .\label{eq:vyjump}
 \end{align}
 The above equations can be further simplified in our case because the pre-shock Alfv\'en velocity, $ v_{A, \pre} $, can be safely neglected in comparison to the upstream gas velocity $v_{\mathrm{gas}}$. Indeed, we verified that for the entire cluster volume the distribution of $ v_{A, \pre}/v_{\mathrm{gas}}$ is well described by a log-normal distribution centred on $v_{A, \pre}/v_{\mathrm{gas}} \approx 0.01$, and extending to beyond 1 only in $\sim 10^{-5}$ of cases.
 Therefore, owing to the low magnetisation of the ICM we can treat our shocks in the (simpler) hydrodynamical regime, in which case the above Eq. \ref{eq:byjump} and \ref{eq:vyjump} reduce to $ \frac{B_{y2}}{B_{y1}} = r$ and $\frac{v_{y2}}{v_{y1}} = 1$. Using these jump conditions, we derive $\theta (M)$ from
  \begin{align}
   \cos \left( \theta_{\post} \right) = \frac{\Delta \boldsymbol{v} \cdot \boldsymbol{B}}{\left| \Delta \boldsymbol{v} \right| \left| \boldsymbol{B}\right|}
  \end{align}
  as
  \begin{align}
   \theta_{\post} \left( M \right) = \arccos \left[ \frac{B_{x1}}{\sqrt{B_{x1}^2 + r^2 B_{y1}^2}} \right]. \label{eq:thetaofM}
  \end{align}
 In Eq. (\ref{eq:thetaofM}), $\theta_{\post}(M)$ only depends on the pre-shock values. $B_{x1}$ and $B_{y1}$ are connected via $\theta_{\pre}$ as $B_{y1} = B_{x1} \cdot \tan\left( \theta_{\pre}\right)$. Therefore, the change of a pre-shock obliquity only depends on the angle itself and the compression ration $r$ as
 \begin{align}
  \theta_{\post} \left( M \right) = \arccos \left[ \frac{1}{\sqrt{1 + r^2 \tan \left( \theta_{\pre}\right)^2}} \right]. \label{eq:thetaoftheta}
 \end{align}
 For any Mach number the distribution is compressed towards $90^{\circ}$, and  the compression is stronger for stronger shocks. In Fig. \ref{fig:thetashock}, we show how the distribution of obliquity changes, once it is passed by a shock. Overall the distribution of pre- and post-shock obliquities in the cluster is strongly linked to the dynamical history of the cluster itself as the pre-shock distribution at later timesteps is a result of the post-shock distribution at earlier timesteps. \\
 We computed the distribution of the pre- and post-shock obliquities in our simulation at $z \approx 0.12$ (see red and blue line Fig. \ref{fig:thetashock_a}). Overall their shapes match the distribution of random angles in 3D well (black line in Fig.~\ref{fig:thetashock_a}). The differences to the distribution of random angles is plotted in Fig. \ref{fig:thetashock_b}. We chose $\theta = 50^{\circ}$ as the threshold angle to mark the division between quasi-parallel and quasi-perpendicular shocks. This choice is based on Fig. 3 of \citet{Caprioli_Spitkovsky_2014_ion_accel_I_eff} as the acceleration efficiency of protons drops significantly beyond this for $M \leq 10$ shocks. In both distributions we observe more quasi-perpendicular shocks and less quasi-parallel shocks than expected. For the post-shock distribution this is expected according to Eq. (\ref{eq:thetaoftheta}). We find that also the pre-shock distribution shows a departure from isotropy, caused by shock compression at the previous epochs. Although turbulent motions in the ICM are expected to distribute angles randomly, the rather continuous crossing by shocks tends to concentrate the angles toward quasi-perpendicular geometry. This makes the pre-shock distribution at all epochs already slightly more concentrated towards perpendicular angles, than expected from isotropy. \\
 This is confirmed by the distribution of pre-shock angles at different redshifts: in Fig. \ref{fig:thetashock_c} we show their differences to the isotropic distribution. Independent of redshift, we observe more quasi-perpendicular shocks than quasi-parallel shocks. Moreover, the distribution tends to concentrate slowly towards $\sim 90^{\circ}$ as a function of time but the effect is very small and by and large the angles are distributed isotropically. \\
 Finally, at $z=0$ we we divided the box into five spherical, concentric, equidistant shells and computed the pre-shock distributions for each shell separately. The differences to the predicted distribution is shown in Fig. \ref{fig:thetashock_d}.  All shells show patterns that are compatible with a random distribution of angles. The central region (blue line in Fig. \ref{fig:thetashock_d}) is most turbulent and the magnetic fields are most isotropic. The distribution shows a larger excess of quasi-perpendicular shocks at larger radii (from light blue to red lines in Fig. \ref{fig:thetashock_d}). Indeed, in cluster outskirts shocks are more frequent and stronger causing a stronger alignment of magnetic fields. In the following sections we will show how this behaviour might have important consequences in the acceleration of cosmic rays by cluster shocks. 
 \subsection{Cosmic-ray electrons \& radio emission}\label{subsec:RadioEmission}
 The cluster studied in this paper has been chosen because it shows two prominent radio relics at $z \approx 0$. These radio relics are produced by shock waves launched by a major merger of three gas clumps and propagate along the horizontal direction in Fig. \ref{fig:tacc_post_sigma} and \ref{fig:bpmap}. We compute the radio emission on shocked tracers using the formula \citep[from][]{2007MNRAS.375...77H}
 \begin{align}
  \begin{split}
   \frac{\mathrm{d} P_{\radio} \left( \nu_{\mathrm{obs}} \right) }{\mathrm{d} \nu} &=  \frac{6.4 \cdot 10^{34} \ \mathrm{erg}}{\sek \cdot \mathrm{Hz}} \frac{A}{\mathrm{Mpc}^2}  \frac{n_e}{10^{-4} \ \cm^{-3}}
   \frac{\xi_e}{0.05}\left( \frac{T_d}{7 \ \keV} \right)^{\frac{3}{2}}  \\
   &\times \left( \frac{\nu_{\mathrm{obs}}}{1.4 \ \GHz} \right)^{-\frac{s}{2}} 
   \frac{\left( \frac{B}{\mu \G} \right)^{1+\frac{s}{2}}}{\left(\frac{B_{\mathrm{CMB}}}{\mu \G} \right)^{2}+ \left(\frac{B}{\mu \G} \right)^{2}} \cdot \eta \left( M \right)
   \end{split}.
 \end{align}
 The quantities in the formula that either have been recorded from the grid or computed with the recorded values, are: $A$ the surface area of a tracer\footnote{The surface area is computed from the volume occupied by a tracer in a gridcell as: $A = \left( V_{\mathrm{cell}} \cdot \frac{m_{\mathrm{cell}}}{m_{\mathrm{tracer}}} \right)^{\frac{2}{3}}$. Here $V_{\mathrm{cell}}$ is the volume of the cell and $m_{\mathrm{cell}}$ is the total mass in that cell. $m_{\mathrm{tracer}}$ is the tracer mass.}, $n_e$ the number density of electrons, $T_d$ the downstream temperature, $B$ the magnetic field strength and the acceleration efficiency $\eta \left(M\right)$ depending on the Mach number $M$. We used the acceleration efficiencies $\eta \left(M\right)$ derived in \citet{2013ApJ...764...95K}. The other quantities are the electron-to-proton ratio, $\xi_e = 0.01$, the equivalent magnetic field of the cosmic microwave background, $B_{\mathrm{CMB}} = 3.2 \cdot (1+z)^2 \ \mu \G$ and the observed frequency band, $\nu_{\mathrm{obs}}=1.4 \ \GHz$ . 
 \begin{figure}\centering
  \includegraphics[width = 0.48\textwidth]{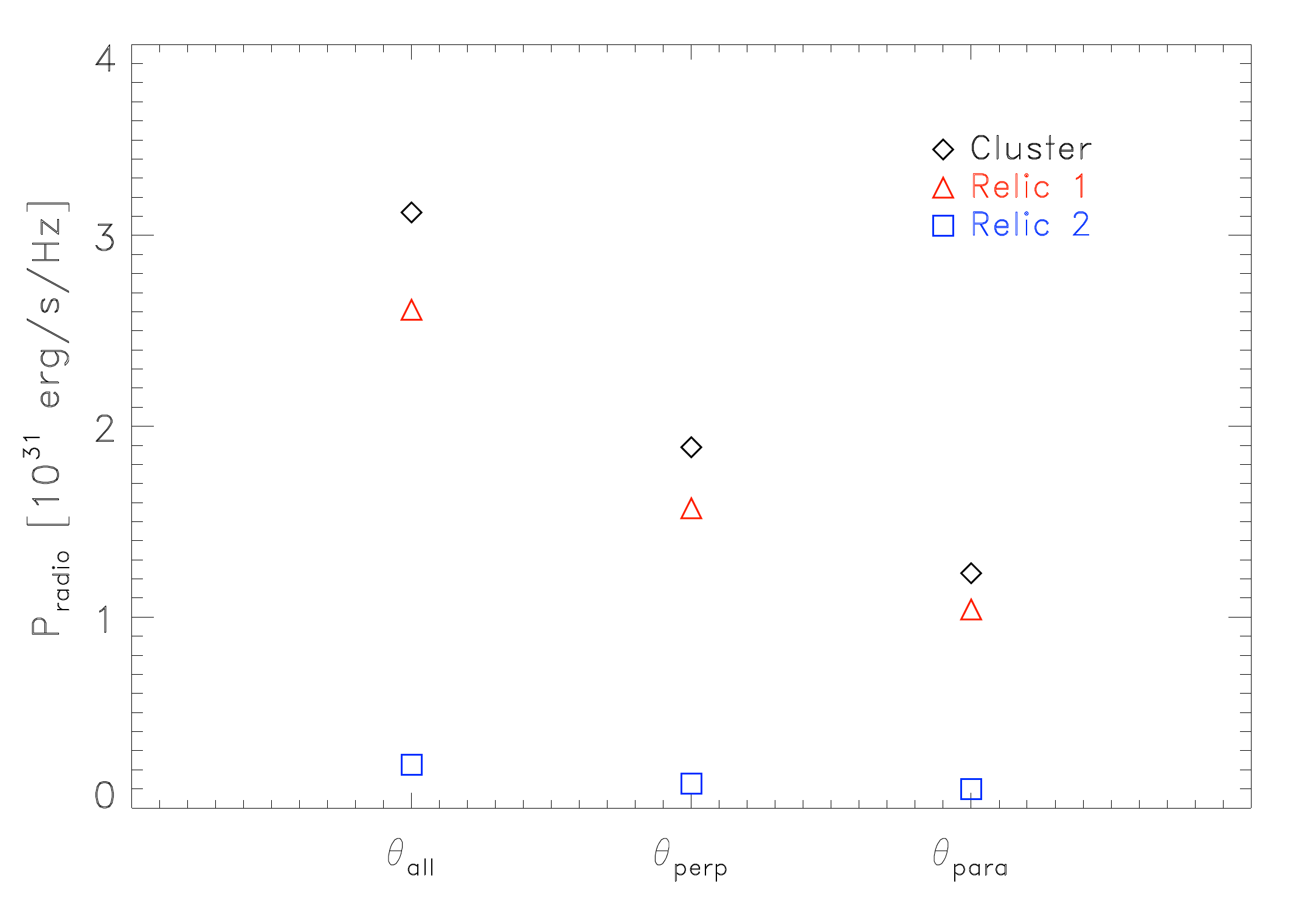}
  \caption{Total radio emission from our simulated cluster at $z \approx 0$, considering the
  total emission from the cluster (diamond) or the emission from  relic one (triangle) or relic two (square), for different selections of the obliquity angle, $\theta$.}
 \label{fig:radio_of_theta}
 \end{figure}
 In Fig. \ref{fig:radiomap} we show the observed radio emission at $z \approx 0$, overlayed with the corresponding density contours, which features two prominent radio relics at opposite sides of the cluster core.  The alignment and morphologies of the two relics indicate that they have been produced by the major merger at $z \approx 0.27$, which happened along the horizontal axis in the image. We measure a Mach number $M \approx 3.5$ for the relic located west of the cluster centre (hereafter relic 1), and $M \approx 2.7$ for the relic located at the opposite side (hereafter relic 2). The total radio emission from the cluster is $P_{\radio} \approx 3.12 \cdot 10^{31} \ \erg \ \sek^{-1} \ \Hz^{-1}$, while the emission from relic 1 is $P_{\radio} \approx 2.61 \cdot 10^{31} \ \erg \ \sek^{-1} \ \Hz^{-1}$ and from the relic 2 is $P_{\radio} \approx 2.27 \cdot 10^{30} \ \erg \ \sek^{-1} \ \Hz^{-1}$ (see Fig. \ref{fig:radio_of_theta}). If the cluster is located at the luminosity distance of $100 \ \Mpc$, relic 1 is bright enough to be  detectable at $1.4 \ \GHz$ by both the JVLA (assuming the $0.45 \ \mathrm{mJy/beam}$ sensitivity of the NVSS survey, \citealt{1998AJ1151693C}) and by ASKAP (assuming a sensitivity of $0.01 \ \mathrm{mJy/beam}$ as in the EMU survey, \citealt{2011PASA28215N}). At the distance of $100 \ \Mpc$,  relic 2 would be too faint for the JVLA, while it would be instead at the edge of detection with ASKAP. \\
 Next, we used the obliquity $\theta$ to limit the injection of cosmic-ray electrons and study its observable effect on the relic emission. In the following we compare the radio emission including all shocked particles to the one produced by particles that have only crossed a quasi-perpendicular\footnote{$\theta \in [50^{\circ}, 130^{\circ}]$} or quasi-parallel\footnote{$\theta \in [0^{\circ}, 50^{\circ}]$ or $\theta \in [130^{\circ}, 180^{\circ}]$} shock. In the following, the subscripts $\all$, $\Perp$ and $\para$ correspond to the cuts mentioned above. \\
 From the total emission shown in Fig. \ref{fig:radio_of_theta}, we can see that relic 1 is still observable even if only quasi-perpendicular shocks are allowed to accelerate the cosmic-ray electron, while relic 2 would remain undetectable. We give a close-up view onto the relic regions in Fig. \ref{subfig:TnvCCpc}, where we show the projected temperatures and the radio contours, with additional vectors of projected magnetic fields. The range of magnetic vectors is too large to allow a clear visualisation, and therefore all vectors have been renormalised to the same unit length while the magnetic intensity is shown through the color coding (with intensity increasing from light to dark blue). The radio emission produced by $\theta_{\all}$ (left column), $\theta_{\Perp}$ (middle column) and $\theta_{\para}$ (right column) is given for relic 1 on the upper row, and fore relic 2 in the lower row. The emission does not dramatically decrease across most of the relic surface when either of the two obliquity cuts is performed. This is because in these regions the angles are distributed close to the random distribution (see Sec.\ref{subsec:obliquity}), and therefore the radio emitting volume in both scenarios is still of the same order of magnitude as in the case without obliquity selection. Based on this test, we conclude that it is possible that observed radio relics are indeed tracing cosmic-ray electrons only accelerated by quasi-perpendicular shocks (and hence, from the combination of SDA and DSA, \citealt{Guo_eta_al_2014_I}).  \\
 We performed the same analysis for a $2.8 \cdot 10^{14} \ \Msun$ cluster (see Appendix \ref{sec:tracA2}) and we found similar results. \\
 \subsection{Cosmic-ray protons \& $\gamma$-rays}\label{subsec:gammaEmission}

 Next we test the time-integrated effects of imposing the same selection as above (see Sec. \ref{subsec:RadioEmission}) in the obliquity of shocks accelerating cosmic-ray protons, following the results of \citet{Caprioli_Spitkovsky_2014_ion_accel_I_eff}, who found an efficient acceleration of cosmic-ray protons only for quasi-parallel shocks. \\
 \begin{figure*} \centering
  \subfigure[]{\includegraphics[width = 0.32\textwidth]{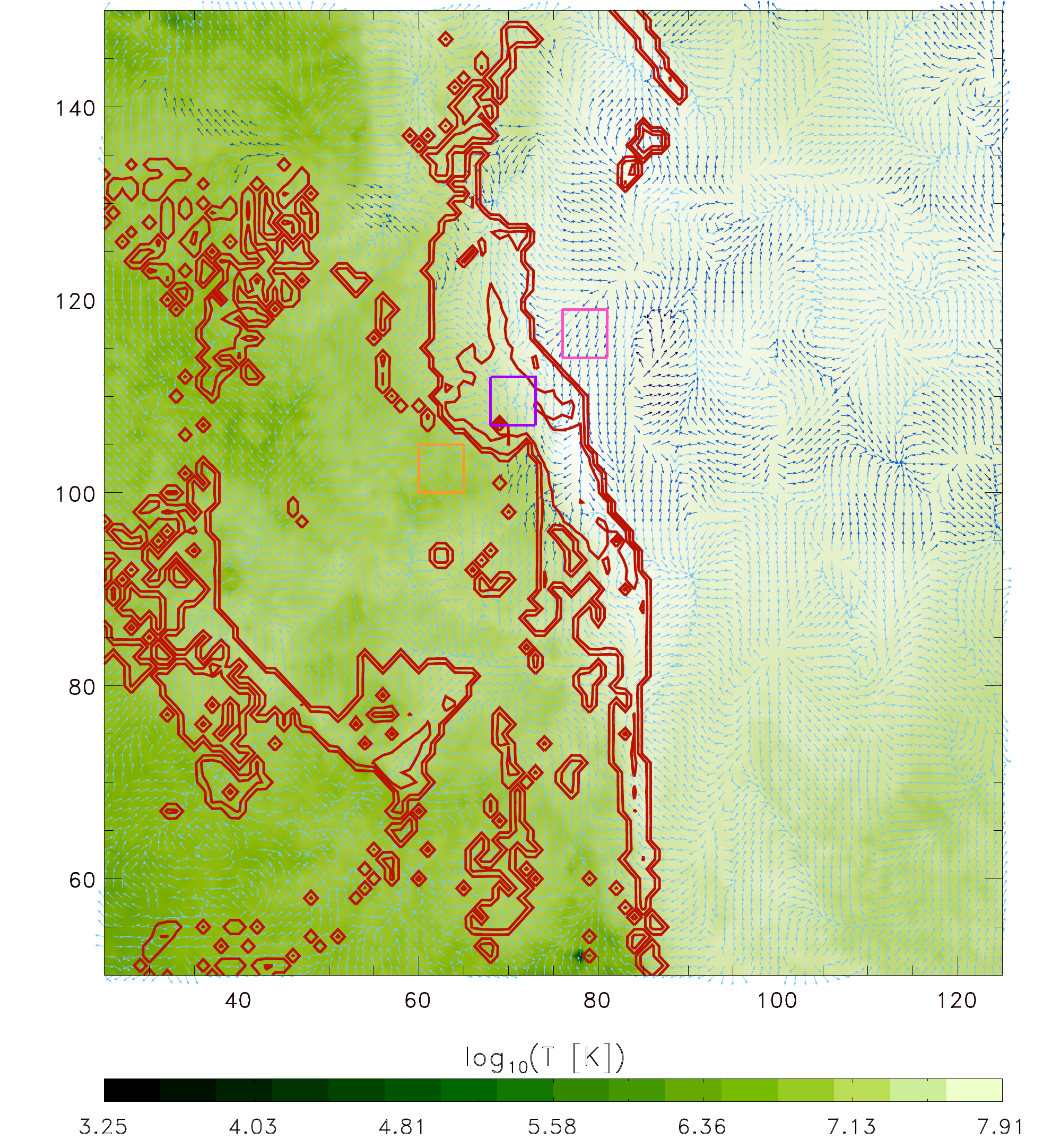} \label{subfig:shocka}}
  \subfigure[]{\includegraphics[width = 0.32\textwidth]{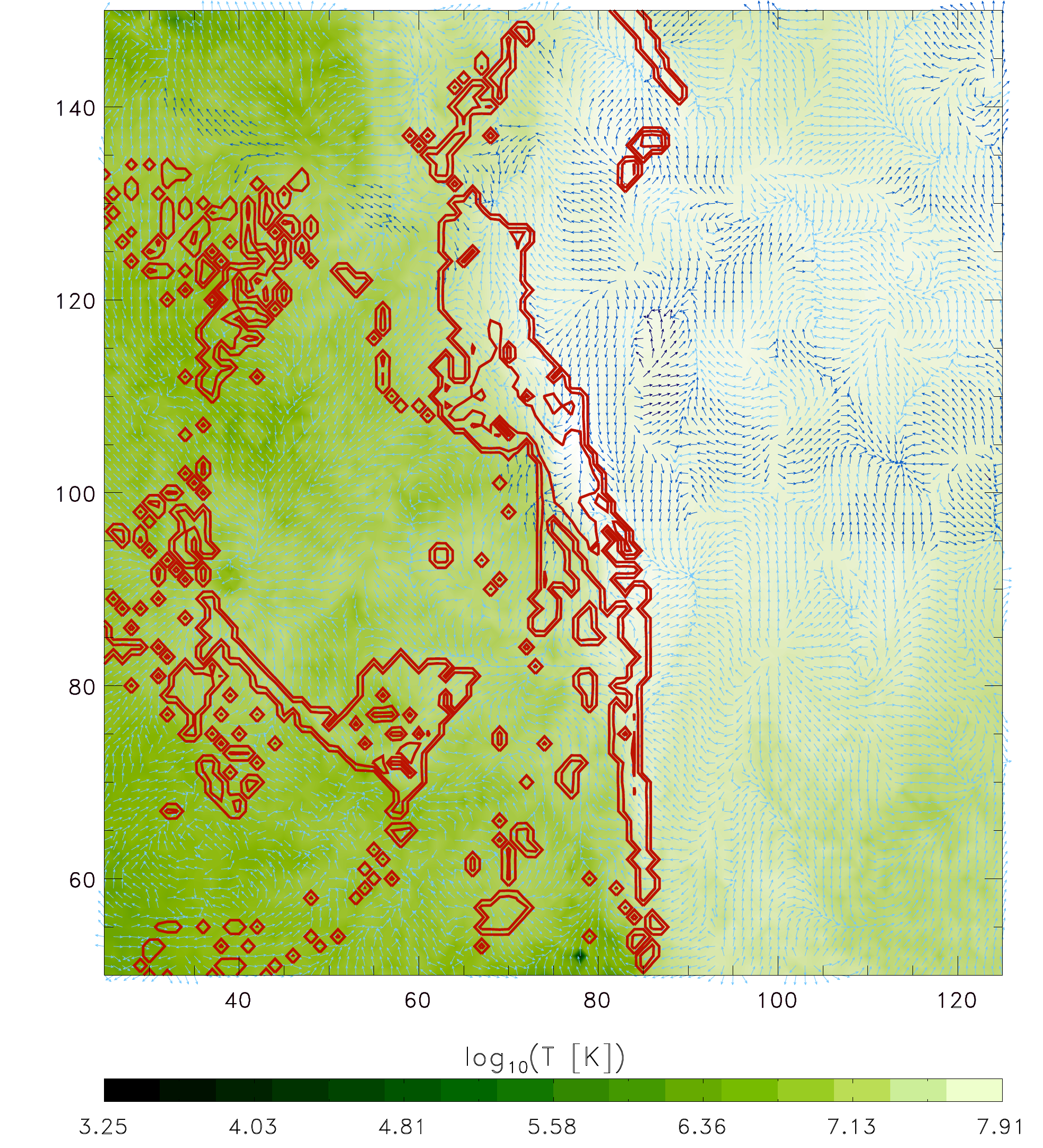}} 
  \subfigure[]{\includegraphics[width = 0.32\textwidth]{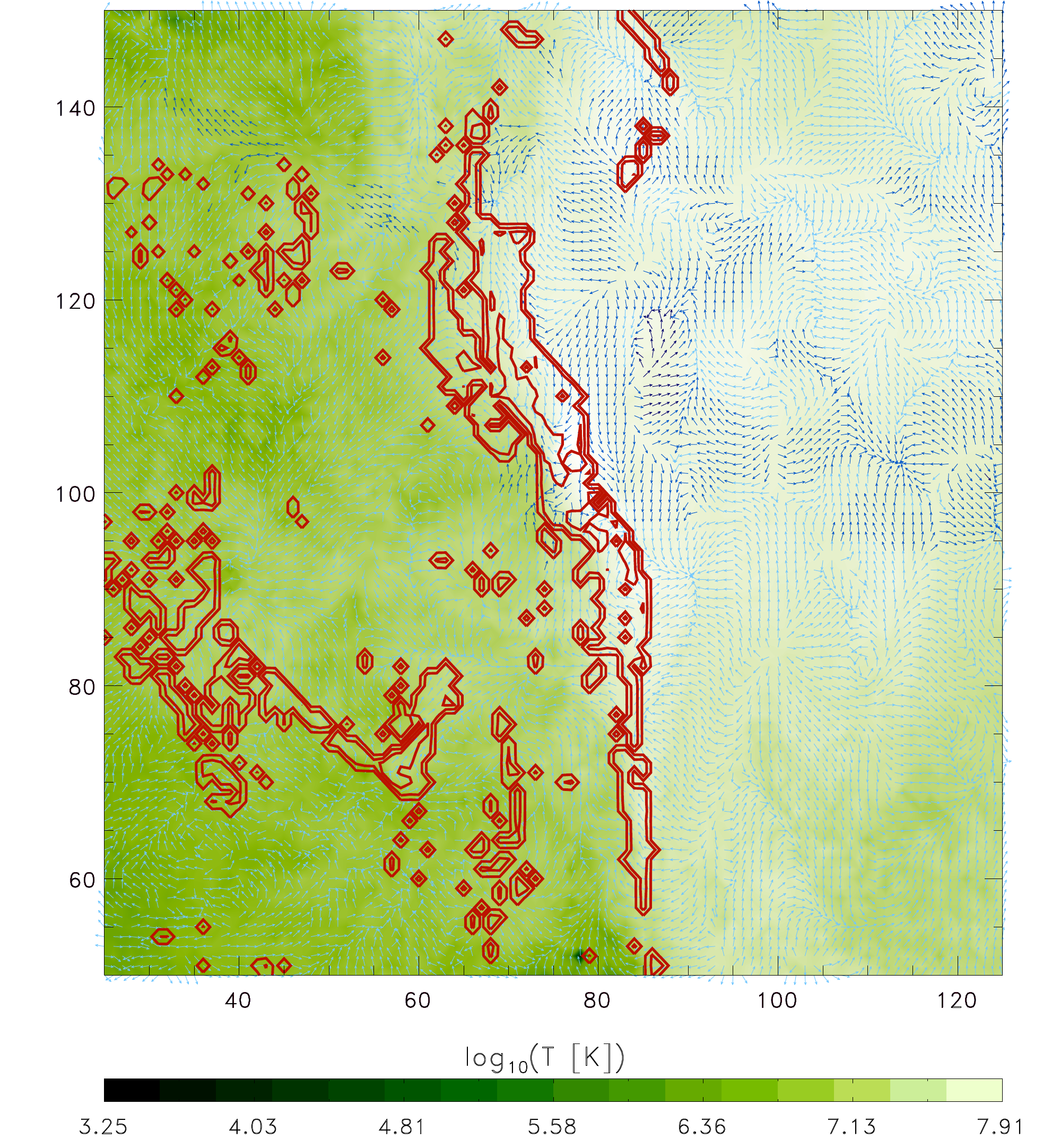}} \\
  \subfigure[]{\includegraphics[width = 0.32\textwidth]{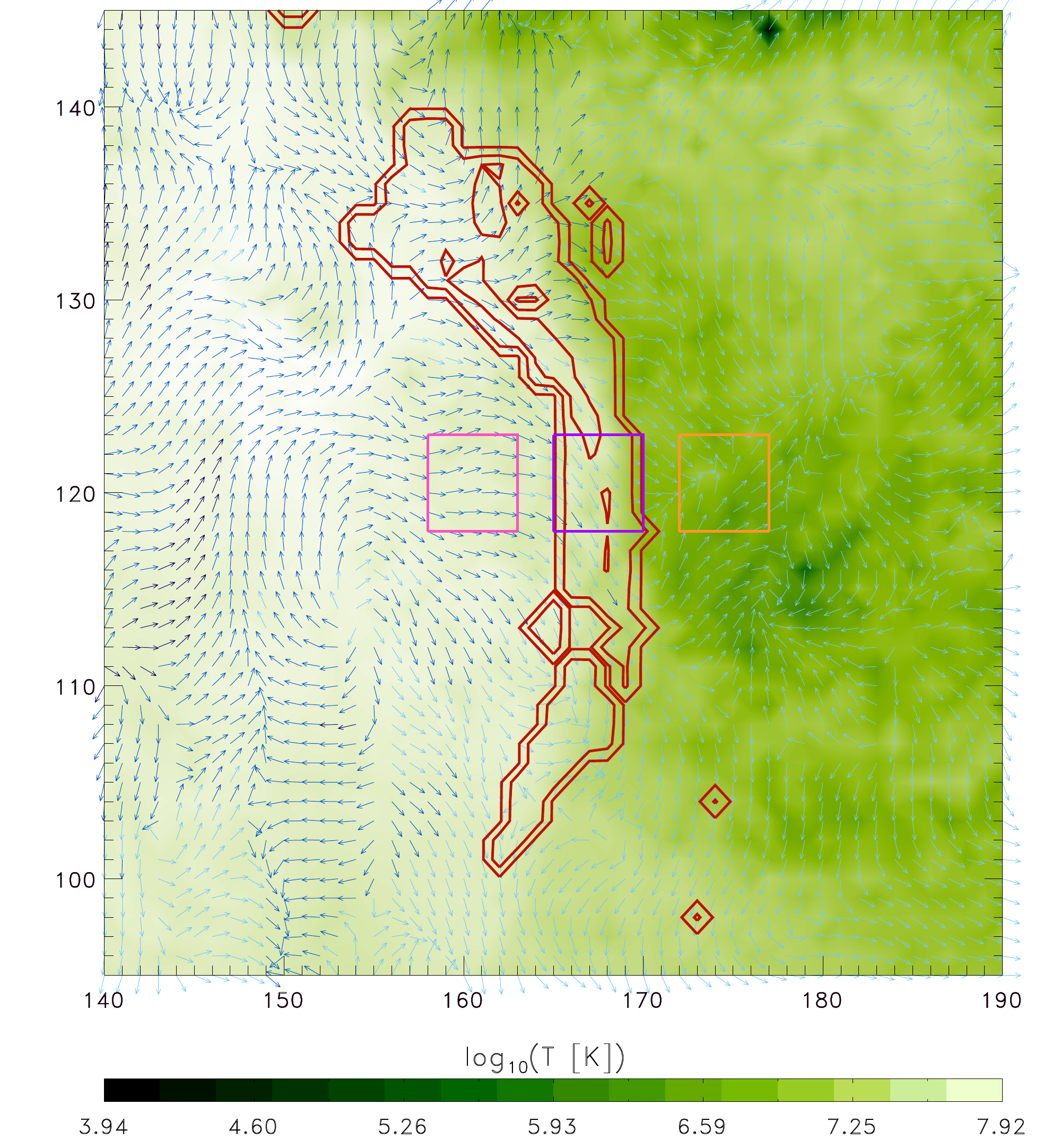} \label{subfig:shockb}}
  \subfigure[]{\includegraphics[width = 0.32\textwidth]{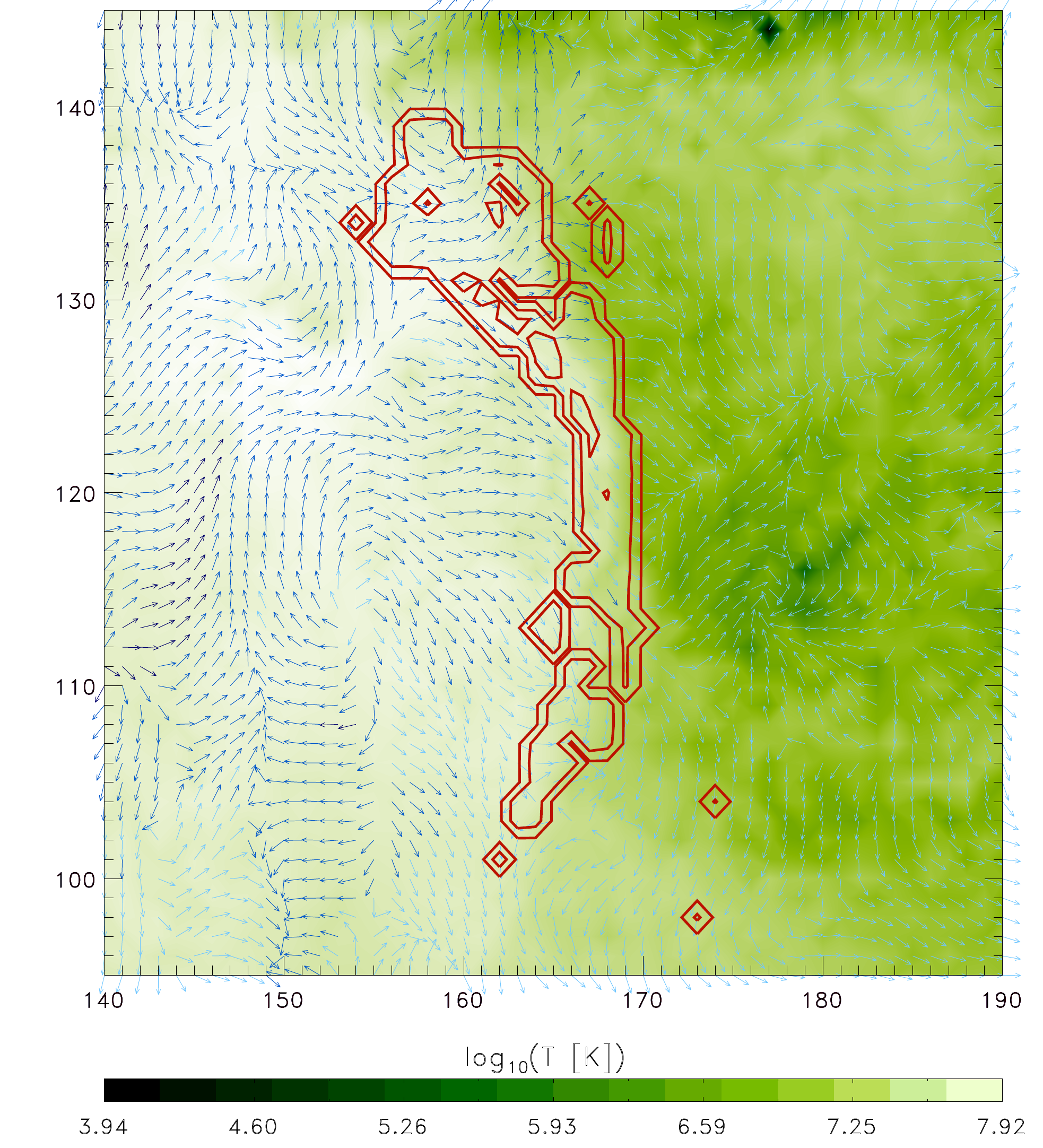}} 
  \subfigure[]{\includegraphics[width = 0.32\textwidth]{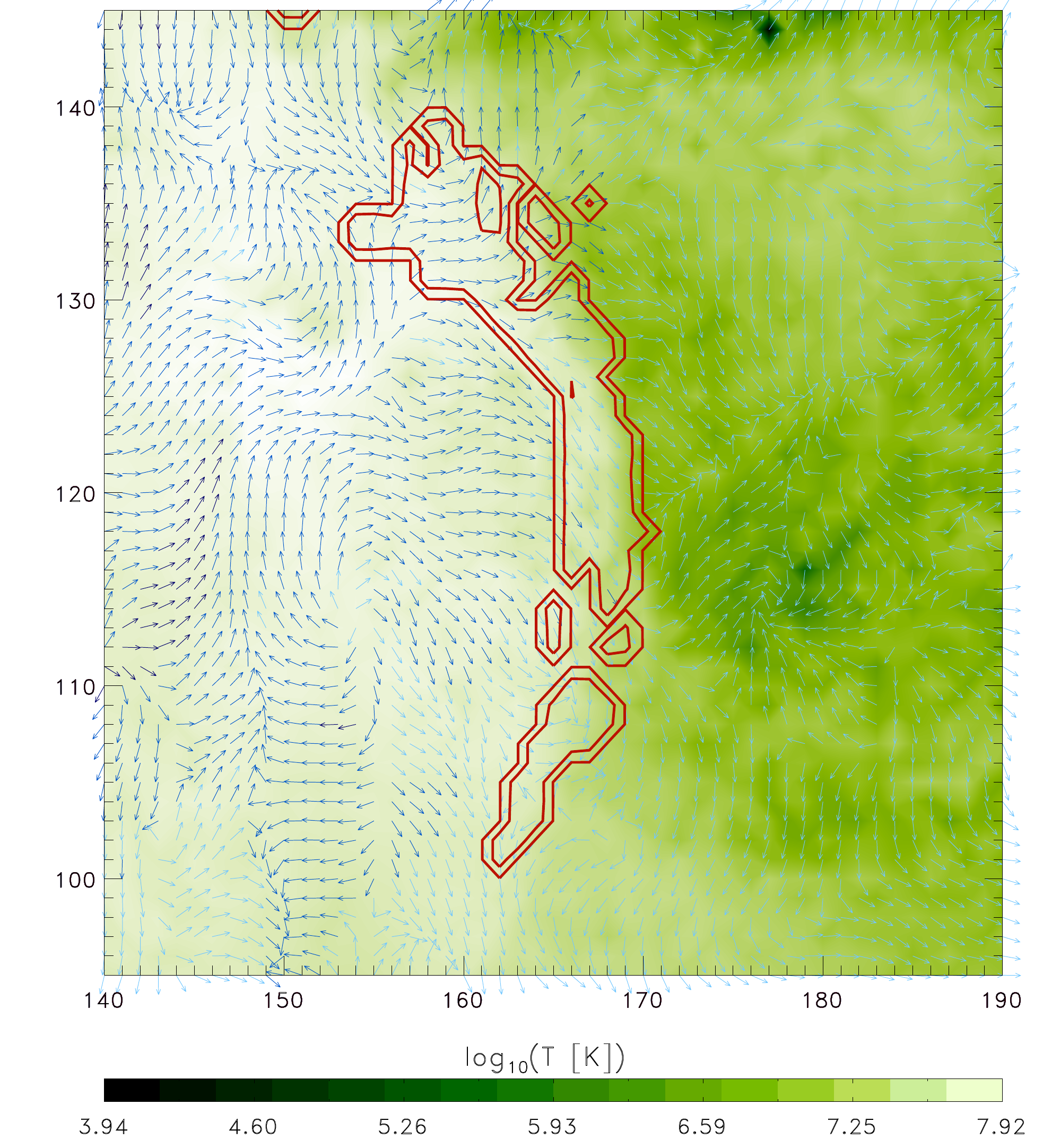}} 
  \caption{Zoomed versions of our simulated radio relics. The upper row (Fig. (a), (b) and (c)) displays relic 1, while relic 2 is shown in the lower row ((d), (e) and (f)). The green colours show the temperature of the ICM. The direction of the arrows indicates the direction of the magnetic field and their colour gives their magnetic field strength, with a logarithmical stretching, while the red contours show the radio emission. The left column shows $\theta_{\all}$. The middle column shows $\theta_{\Perp}$ and the right column shows $\theta_{\para}$. The axis are in $\dd x = \ 31.7 \ \kpc$ units. The squares in Fig. (a) and (d) mark the regions of the tracers selected in Sec. \ref{subsec:closeup}. The colours orange, purple and pink mark the regions in front of, on top of and behind the relic respectively.}
  \label{subfig:TnvCCpc}
 \end{figure*}
 The total energy budget in cosmic ray protons as a function of redshift is obtained by integrating
 \begin{align}
   E_{\mathrm{CR}}  = \int\limits_{z = 1}^{0} \sum\limits_{i = 1}^{N_p} F_{k,i} \Delta t (z_1, z_2) \xi \left( \theta_i \right) \dd z . \label{eq:ECR}
 \end{align}
 over all timesteps. In Eq. \ref{eq:ECR} $F_{k,i}$ is the kinetic or cosmic-ray energy flux, $k \in [\mathrm{CR}, \ \mathrm{gas}]$ and $i \in [\all, \ \para, \ \Perp]$. For simplicity, we neglect energy losses (which is reasonable in the case of this perturbed cluster, which is not characterised by $\geq 10^{-2} \ \mathrm{part}/\cm^3$ gas densities),  and therefore our values represent an upper limit on the cosmic-ray energy at all time steps. In the equations above $\xi \left( \theta_i \right)$ is a Heaviside function which allows us to compute only the energy content for specific obliquities. Therefore we applied $\xi \left( \theta_{\all}\right)$ to let cosmic-rays to be accelerated in all shocked tracers, $\xi \left( \theta_{\para}\right)$ for only parallel shocks and $\xi \left( \theta_{\Perp}\right)$ that only accounts perpendicular shocks. In the following the subscripts $\all$, $\Perp$ and $\para$ will correspond to the above selections. The acceleration efficiencies (see Eq. \ref{eq:eta_of_M}) have been further  reduced by a factor of $2$ in the case of quasi-parallel shocks, following the recent results by \citet{Caprioli_Spitkovsky_2014_ion_accel_I_eff}. \\ 
 Finally, we note that it in the complex flows in galaxy clusters the identification of weak shocks, e.g.  $M \le 1.5$, is made uncertain by numerics, while the injection of cosmic rays is expect to be dominated by $M \gg 2$ in DSA \citep[e.g.][]{Ryu_et_al_2003_shock_waves_large_scale_universe}. For these reasons, we only include shocks with $M > 2$ in the following analysis. \\
 The evolution of the cosmic-ray energy across all tracers is shown in Fig. \ref{fig:evoeg_a}. At $z \approx 0$ the cosmic-ray energy for $\theta_{\all}$ is $\sim 8.9 \%$ of the thermal energy of the gas. The cosmic-ray energy of $\theta_ {\Perp}$ is about $\sim 6.2 \%$ of the gas energy and for $\theta_{\para}$ the cosmic-ray energy is about $\sim 2.6\%$ of the thermal gas energy. Most of the cosmic-ray energy is stored in the particles that have crossed a quasi-perpendicular shock, about $\sim 71\%$. The ratio of $E_{\mathrm{CR}}\left( \theta_{\Perp} \right)$ to $E_{\mathrm{CR}}\left( \theta_{\para} \right)$ is $\sim 2.5$. This ratio stays constant over time. At early redshifts $z  > 0.4$ a higher kinetic energy flux is injected by $\theta_{\Perp}$ due to more cosmic-ray injection by quasi-perpendicular shocks. Between $z \sim 0.6$ and $z \sim 0.25$ the injected energy is about the same for $\theta_{\Perp}$ and $\theta_{\para}$. At $z \sim 0.2$ the kinetic energy is higher for $\theta_{\Perp}$. After $z \sim 0.2$ the injected energy is about the same again for $\theta_{\Perp}$ and $\theta_{\para}$. The ratio of cosmic-ray energy injected by $\theta_{\Perp}$ and $\theta_{\para}$ is, except for a few exceptions, in the range of $\sim 1-5$.
 We computed  the $\gamma$-ray emission following the standard approach described, e.g., in \citet{2010MNRAS.407.1565D}, \citet{2013A&A...560A..64H} and \citet{2015MNRAS.451.2198V} (see also the Appendix \ref{sec:gammaray}). For every shocked tracer we compute the spectral index of the momentum distribution of accelerated cosmic rays as $s = -2 \cdot \frac{M^2 + 1}{M^2 - 1}$. At each timestep we compare the injection spectrum to the spectrum of the existing distribution of cosmic-rays (in case the tracers have been previously shocked already) and the current spectral index is set to the flatter among the two. Averaged over the tracer population, we observe a continuous decrease in the average spectral index until $z \approx 0.25$, indicating that the shocked population of tracers is progressively dominated by weaker and energetic shocks. However, the spectral index experiences a new steep increase caused by a strong shock event, corresponding to the time of the major merger in our simulation. Overall the spectral index varies only modestly, $s_{\max}-s_{\min} \approx 0.15$, across the investigated cluster evolution from $z=1$. \\
 To compare in detail  with the limits set by \textit{Fermi}-LAT, we selected four clusters (A1795, A2065, A2256 and ZwCl1742) of similar masses given in \citet{2014ApJ78718A}, and the Coma cluster \citep{2016ApJ819149A}, all located in the redshift range $z \in [0.02, 0.08]$. We used all upper limits at the low energy range of  $500 \ \MeV$\footnote{In the case of Coma, the limits given by \citet{2016ApJ819149A} are given for the lower energy limit of $100\ \MeV$. Therefore, we rescaled this value to the higher low-energy range of $500\ \MeV$ used in our comparison, as $F (E>500\ \MeV) = F(E>100\ \MeV) \frac{\int\limits^{E_1}_{E_3}A\cdot E^{-\Gamma} \dd E}{\int\limits^{E_2}_{E_3}A\cdot E^{-\Gamma} \dd E}$ assuming a photon index $\Gamma = 2$, $E_1 = 500 \ \MeV$, $E_2 = 100 \ \MeV$ and $E_3 = 1 \ \TeV$.}. Table \ref{tab:refcluster} shows the main properties of those galaxy clusters. \\
 Our simulated  $\gamma$-ray emission (for the energy range of $E \in [0.5, 200] \ \GeV$)  and the observed upper limits derived are given in the first panel of Fig. \ref{fig:gamma}. The $\gamma$-ray emission of our cluster for $\theta_{\all}$ is  $\approx 0.64 \cdot 10^{45}\frac{\ph}{\sek}$ and is above the upper limit for the Coma cluster. If we only use the energy of the cosmic rays gained by crossing parallel shocks (see Eq. \ref{eq:ECR}), the $\gamma$-ray emission is lowered by a factor of $\sim 3.4$. However, even in this case the $\gamma$-ray emission for $\theta_{\para}$ exceeds the observed limits for the Coma cluster.
 \begin{figure}\centering
  \includegraphics[width = 0.5\textwidth]{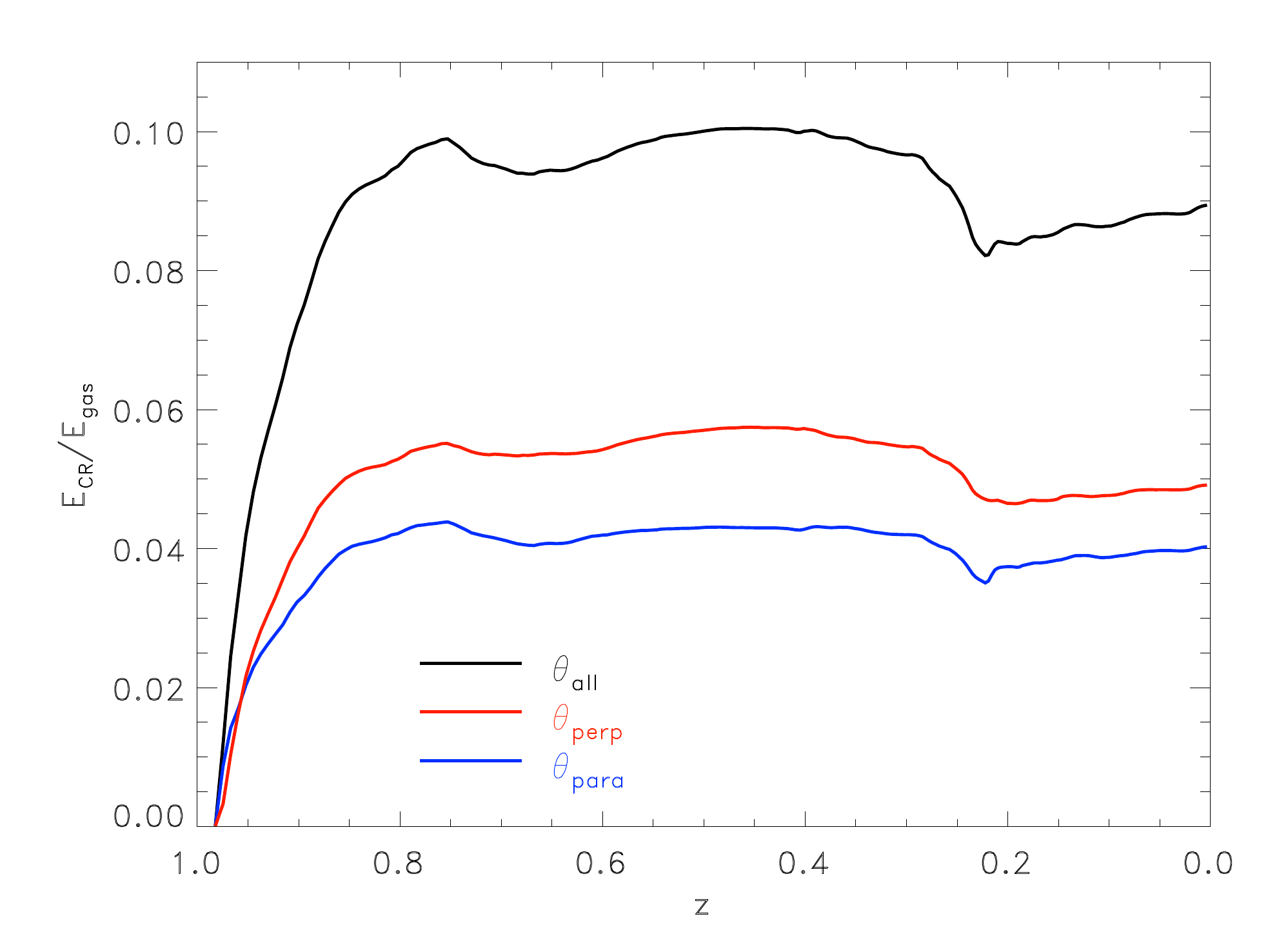}
  \caption{Evolution of the ratio of cosmic-ray to thermal gas energy for $\theta_{\all}$ (black), $\theta_{\Perp}$ (red) and $\theta_{\para}$ (blue) across all tracers.}
  \label{fig:evoeg_a}
 \end{figure}
 \begin{table}\centering
  \begin{tabular}{l | c   | c | c}     
   Name 	& z 	  &  $\Mth  $ & $F_{\gamma}^{\mathrm{UL}}(E > 500 \ \MeV)  $ \\ 
     & 	& $\left[ 10^{15} \ \Msun  \right]$ & $\left[ 10^{45}\frac{\ph}{\sek}\right]$ \\
   \hline \hline
   $\theta_{\all}$ 			& 0.00	 	& 0.97	& 0.640 \\ \hline
   $\theta_{\para}$ 			& 0.00	 	& 0.97	& 0.190 \\ \hline\hline
   $\theta_{\all}$, $B>0.1\mu \G$	& 0.00		& 0.97	& 0.512  \\ \hline
   $\theta_{\para}$, $B>0.1\mu \G$	& 0.00	 	& 0.97	& 0.102  \\ \hline
   $\theta_{\all}$, $B>0.5\mu \G$	& 0.00		& 0.97	& 0.411  \\ \hline
   $\boldsymbol{\theta_{\para}}$, $\boldsymbol{B>0.5\mu \G}$	& \textbf{0.00} & \textbf{0.97}	& \textbf{0.011} \\ \hline\hline
   $\theta_{\all}$, 			& 0.20	 	& 0.66	& 1.047 \\ \hline
   $\theta_{\para}$ 	 		& 0.20	 	& 0.66	& 0.336 \\ \hline
   $\theta_{\all}$, 			& 0.34		& 0.12	& 0.498 \\ \hline
   $\theta_{\para}$ 			& 0.34 		& 0.12 	& 0.152 \\ \hline \hline
   $\theta_{40}$, 			& 0.00	 	& 0.97	& 0.136  \\ \hline
   $\theta_{30}$ 	 		& 0.00	 	& 0.97	& 0.084  \\ \hline
   $\theta_{20}$, 			& 0.00 		& 0.97	& 0.041 \\ \hline\hline
   A1795				& 0.06	 	& 0.95	& 6.068 \\ \hline
   A2065				& 0.07		& 1.09	& 5.256  \\ \hline
   A2256				& 0.06		& 1.18	& 1.075  \\ \hline
   Coma					& 0.02		& 0.96	& 0.035  \\ \hline
   ZwCl1742				& 0.08		& 0.98	& 2.560 \\ 
  \end{tabular}
  \caption{Comparison of the total integrated $\gamma$-ray emission of our different models and a number of observed clusters at the bottom. For each cluster we give the name, redshift $z$, mass $\Mth$ and upper $\gamma$-flux $F_{\gamma}^{\mathrm{UL}}$. The last five rows show the reference clusters taken from \protect\citet{2014ApJ78718A}. The first two rows show our simulations for $\theta_{\all}$ and $\theta_{\para}$. The following rows show the results for the different simulations depending on $B_{\min}$, $z$ and different selections of $\theta$. Our cluster simulation compatible with the \textit{Fermi}-limits for the Coma cluster is highlighted in boldface.
}
  \label{tab:refcluster}  
 \end{table} \\
 \begin{figure}\centering
  \includegraphics[width = 0.5\textwidth]{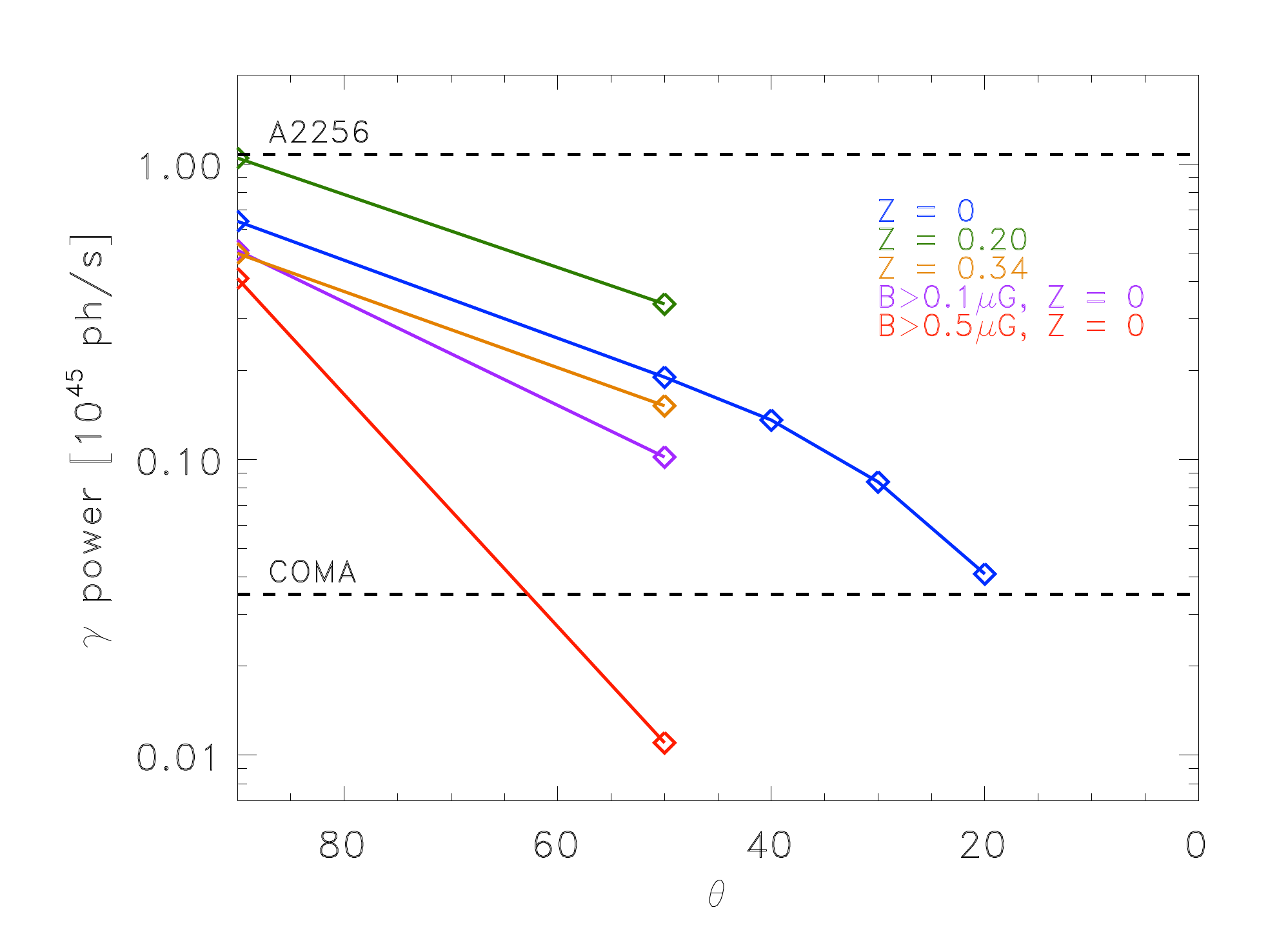}
  \caption{Total integrated $\gamma$-ray emission of our different models (color). The dashed lines show the \textit{Fermi}-limits of A2256 and the Coma cluster taken from \protect\citet{2014ApJ78718A} and \protect\citet{2016ApJ819149A}.}
  \label{fig:gvt}
 \end{figure}
 In the framework of the DSA theory, it is uncertain whether a specific minimum value of upstream magnetization is necessary to scatter the particles enough to enter the DSA acceleration loop. On the one hand, the extrapolation of DSA from the highly magnetized regime of supernova remnants ($\sim 1-100 \ \mu \G$) to the cosmic web is very uncertain. On the other hand, several papers have suggested that collisionless shocks can significantly amplify the upstream magnetic field independently of the initial conditions (e.g. \citealt{2012MNRAS.427.2308D}, \citealt{2013MNRAS.436..294B}, \citealt{2014ApJ...794...46C}). As an explorative study, we investigated the effect of a minimum magnetisation level to allow for DSA, by limiting the acceleration of cosmic-ray protons to upstream fields $B_{\mathrm{up}}  > B_{\min}$. Here we tested the cases of $B_{\min} \geq 0.1 \ \mu \G$ and $B_{\min}  \geq 0.5 \ \mu \G$. The results are shown in \ref{fig:gamma_a}. 
 In both cases the hadronic $\gamma$-ray emission is lowered towards the observed upper limits of the Coma cluster. The emission is significantly lowered towards the upper limit of Coma, if additionally only quasi-parallel shocks with an upstream magnetic field larger than $0.5 \ \mu \G$ inject cosmic rays. In this case the $\gamma$-ray emission drops significantly below the observed upper limit of the Coma cluster by a factor of $\sim 117$.  \\
 The $\gamma$-ray properties discussed above are also present $0.5 \ \Gyr$ before and after the major merger (see second panel of Fig. \ref{fig:gamma}). The role played by shock obliquity on the injection of cosmic rays is found to be as strong as at $z=0$: in both cases the $\gamma$-ray emission drops by a factor of $\sim 3.1-3.3$. But in neither of the cases the hadronic $\gamma$-ray emission is below the upper limit of the Coma cluster.\\
 The $\gamma$-ray emission depends on the value chosen for $\theta_i$ in Eq. \ref{eq:ECR}. We conducted the same experiment using different ranges for $\theta_{\para}$:
 \begin{itemize}
  \item  $\theta_{50}$: $\theta \in [0^{\circ}, 50^{\circ}]$ and $\theta \in [130^{\circ}, 180^{\circ}]$
  \item  $\theta_{40}$: $\theta \in [0^{\circ}, 40^{\circ}]$ and $\theta \in [140^{\circ}, 180^{\circ}]$ 
  \item  $\theta_{30}$: $\theta \in [0^{\circ}, 30^{\circ}]$ and $\theta \in [150^{\circ}, 180^{\circ}]$  
  \item  $\theta_{20}$: $\theta \in [0^{\circ}, 20^{\circ}]$ and $\theta \in [160^{\circ}, 180^{\circ}]$ 
 \end{itemize}
 The $\gamma$-ray emission is reduced every time we restrict the shocks to a smaller range of obliquities (see third panel of Fig. \ref{fig:gamma}). Only in the case of $\theta_{20}$ the $\gamma$-ray emission is close to the limit of the Coma cluster. Therefore, the hadronic $\gamma$-ray emission is not very sensitive to the selection of $\theta_{\para}$. \\
 In summary, with our tracer-based method we tested two possible scenarios to reconcile the hadronic $\gamma$-ray emission from protons accelerated by cluster shocks with the observed upper limits for galaxy clusters \citep[][]{2014ApJ78718A,2016ApJ819149A}. First, we tested how an obliquity switch affects the $\gamma$-ray emission. Second, we studied the effect of a minimum magnetic field strength on the acceleration of cosmic-ray protons and on their hadronic emission. In both cases the $\gamma$-ray emission was reduced, yet the fluxes were not reduced below the limits established by the \textit{Fermi}-LAT observation of the COMA cluster \citep[see][]{2016ApJ819149A}. A combination of both might be a possible explanation for the missing $\gamma$-ray emission as it reduces the fluxes below the \textit{Fermi}-limits. The results of our different simulations are summarized in Table \ref{tab:refcluster} and plotted in Fig. \ref{fig:gvt}. \\
\subsection{Close-up view of the relic regions}\label{subsec:closeup}
 Finally, we take a closer look at the thermodynamical and magnetic properties of particles in the relic regions. We selected three sets of particles in front of (i.e. upstream), on top of and behind (i.e. downstream) the relics seen in figure \ref{subfig:shocka} and \ref{subfig:shockb}. The selected regions are of the size of $158.8 \cdot 1268 \cdot 1268 \ \kpc^3$ for both relics. The number of particles per selection are about $1-9 \cdot 10^3$. \\
 \begin{figure}\centering
      \includegraphics[width = 0.48\textwidth]{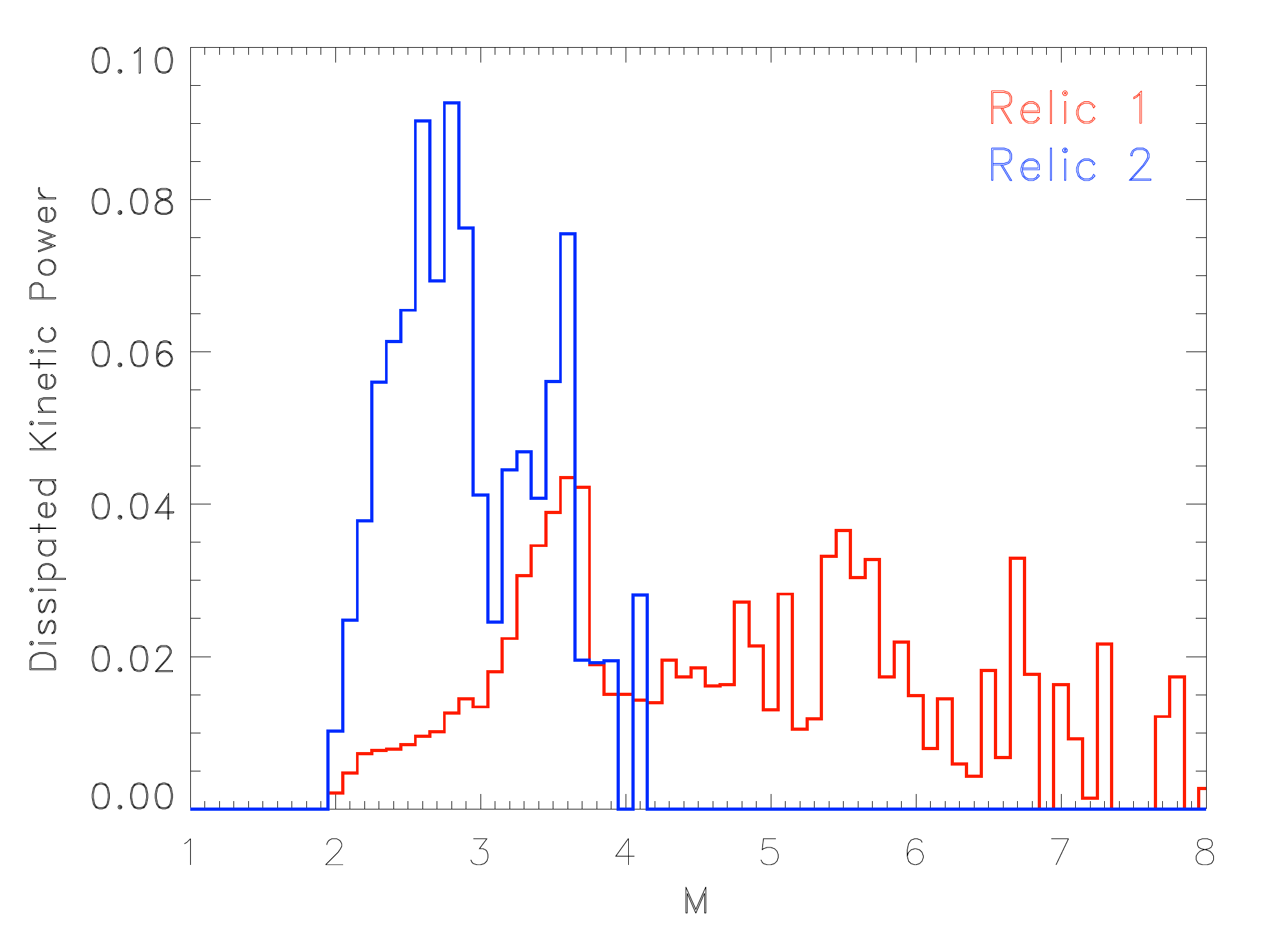}
  \caption{Dissipated kinetic power weighted distribution of the Mach numbers across the two relics.}
 \label{fig:Mdis}
 \end{figure}
 In Fig. \ref{fig:rels_sel} we show the evolution of temperature, magnetic field strength and ratio of compressive to solenoidal energy (using Eq. \ref{eq:div} and Eq. \ref{eq:curl}) across the last two $\Gyr$. The evolution of the temperature and magnetic field strength in all sets of particles is similar at early times. Later on,  the tracers selected to be in the post-shock region at $z \approx 0$ show a significant jump in temperature and magnetic field strength, owing to the compression by the shocks they experienced. The Mach numbers producing the radio emission cover a big range in both relics \citep[consistent with the findings of][]{2010AAS...21543630S}, yet the bulk of radio emission comes from the $M \sim 3.5$ (relic 1) and $M \sim 2.7$ (relic 2) shock (see Fig. \ref{fig:Mdis}). However, the magnetic field varies more than the temperature because of the chaotic evolution downstream of the two shocks. On average, the amplification of the downstream magnetic fields is $\sim 2-3$ at most. This is in line with recent results based on tailored MHD simulations of shocks by \citet{2016arXiv160308518J}, who reported a similar amplification downstream of magnetic fields, mostly due to compressive turbulent motions of $M = 4$ shocks. 
 In the lower panel of Fig. \ref{fig:rels_sel} we show the modal decomposition of small-scale turbulence measured by the tracers: for most of their evolution, the solenoidal velocity is found to be predominant, $\sim 3 - 10 $ times larger than the compressive component. However, relic 1 shows more compressive turbulence from $z \approx 0.15$. Based on Fig. \ref{fig:tacc_post_sigma}, this is likely due to the fact that a large fraction of the gas ending up in relic 1 has crossed the central cluster region, where shocks launched by the major merger have increased the compressive energy component. The tracers connected to relic 1 also seem to have been subjected to a significant injection of cosmic rays by previous shocks. \\
 In Fig. \ref{subfig:mean_mach_ew} we show the evolution of the mean Mach number (weighted for the injected cosmic-ray energy) for the particles ending up behind the relics. We see that up to redshift $z \approx 0.35$ the particles have crossed several weak shocks with values about $\langle M \rangle_{\mathrm{ECR}} \approx 2.5$. In the redshift range $z \approx 0.35 - 0.30$ the particles are exposed to stronger shocks, $\langle M \rangle_{\mathrm{ECR}} \approx 4$. These events correlate with the time of the major merger observed in our cluster. After $z \approx 0.2$ the tracers ending up in relic 1 have been crossed by several strong shocks, whereas the particles connected to relic 2 have only been crossed by a strong shock close to the
 major merger,  at $z \approx 0.2${\footnote{We notice that the apparent floor of $\langle M \rangle_{\mathrm{ECR}} \approx 2$ is a result of restricting only $M>2$ shocks to inject cosmic rays (see Sec. \ref{subsec:tracer})}}. \\
 We also study the occurrence of multiple shocks on the particles swept by relics, by computing the average number of times each tracer has been crossed by shocks of a given Mach number shown in Fig. \ref{subfig:shocked_post_sel}, and the corresponding shocked mass fraction shown in Fig. \ref{subfig:rho_frac_post_sel}. For both relics we observe a continuous increase in the average number of particles shocked by $M > 1.5$ shocks and by $z=0$  basically all particles have been shocked at least once by a $M \geq 1.5$ shock.  Less particles are crossed by $M>2$ or $M>3$ shocks, especially before the last major merger.  By $z \sim 0.1$, $\sim 40-60 \%$ of particles in both selected regions have been already shocked by $M \geq 2$ shocks, while only $\sim 10- 20 \%$ of the particles have been shocked by $M \geq 3$ shocks. This finding suggests that a large fraction of radio emitting particles present in relics may have been subject to several cycles of DSA (re)acceleration over their lifetime.\\
Finally, we found no evidence supporting the possibility of significant turbulent re-acceleration \citet{2015arXiv151101897F} of radio emitting electrons neither in the upstream nor downstream of relics, owing to the  typically long ($\geq 1-10 \ \Gyr$) acceleration time on our tracers, which are much larger than the 
typical radiative cooling time of these particles. However,  we defer to future work a more systematic analysis of this scenario, which also requires to carefully model the balance of energy gain and losses of radio emitting electrons in a time-dependent way \citep[e.g.][]{2014MNRAS.443.3564D}.
 \begin{figure}\centering
   \subfigure[]{\includegraphics[width = 0.5\textwidth]{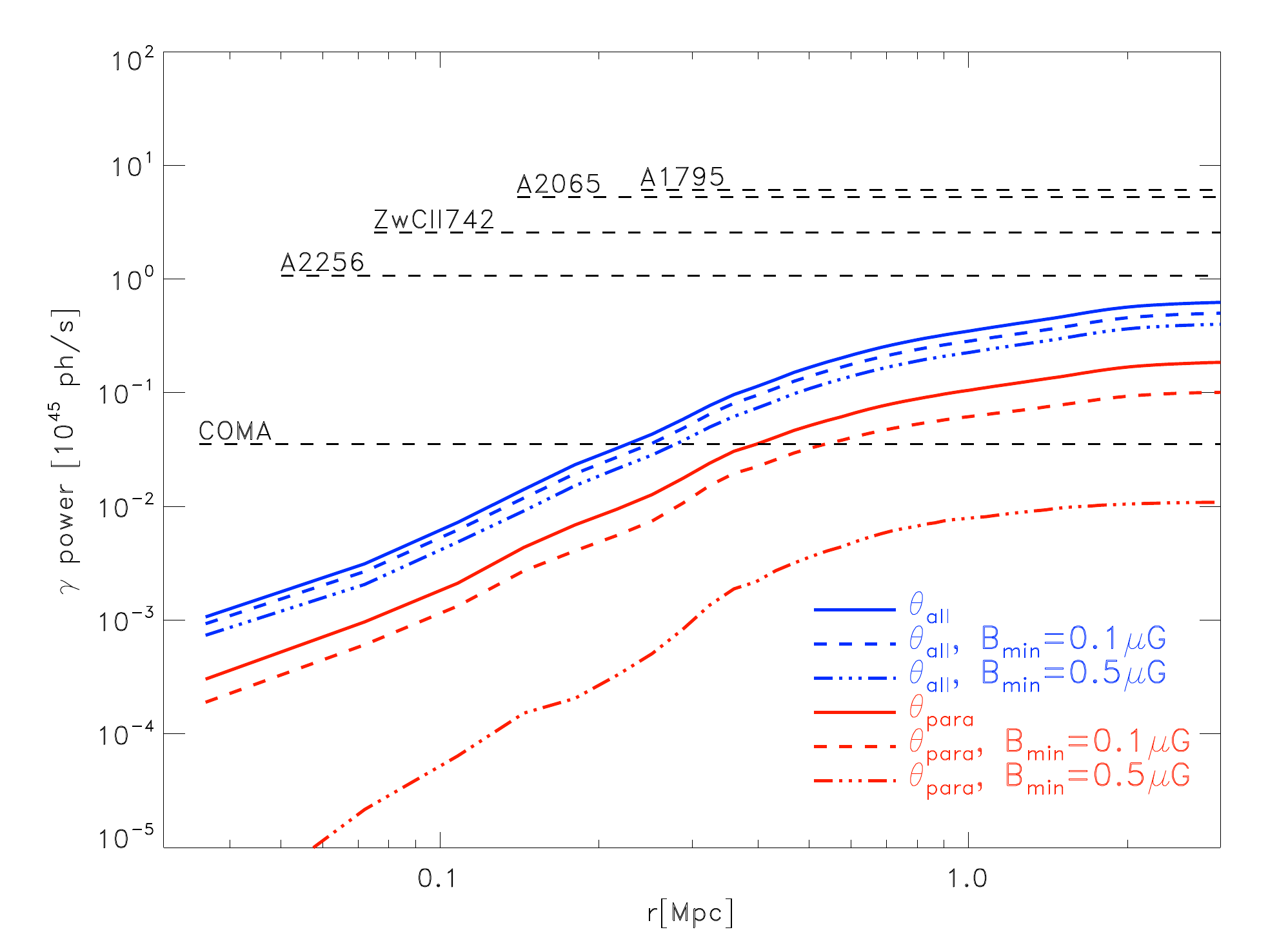}\label{fig:gamma_a}} 
   \subfigure[]{\includegraphics[width = 0.5\textwidth]{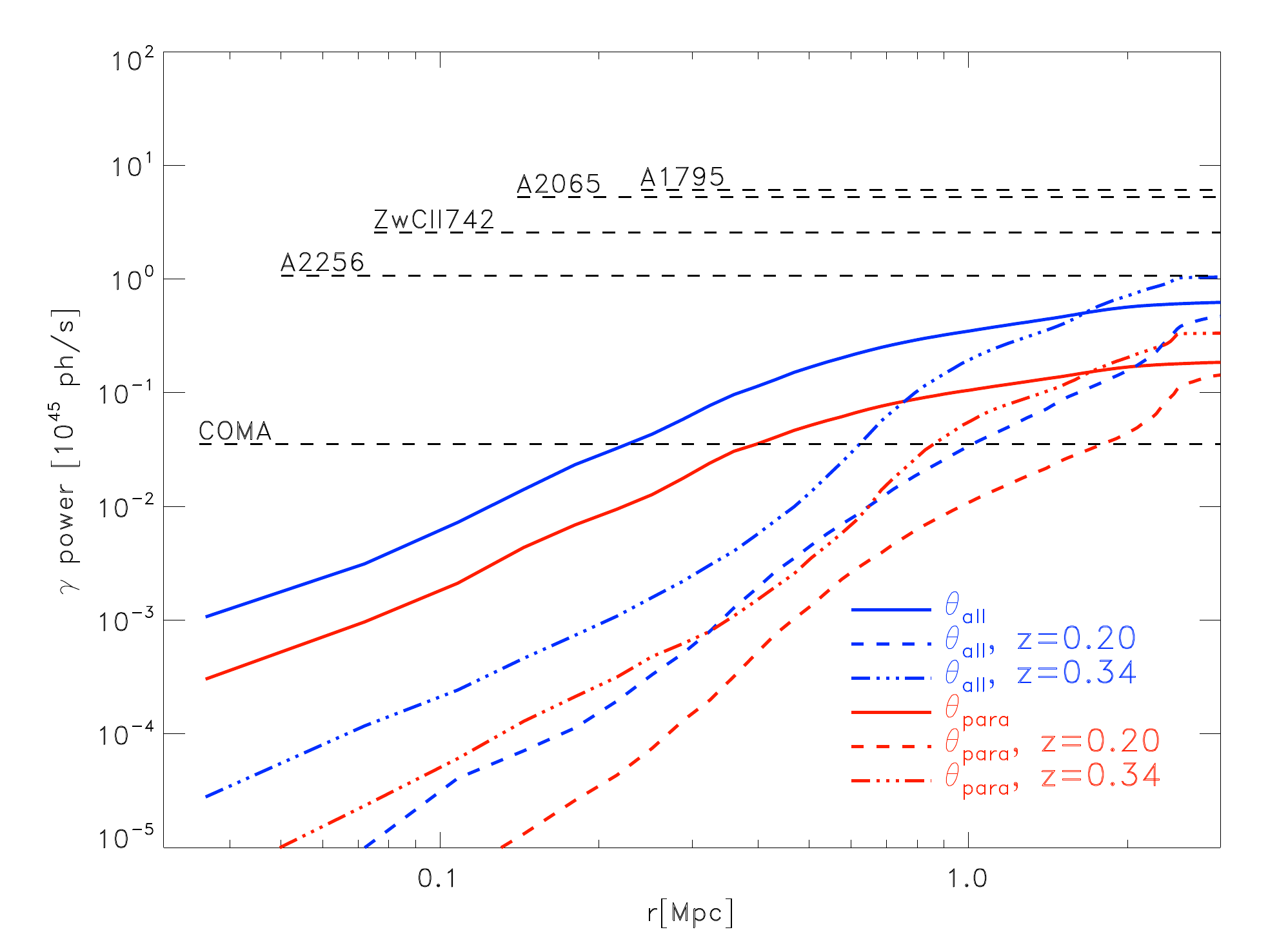}\label{fig:gamma_b}} \\
   \subfigure[]{\includegraphics[width = 0.5\textwidth]{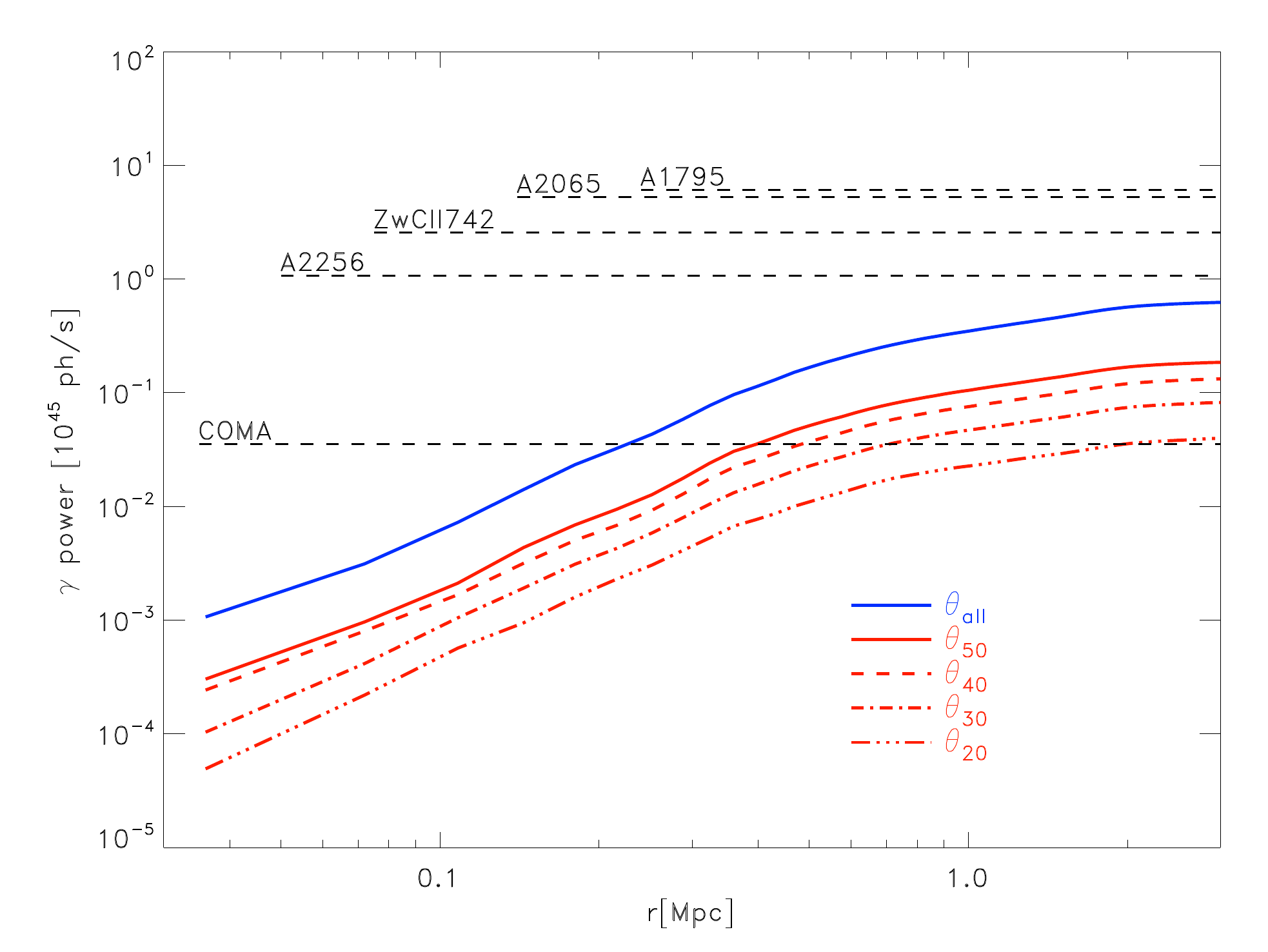}\label{fig:gamma_c}} \\
   \caption{Integrated $\gamma$-ray emission for all tracers (blue, solid line) and for the tracers that only experienced quasi-parallel shocks (red, solid line). The dashed lines in panel (a) show the results for the additional requirement of a minimum magnetic field. Panel (b) gives the results at different redshifts. Panel (c) shows the results for different ranges of $\theta$. In all plots the horizontal dashed black lines give the \textit{Fermi}-limits derived by \protect \citet{2014ApJ78718A}.}
  \label{fig:gamma}
 \end{figure}
 \begin{figure}
   \subfigure[]{\includegraphics[width = 0.5\textwidth]{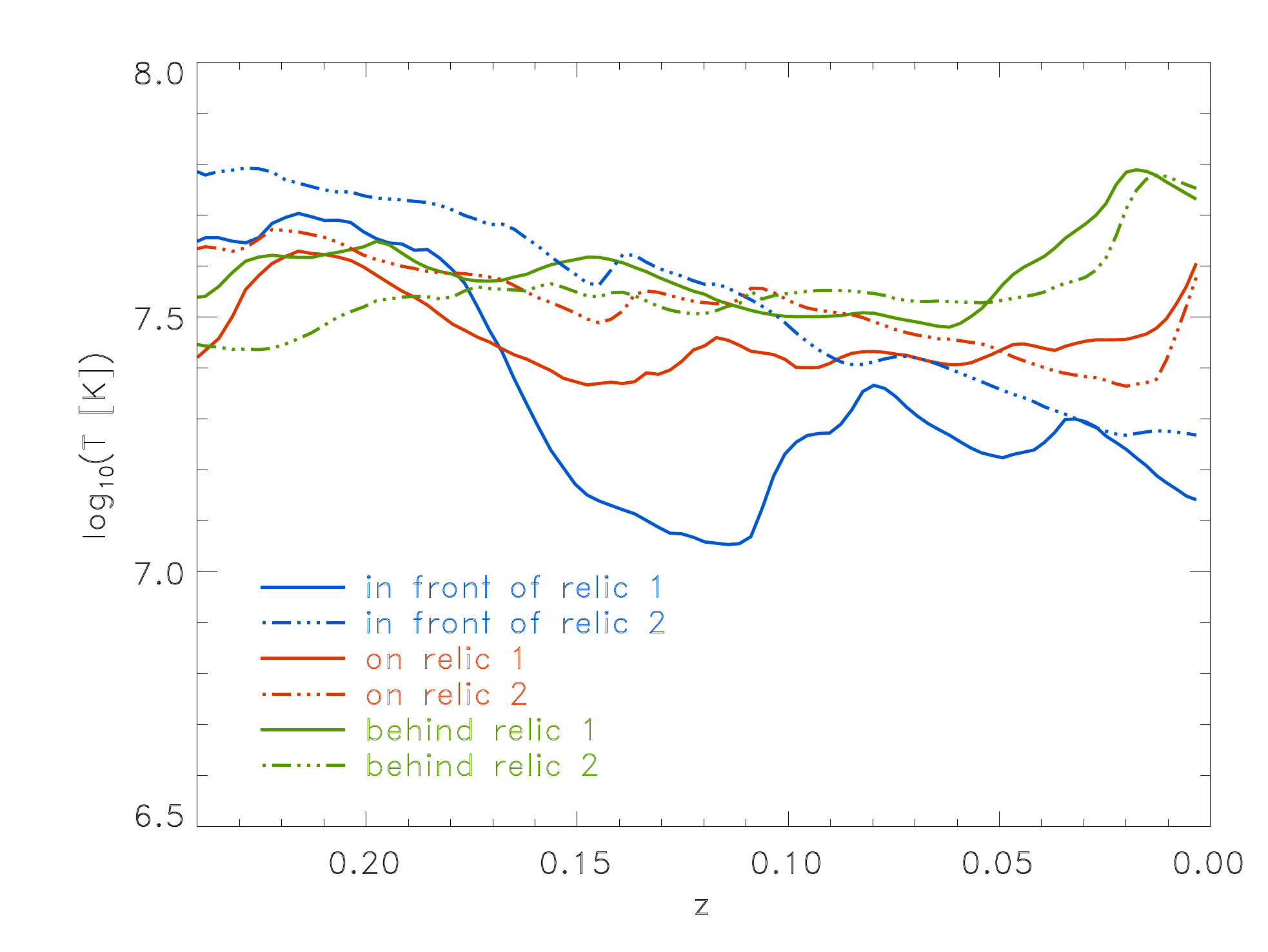}\label{subfig:T_rels_sel}} \\
   \subfigure[]{\includegraphics[width = 0.5\textwidth]{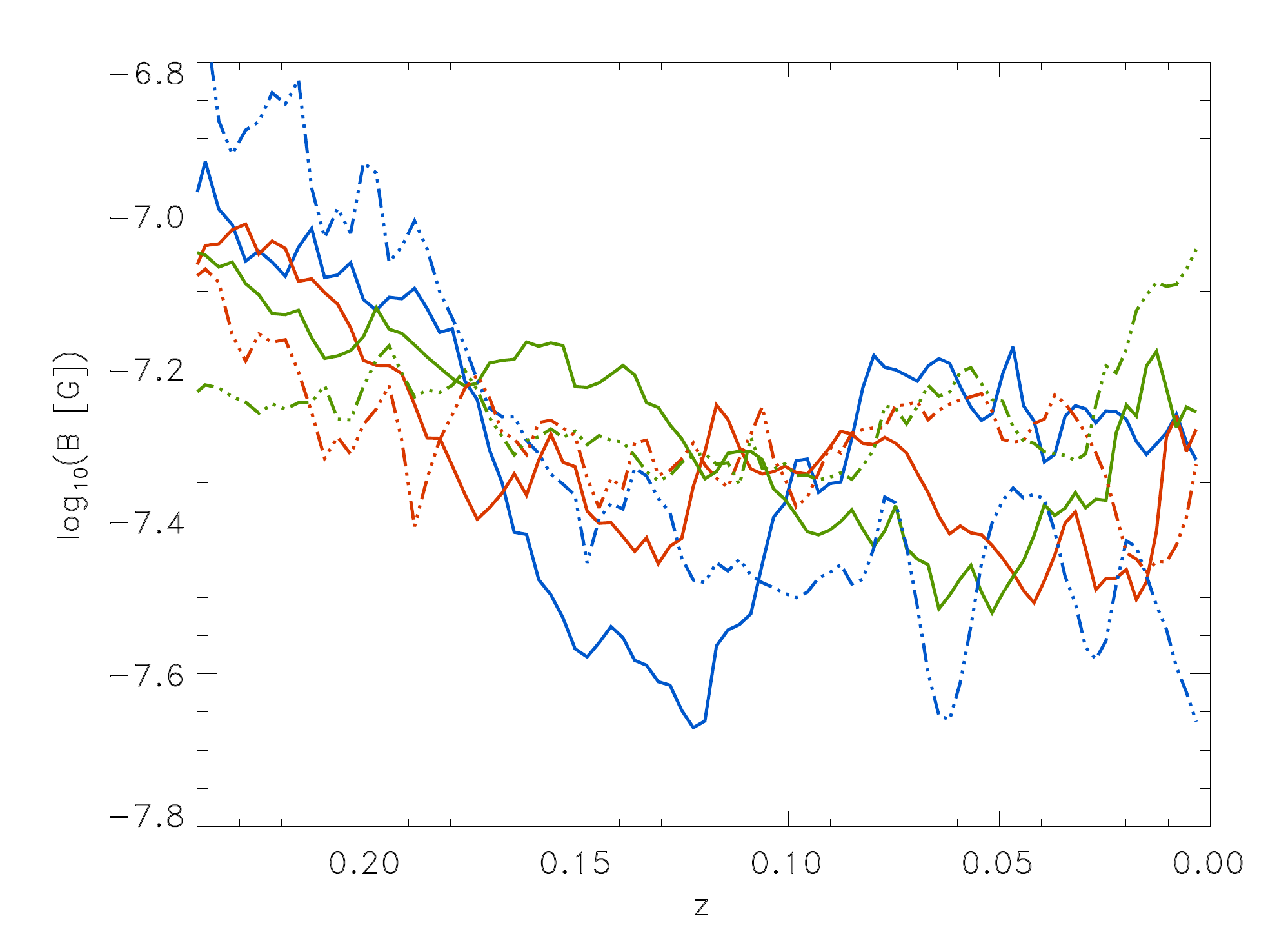}\label{subfig:B_rels_sel}} \\
   \subfigure[]{\includegraphics[width = 0.5\textwidth]{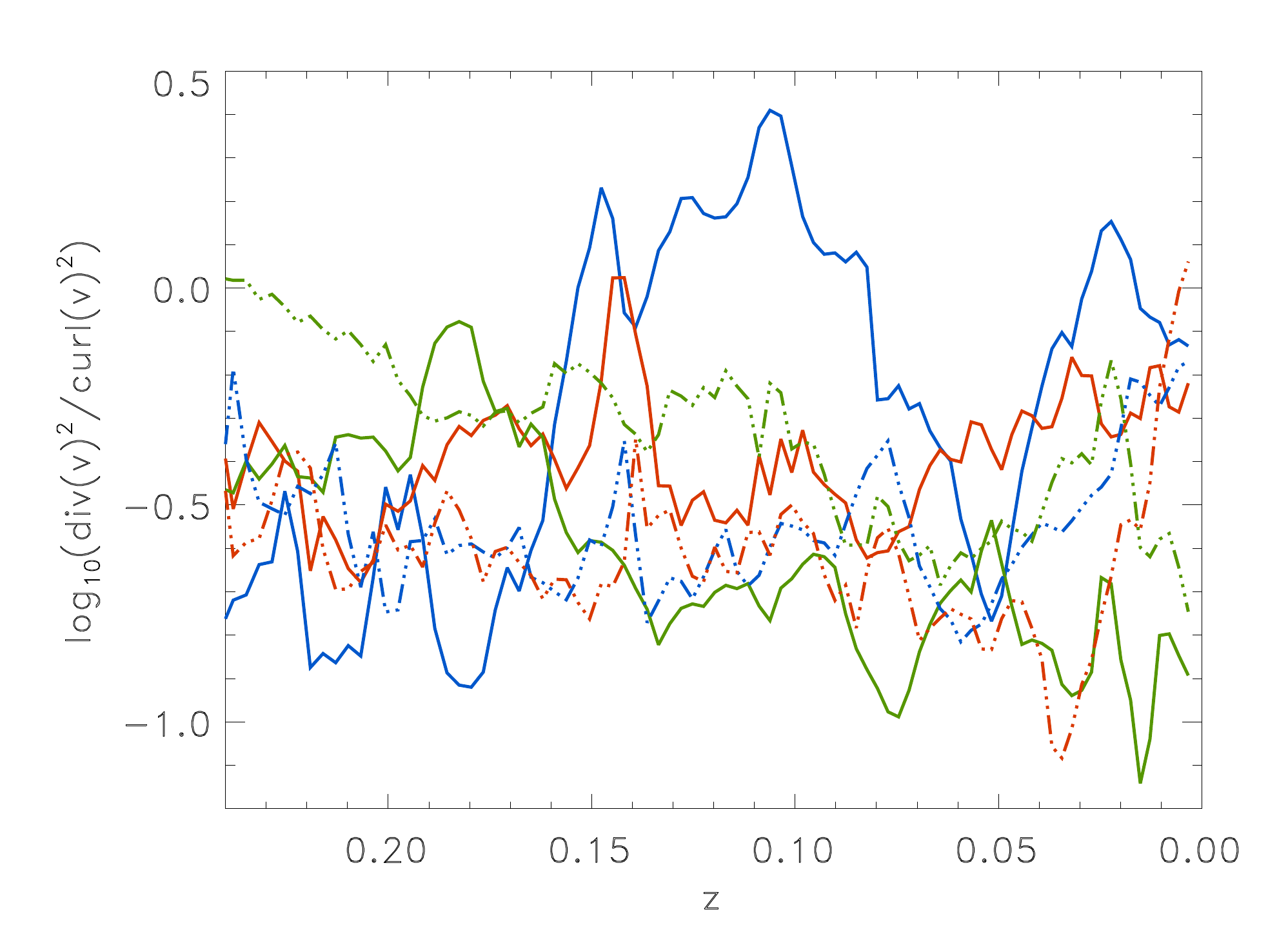}\label{subfig:dc_rels_sel}}
   \caption{Evolution of (a) the temperature, (b) the magnetic field and (c) the ratio of compressive and solenoidal turbulent energy of the selected tracers over the last two Gyr. The solid lines show the selection of relic 1 and the dashed lines show the selection of relic 2. The colours indicate if the selection is upstream of the relic (green), on top of the relic (red) or downstream of the relic (blue).}
   \label{fig:rels_sel}
 \end{figure}
\begin{figure}
  \subfigure[]{\includegraphics[width = 0.5\textwidth]{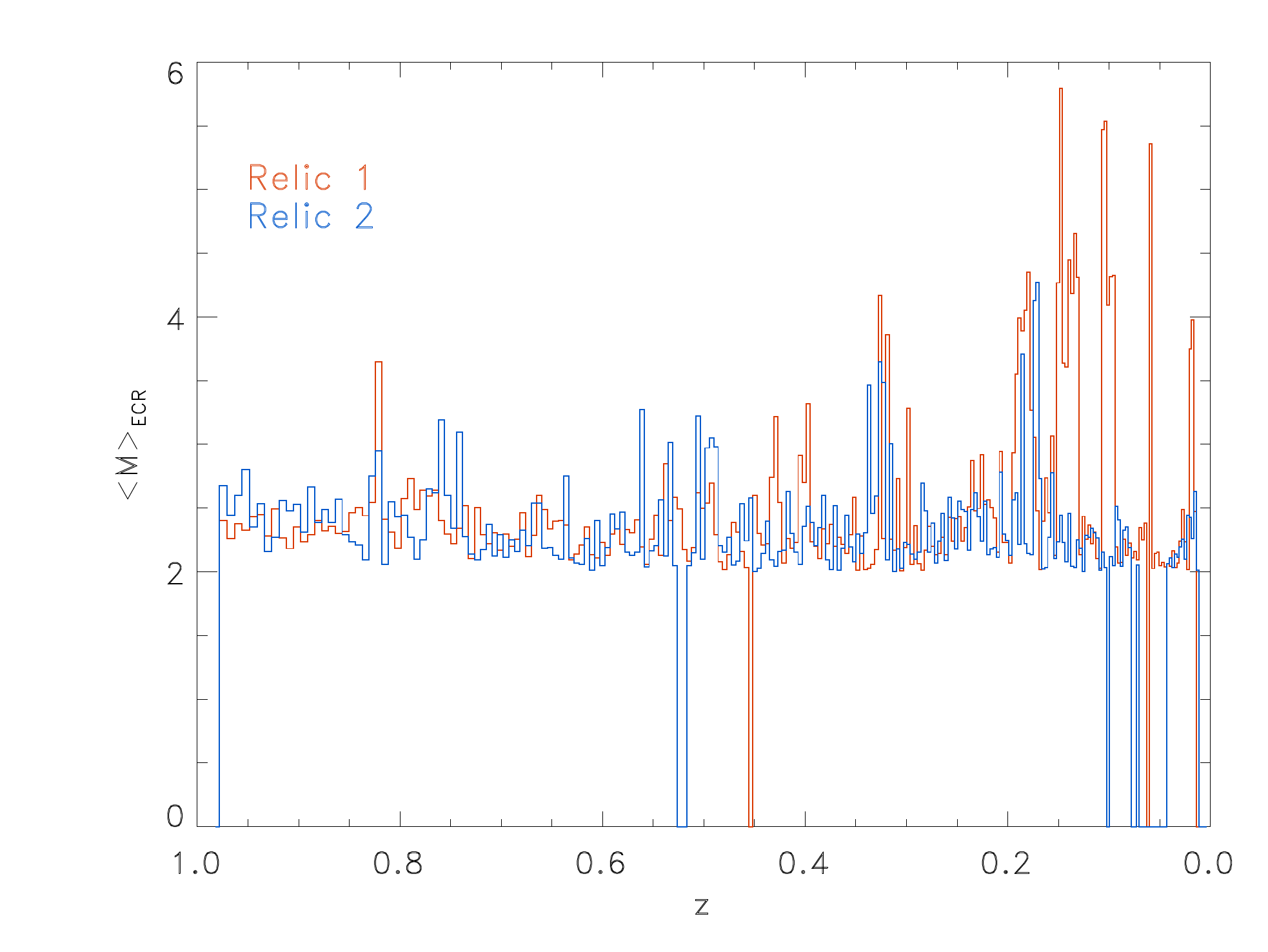}\label{subfig:mean_mach_ew}} \\
  \subfigure[]{\includegraphics[width = 0.5\textwidth]{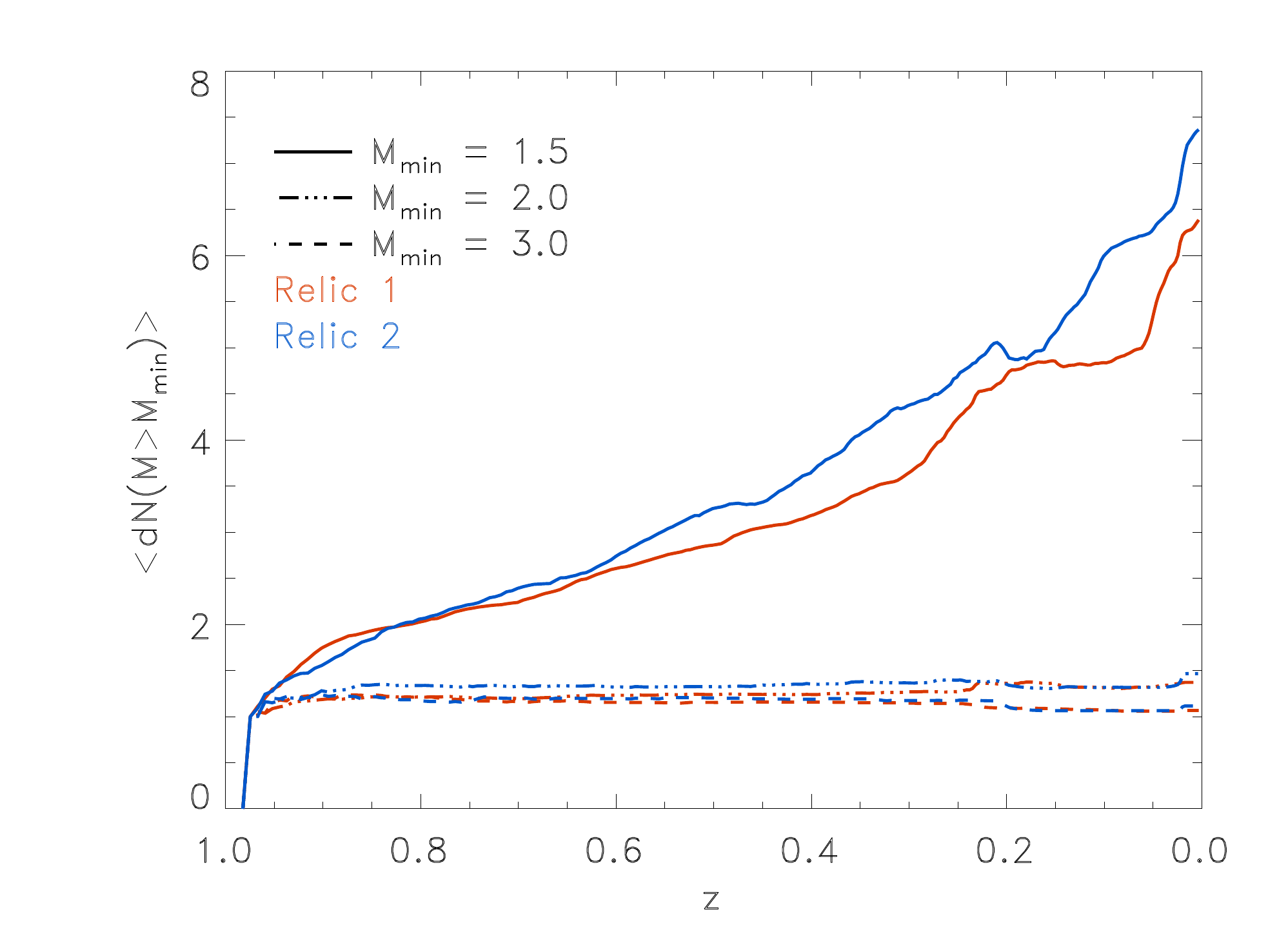}\label{subfig:shocked_post_sel}} \\
  \subfigure[]{\includegraphics[width = 0.5\textwidth]{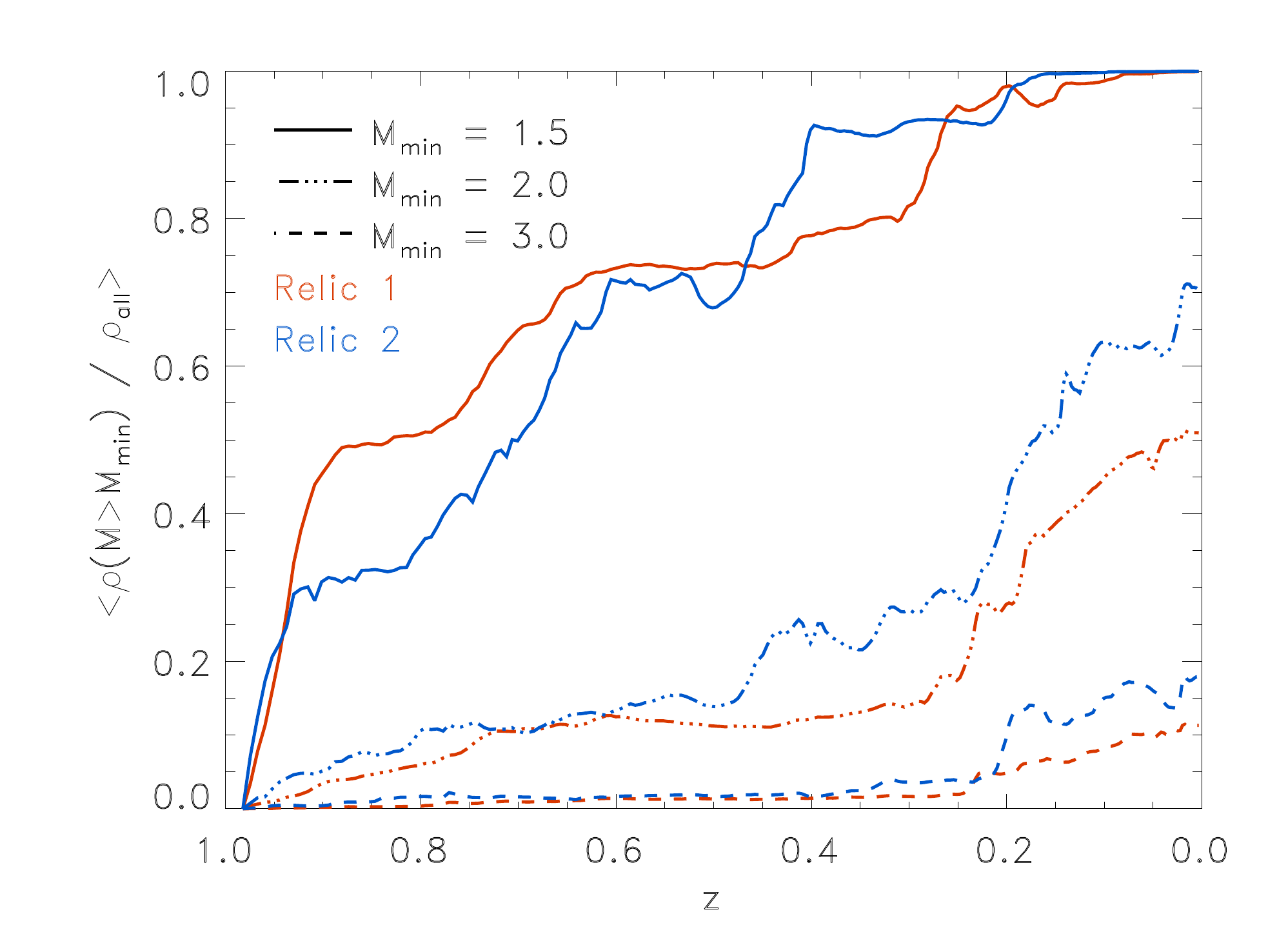}\label{subfig:rho_frac_post_sel}}
  \caption{Evolution of (a) the cosmic-ray energy weighted Mach number, (b) frequency of $M>M_{\min}$ shocks sweeping tracers and (c) mass fraction for the particles ending up behind the relics. The red lines show the first relic and the blue lines show the second relic. In panels (b) and (c) the solids line show $M_{\min} = 1.5$, the long dashed lines show $M_{\min} = 2.0$ and the short dashed lines show $M_{\min} = 3.0$.}
\end{figure}
\section{Discussion \& Conclusions}\label{sec:summary} 
 We have studied the Lagrangian properties of gas in a galaxy cluster over the course of its history and tracked the energy evolution of cosmic rays accelerated by shock waves. Thus, we could test a few  variations of the DSA picture for the acceleration of cosmic rays at shocks, with the aim of reproducing, both, the observed occurrence of radio emission in radio relics and the lack of $\gamma$-ray emission from galaxy clusters. Focusing on the evolution of a massive cluster with a major merger at low redshift, we obtained the following results:  
 \begin{itemize}
  \item We measured the distribution of shock obliquities across the cluster, which is very nearly consistent with an isotropic field distribution. After shock passage, the pre-shock distribution gets progressively more concentrated towards $90^{\circ}$. (see Sec. \ref{subsec:obliquity})
 
  \item We studied how the radio emission changes if only quasi-perpendicular shocks are able to accelerate particles \citep[e.g.][]{Guo_eta_al_2014_I, Guo_eta_al_2014_II}. The radio emission is not much affected by any obliquity switch, i.e. it drops by $\sim 40\%$ if only quasi-perpendicular shocks are taken into account, still producing detectable radio relics. (see Sec. \ref{subsec:RadioEmission})
 
  \item We used a similar restriction on obliquity to limit the acceleration of cosmic-ray protons to 
  quasi-parallel shocks \citep[e.g.][]{Caprioli_Spitkovsky_2014_ion_accel_I_eff} and we computed the resulting hadronic $\gamma$-ray emission.  Over the cluster, the injected cosmic-ray proton energy is on average reduced  by $\sim 3.6$  if DSA is allowed only for quasi-parallel shocks. For the investigated cluster, this is still not enough to decrease the predicted $\gamma$-ray flux below the present constrains by \textit{Fermi}-LAT on the Coma cluster. (see Sec. \ref{subsec:gammaEmission})

 \item Only by limiting the acceleration of cosmic-ray protons to shocks with $\theta \leq 20^{\circ}$ the hadronic emission from our cluster is found to be close to the upper limits of the Coma cluster.
 
  \item We have tested the reduction in cosmic-ray proton acceleration resulting from imposing a minimum magnetization level. Only for a minimum magnetic field $\geq 0.5 \mu G$ the $\gamma$-ray emission decreases below the \textit{Fermi}-LAT limits, also for the Coma cluster. Combining the requirement of a minimum magnetic field and only using proton injection by quasi-parallel shocks, the $\gamma$-ray emission decreases by a factor of $\sim 117$ and is below the \textit{Fermi}-limits. In this case, the predicted hadronic $\gamma$-ray emission should lie only a factor of a few below the limits by \textit{Fermi}. (see Sec. \ref{subsec:gammaEmission})
  
  \item The gas in the post-shock region of relics has been shocked about $7-8$ times more often by a $M=1.5$ shock than by  $M=3$ shocks. While the observed $\gamma$-ray spectrum is dominated by the few strong shocks observed in the past, the cosmic-ray energy is dominated by re-acceleration of weak shocks at lower redshift. (see Sec. \ref{subsec:gammaEmission} and Sec. \ref{subsec:closeup})
  
  \item We did not find evidence supporting acceleration of electrons via \textit{Fermi}-II re-acceleration, neither upstream nor downstream of relics. (see Sec. \ref{subsec:closeup})

 \end{itemize}

 Our study has shown that if DSA operates very different for different shock obliquities the acceleration of cosmic-rays in the ICM can be modified at a significant level compared to what has been assumed so far. If the acceleration of cosmic-ray protons is limited to {\it quasi-parallel} shocks \citep[e.g.][]{Caprioli_Spitkovsky_2014_ion_accel_I_eff} the resulting hadronic $\gamma$-ray emission decreases towards the upper limits by \textit{Fermi}, alleviating the reported tension with observations \citep[][]{2014MNRAS.437.2291V, 2015MNRAS.451.2198V}. It is not possible to make any conclusive assessment based on our comparison with the Coma cluster, because the Coma cluster is in a different dynamical state, minor merger, than our simulated cluster, major merger. Conversely, the radio emission from merger shocks (i.e. radio relics) is changed at a level which is still compatible with observations if only  {\it quasi-perpendicular} shocks can accelerate the cosmic-ray electrons (e.g.\citealt{Guo_eta_al_2014_I} and \citealt{Guo_eta_al_2014_II}). This is because in the regions where radio relics are typically formed, the magnetic field is so tangled that the distribution of angles closely follows the random distribution, which peaks towards $90^\circ$. \\
 As a concluding caveat, in this work we did not include any microphysics such as microscopic magnetic field generation in a shock \citep[e.g.][]{2013MNRAS.436..294B} or microscopic plasma instabilities \citep[e.g.][]{2014PhRvL.112t5003K}. Therefore, we restricted ourselves to the assumption that the magnetic field obliquity (and strength) observed at the scales resolved in this simulation are preserved down to much smaller scales where cosmic rays are accelerated via DSA and SDA. The validity of this assumption can only be tested in future work, where we plan to combine these results with tailored PIC simulations of cosmic shocks. \\
\section*{acknowledgements}
 The cosmological simulations were performed using the {\enzo} code (http://enzo-project.org) and were partially produced at Piz Daint (ETHZ-CSCS, http://www.cscs.ch) in the Chronos project ID ch2 and s585, and on the JURECA supercomputer at  the NIC of the Forschungszentrum J\"{u}lich,  under allocations no. 7006 and 9016 (FV) and 9059 (MB). DW acknowledges support by the Deutsche Forschungsgemeinschaft (DFG) through grants SFB 676 and BR 2026/17. FV acknowledges personal support from the grant VA 876/3-1 from the DFG. FV and MB also acknowledge partial support from the grant FOR1254 from DFG.  \\
 The distances in Sec. \ref{subsec:gammaEmission} and Appendix \ref{sec:gammaray} have been computed using the Ned Cosmology Calculator \citep{2006PASP..118.1711W}.  \\
We thank our anonymous referee for the useful feedback, that helped improving the final quality of this paper.  We also acknowledge fruitful discussions with T. Jones, K. Dolag and C. Gheller. \\
 \bibliographystyle{mnras}
 \bibliography{mybib}
 \appendix
\section{Density distribution}\label{sec:density}
\begin{figure}
  \includegraphics[width = 0.49\textwidth]{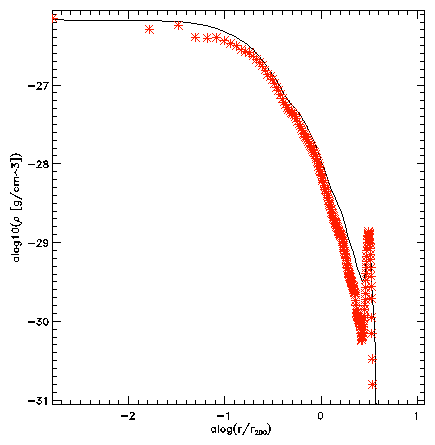}
  \caption{Density profile at $z \approx 0$ of the simulated cluster. The solid black line shows the \enzo-profiles and the red asterixs show the profile computed with the tracers.}
  \label{fig:densprof}
 \end{figure}

\begin{figure*}
  \includegraphics[width = 0.49\textwidth]{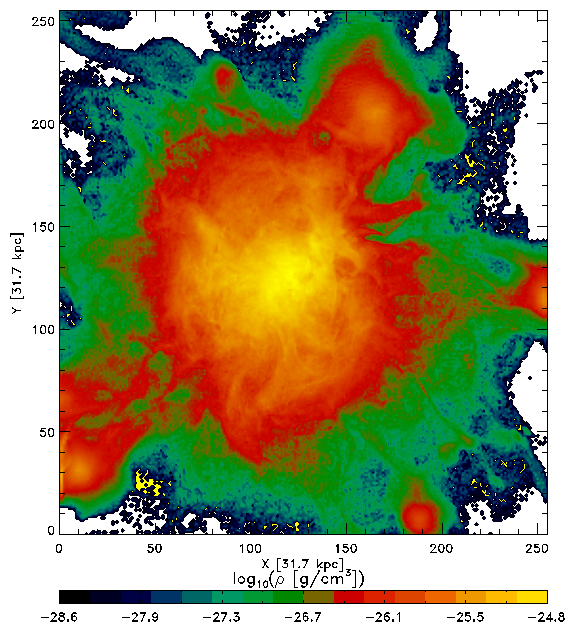}
  \includegraphics[width = 0.49\textwidth]{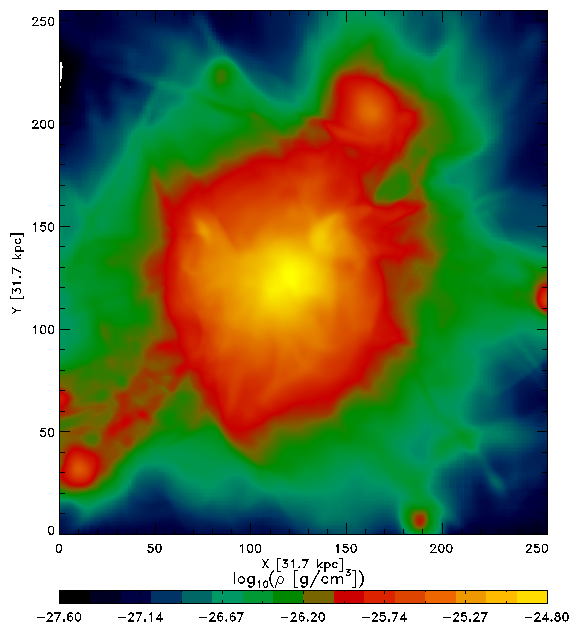}
  \caption{Density projections at $z \approx 0$. The left panel shows the projected density computed with tracers, while the right panel shows the projected density directly simulated in the \enzo \ run.}
 \label{fig:densmap}
 \end{figure*}

 In order to use Lagrangian tracers for analysing the properties of the \enzo-simulation we need to verify that the tracers accurately follow the mass distribution of the cluster. Fig. \ref{fig:tacc_post_sigma} only gives a visual impression that the tracers are following the mass in a correct way. 
 The density profiles are shown on Fig. \ref{fig:densprof}. Especially in the cluster outskirts, where most of the cosmic rays are accelerated, the profile is very well sampled. Only in the cluster centre the tracers have a lower density profile compared to the real gas density, which can be explained by numerical diffusion  (see \ref{subsec:tracer}) as well as by the finite mass resolution of the tracers. In Fig. \ref{fig:densmap} we compare the projected density of the \enzo-simulation and the tracers at redshift $z \approx 0$. The tracers show more detailed structures than the \enzo-simulation. We also observe that the tracers cannot resolve the masses smaller than their threshold (e.g. empty regions in the right panel). 
  \\
\section{Accretion shocks and filaments}\label{sec:tracA2}
 We ran the same analysis of the main paper also for a smaller cluster with a mass of $2.8 \cdot 10^{14} \ \Msun$. This cluster is part of an other \enzo-simulation with a root grid consisting of $256^3$ cells with a resolution of $\dd x = 292\ \kpc$. The used cosmological parameters are: $H_0 = 70.2 \ \km \ \sek^{-1} \ \Mpc^{-1}$, $\Omega_\mathrm{M} = 0.272$ and $\Omega_{\mathrm{\Lambda}} = 0.728$ and $\sigma_8 = 0.8$. The tracers were evolved on a $128^3$ subgrid with resolution $\dd x = 36.62 \ \kpc$ from $z = 1$ to $z = 0$ for a total of 192 timesteps. The final number of tracers is $N_p(z = 0) \approx 6 \cdot 10^7$. 
  \begin{figure} \centering
   \includegraphics[width = 0.49\textwidth]{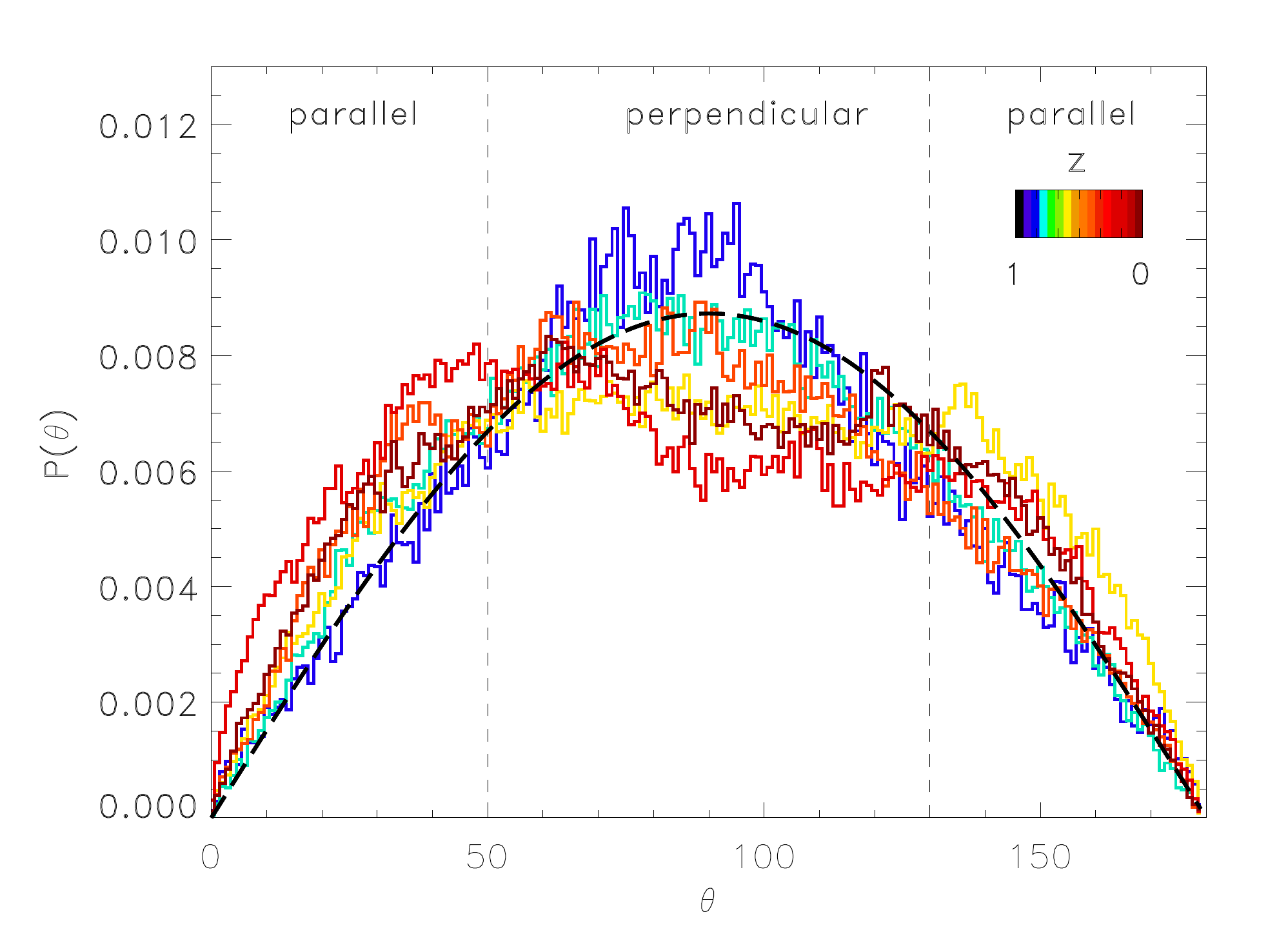}\label{subfig:tracA2_2_1}
   \caption{Distribution of pre-shock obliquities at different redshifts. The solid lines show the results from our simulation. The redshift is colour-coded going from black $z = 1$ to red $z = 0$. The dashed line shows the expected distribution of angles for a random distribution.}
   \label{fig:tracA2_2}
  \end{figure}
 \begin{figure*} \centering
  \subfigure[]{\includegraphics[width = 0.32\textwidth]{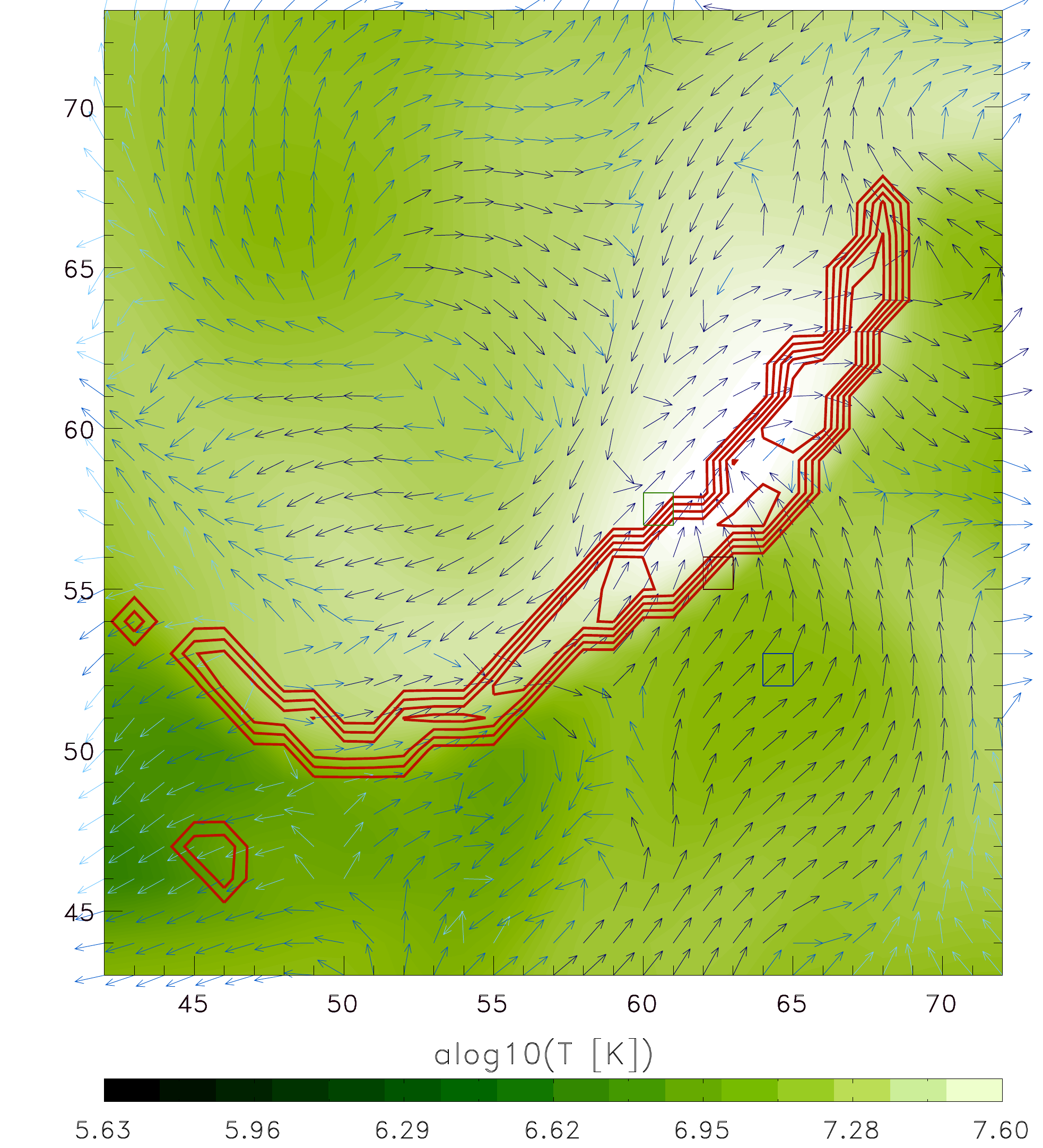} \label{subfig:tracA2_1_1}}
  \subfigure[]{\includegraphics[width = 0.32\textwidth]{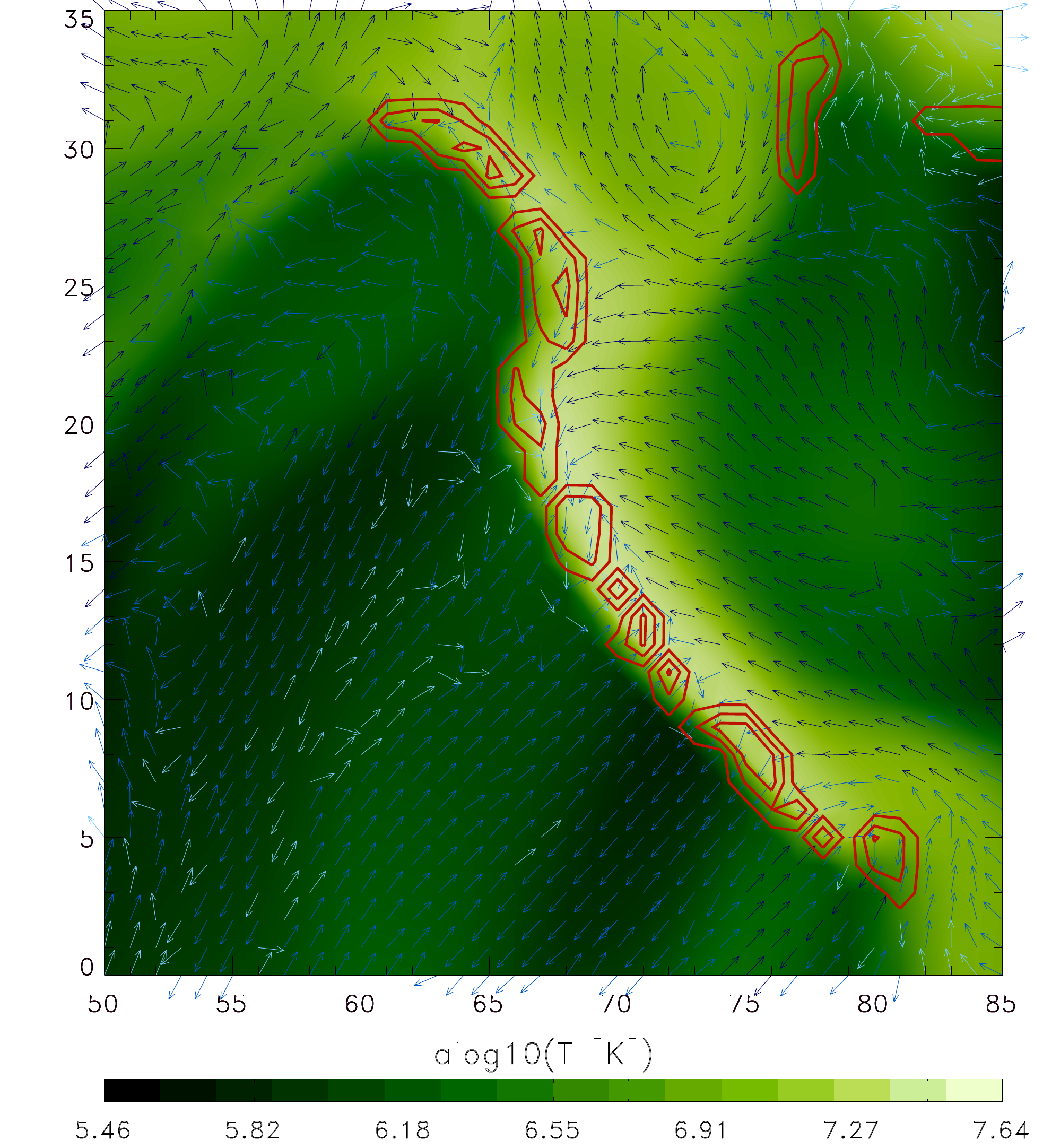} \label{subfig:tracA2_1_2}} 
  \subfigure[]{\includegraphics[width = 0.32\textwidth]{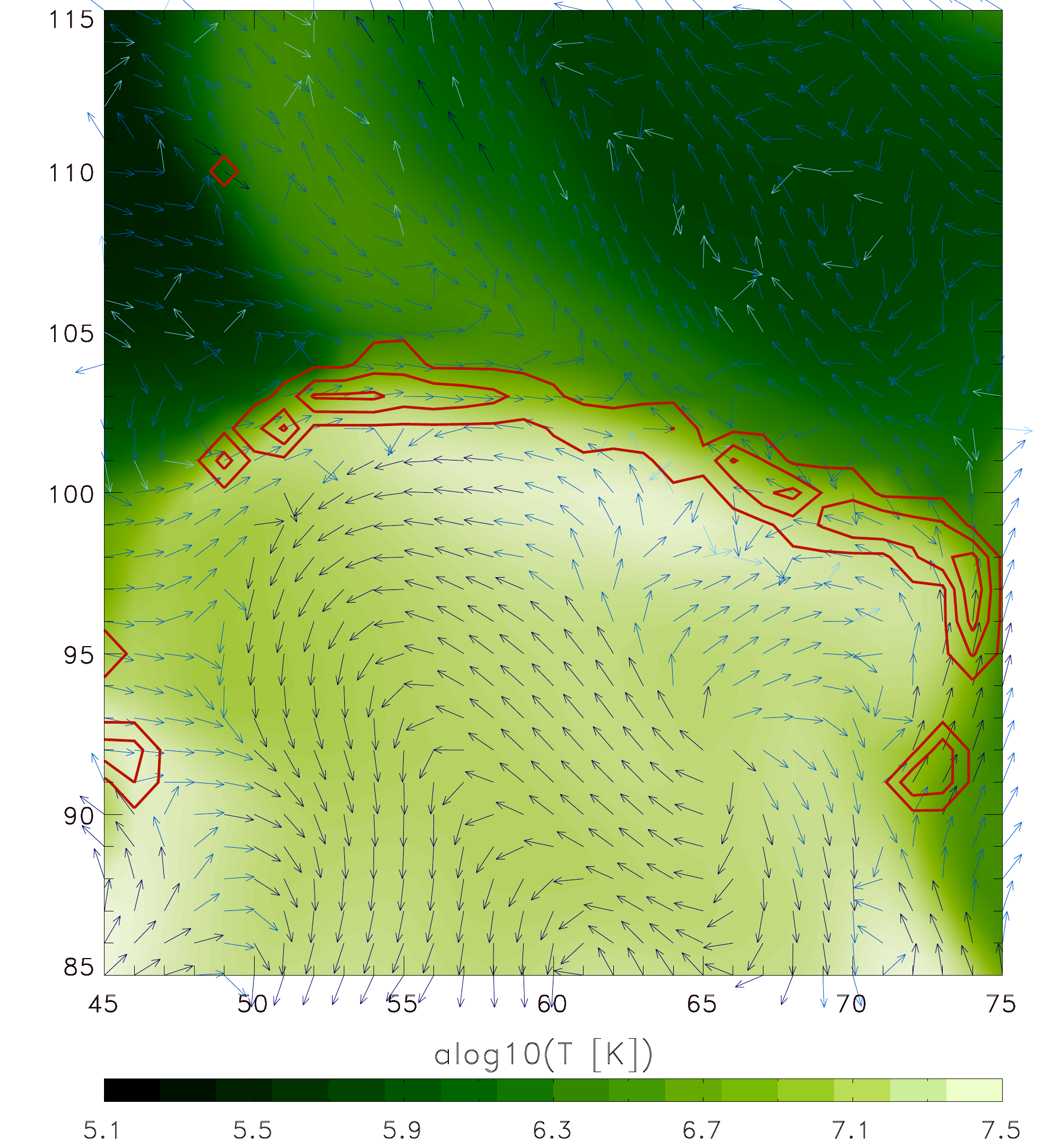} \label{subfig:tracA2_1_3}} \\ 
  \subfigure[]{\includegraphics[width = 0.32\textwidth]{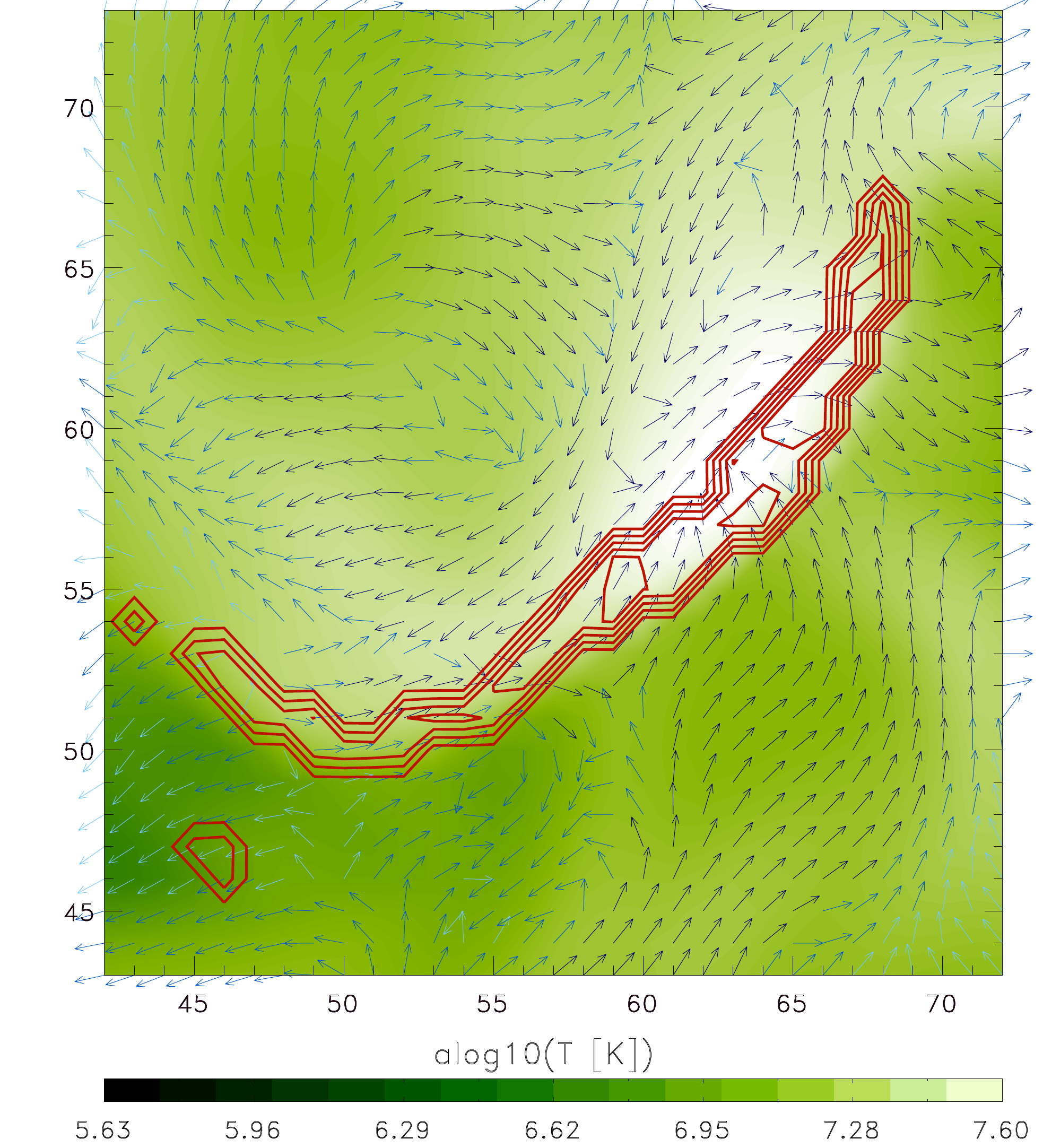} \label{subfig:tracA2_1_4}} 
  \subfigure[]{\includegraphics[width = 0.32\textwidth]{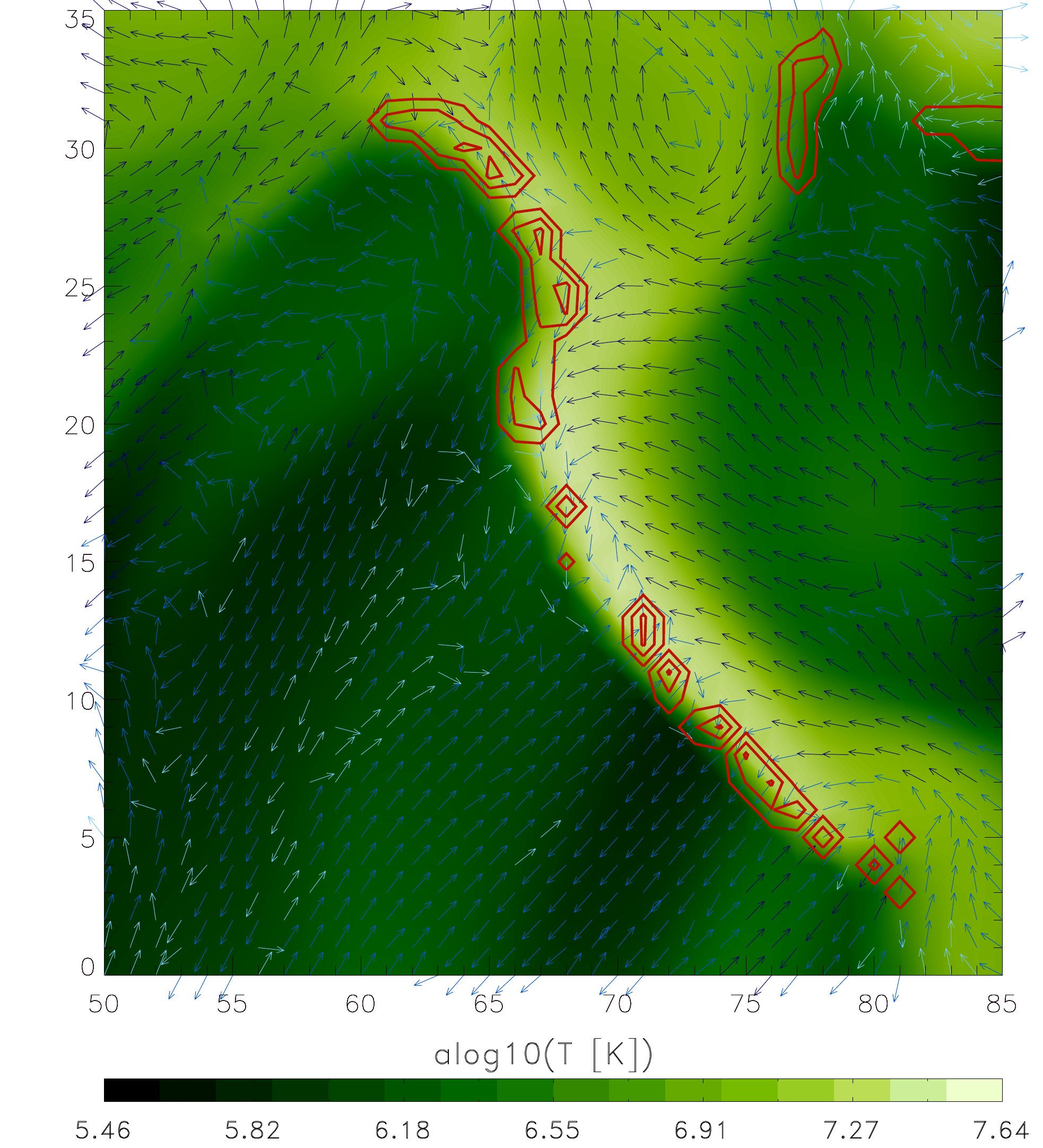}  \label{subfig:tracA2_1_5}}
  \subfigure[]{\includegraphics[width = 0.32\textwidth]{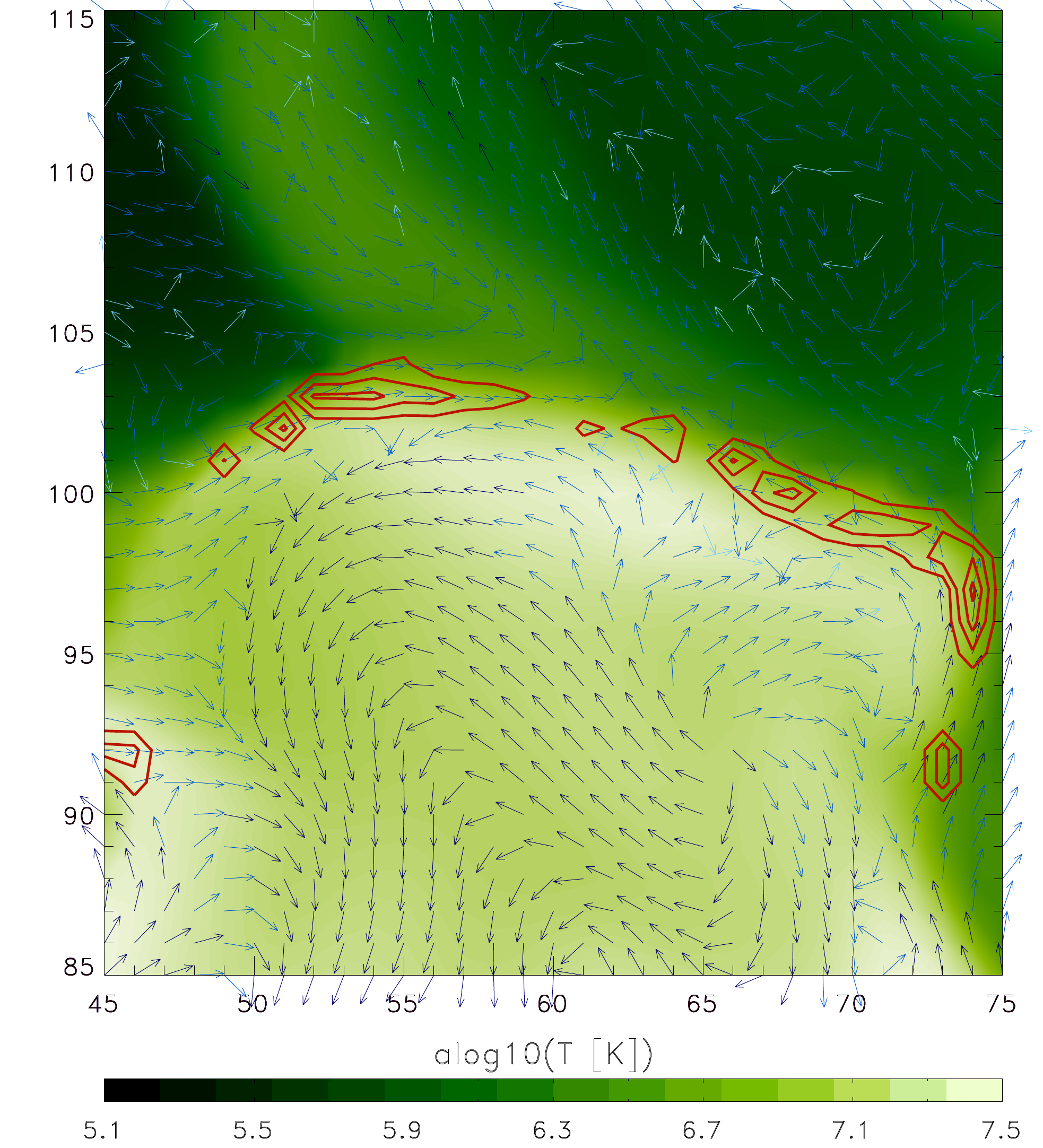} \label{subfig:tracA2_1_6}} 
   \caption{Zoomed versions of the three radio emitting regions are shown at $z \approx 0$. The colour shows the temperature of the ICM. The direction of the arrows indicates the direction of the magnetic and their colour gives their magnetic field strength in logarithmical units. The contours show the radio emission. Plots (a) and (d) show the central relic, plots (b) and (e) show the accretion shock and plots (c) and (f) show the filament. The top row ((a), (b) and (c)) shows the radio emission for $\theta_{\all}$. The bottom row ((d), (e) and (f)) shows the radio emission for $\theta_{\Perp}$.}
  \label{fig:tracA2_1}
 \end{figure*}
  \begin{figure} \centering
   \includegraphics[width = 0.49\textwidth]{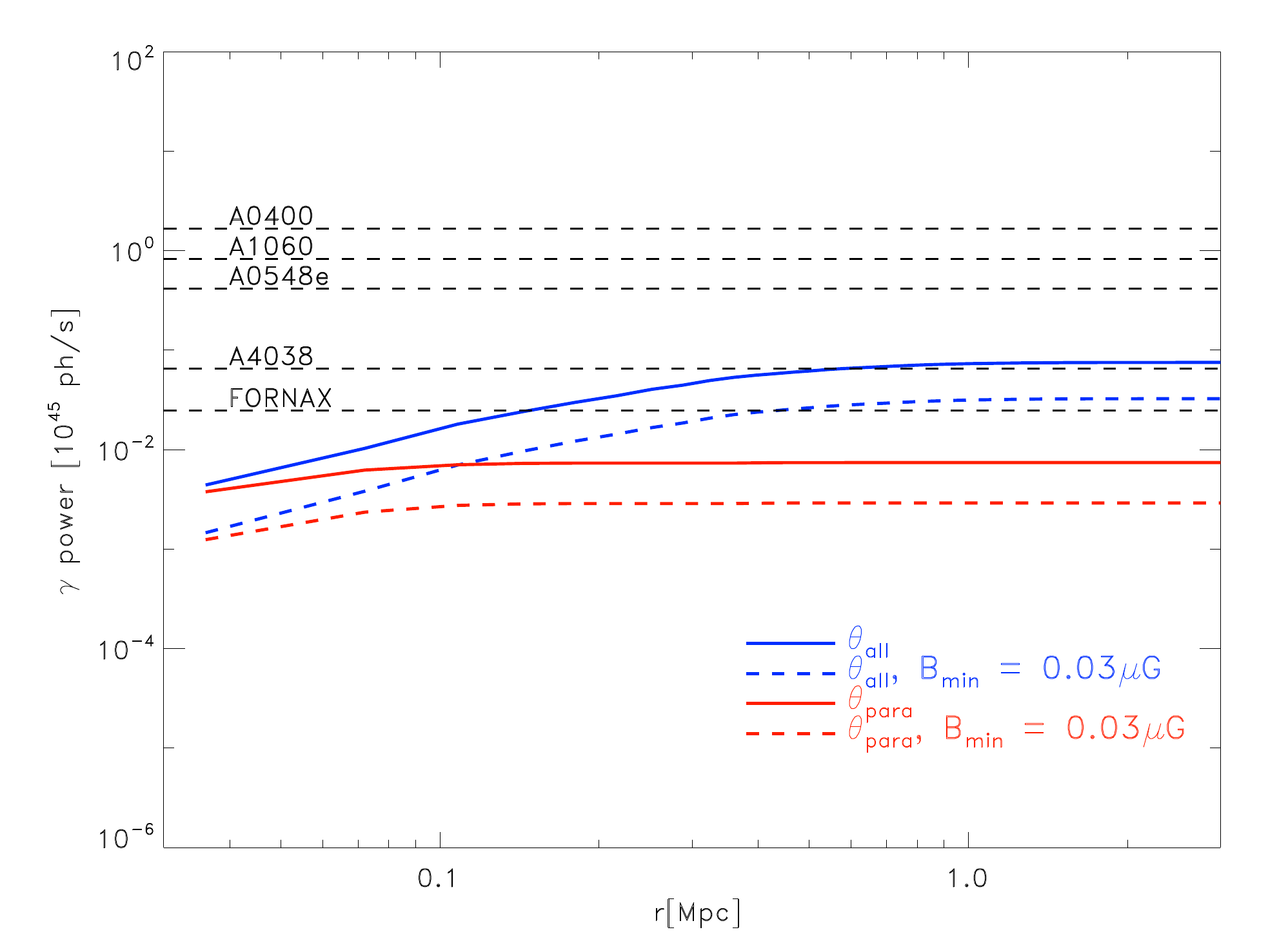}\label{subfig:tracA2_2_2}
   \caption{The $\gamma$-ray emission for all tracers (blue, solid line) and for the tracers that experienced quasi-parallel shocks (red, solid line). The dashed lines show the result for the additional requirement of a minimum magnetic field of $B_{\min} = 0.03 \ \mu \G$ to accelerate cosmic rays.}
   \label{fig:tracA2_3}
  \end{figure} \\
 The smaller size of the cluster allows us to analyse filaments and accretion shocks because the cluster is well contained in the support grid used to evolve tracers. Fig. \ref{fig:tracA2_2} shows the distribution of angles across time. Also in this object and at all epochs, we observe more quasi-perpendicular angles than quasi-parallel angles, in agreement with the results of the main paper. We also observe small excesses for quasi-parallel shocks. These are the result of a preferred parallel alignment of the magnetic field in filaments.  \\
 Following the approach described in Sec. \ref{subsec:RadioEmission} we investigated three regions of the cluster. We investigated a central relic (see Fig. \ref{subfig:tracA2_1_1} and Fig. \ref{subfig:tracA2_1_4}), a filament (see Fig. \ref{subfig:tracA2_1_2} and Fig. \ref{subfig:tracA2_1_5}) and an accretion shock (see Fig. \ref{subfig:tracA2_1_3} and Fig. \ref{subfig:tracA2_1_6}). As we see in Fig. \ref{fig:tracA2_1} for all three regions the bulk of radio emission caused by quasi-perpendicular shocks.  \\
 Similar to Sec. \ref{subsec:gammaEmission}, we computed the $\gamma$-ray emission from shock accelerated cosmic-ray protons. We compared the emission to real clusters\footnote{These clusters are: A0400, A1060, A0548e, A4038 and FORNAX.} with similar masses taken from \citet{2014ApJ78718A}. The $\gamma$-ray emission coming from $\theta_{\all}$ is above the limits of A4038 and FORNAX. The $\gamma$-ray emission drops by a factor of $ 2.3$ if only quasi-parallel shocks accelerate the cosmic rays. Finally, we added the requirement in the minimum magnetic field strength. For this object, a limiting magnetic field of $B_{\min} = 0.03 \mu \G$ is necessary to reduce the hadronic emission at the level of the \textit{Fermi}-limits. Using this additional switch the $\gamma$-ray emission is reduced by a factor of $ 10.1$ for $\theta_{\all}$ and by a factor $ 25.8$ for $\theta_{\para}$.  \\
 We conclude that the distribution of shock obliquities and the resulting effects on particle acceleration are rather similar in groups and galaxy clusters. \\ 
\section{Computing the $\gamma$-ray emission}\label{sec:gammaray}
 To compute the $\gamma$-ray emission we used the same approach as in \citet{2015MNRAS.451.2198V}, \citet{2010MNRAS.407.1565D} and \citet{2013A&A...560A..64H}. The total emission is given by the integral
 \begin{align}
  I_{\gamma}= \int\limits_r \lambda_{\gamma} \left( r \right) S \left( r \right) \dd r.
 \end{align}
 Has to be computed using the emission per unit of volume $\lambda_{\gamma}(r)$: 
 \begin{align}
  \lambda_{\gamma} (r) &= \int\limits_E \dd E_{\gamma} q_{\gamma} \left( E_{\gamma} \right) \nonumber \\
    &= \frac{\sigma_{\mathrm{pp}} m_{\pi} c^3}{3\alpha_{\gamma}\delta_{\gamma}}
       \frac{n_{th} K_p}{\alpha_{\mathrm{p}}-1}
       \frac{\left(E_{\min}\right)^{-\alpha_{\mathrm{p}}}}{2^{\alpha_{\gamma}-1}}
       \frac{E_{\min}}{\GeV} \label{eq:EmissPerUnitVol}\\
       &\ \ \ \ \ \times \left( \frac{m_{\pi_0} c^2}{\GeV} \right)^{-\alpha_{\gamma}}  \left[\mathcal{B_X} \left( \frac{\alpha_{\gamma}+1}{2\delta_{\gamma}} , \ \frac{\alpha_{\gamma}-1}{2\delta_{\gamma}} \right) \right]_{x_2}^{x_1} \nonumber.
 \end{align}
 Here $\mathcal{B_X} (a,b)$ is the incomplete $\beta$-function, $n_{\mathrm{th}}$ is the number density of the protons, $\alpha_{\mathrm{p}}$ and $\alpha_{\gamma}$ are the spectral index for the protons and the $\gamma$-rays, $\delta_{\gamma}  = 0.14 \alpha_{\gamma}^{-1.6}+0.44$ is the shape parameter, $E_{\min}$ is the proton energy threshold, $K_p$ is the normalization, $c$ is the speed of light, $m_{\pi}$ and $m_{\pi_0}$ are the pion-mass. For the effective cross-section $\sigma_{\mathrm{pp}}$ we used Eq. (79) given in \citet{2006PhRvD..74c4018K}
 \begin{align}
  \sigma_{\mathrm{pp}} (E) = &\left( 34.3 + 1.88L + 0.25L^2\right) \left[1- \left( \frac{E_{\mathrm{th}}}{E}\right)^2 \right]^4 \mathrm{mb} \nonumber \\
  &\mathrm{with} \ \ \ \ \ L = \ln \left(\frac{E}{1 \ \TeV} \right)  \\
  &\mathrm{and} \ \ \ \ \ E_{th} = m_p + 2 m_{\pi} + \frac{m_{\pi}^2}{2m_p} \sim 1.22 	\ \GeV \nonumber.
 \end{align}
 In the cross-section $E_th$ is the threshold energy of the production of a $\pi^0$-meson. $m_p$ and $m_{\pi}$ are the proton and pion masses.
\end{document}